\documentclass[twocolumn,english,superscriptaddress,showpacs,floatfix,aps,prb]{revtex4-1}
\usepackage[utf8]{inputenc}
\usepackage[english]{babel}
\usepackage{amsmath}
\usepackage{amssymb}
\usepackage{graphicx}
\usepackage[unicode=true, bookmarks=false, breaklinks=false, pdfborder={0 0 1},colorlinks=false]{hyperref}
\usepackage{breakurl}
\usepackage{natbib}

\makeatletter

\usepackage{dcolumn}
\usepackage{units}
\usepackage{color}
\usepackage{xcolor}
\usepackage{units}
\usepackage[normalem]{ulem}
\usepackage{verbatim}
\usepackage{url}
\usepackage{multirow}
\usepackage{float}
\usepackage{nicefrac}
\usepackage{tabularx}

\makeatother

\begin{document}

\title{Heisenberg and anisotropic exchange interactions in magnetic materials with correlated electronic structure and significant spin-orbit coupling}

\author{Vladislav Borisov}
\affiliation{Department of Physics and Astronomy, Uppsala University, Box 516, SE-75120 Uppsala, Sweden}
\email[Corresponding author:\ ]{vladislav.borisov@physics.uu.se}

\author{Yaroslav O. Kvashnin}
\affiliation{Department of Physics and Astronomy, Uppsala University, Box 516, SE-75120 Uppsala, Sweden}

\author{Nikolaos Ntallis}
\affiliation{Department of Physics and Astronomy, Uppsala University, Box 516, SE-75120 Uppsala, Sweden}

\author{Danny Thonig}
\affiliation{School of Science and Technology, \"Orebro University, SE-701 82, \"Orebro, Sweden}

\author{Patrik Thunstr\"om}
\affiliation{Department of Physics and Astronomy, Uppsala University, Box 516, SE-75120 Uppsala, Sweden}

\author{Manuel Pereiro}
\affiliation{Department of Physics and Astronomy, Uppsala University, Box 516, SE-75120 Uppsala, Sweden}

\author{Anders Bergman}
\affiliation{Department of Physics and Astronomy, Uppsala University, Box 516, SE-75120 Uppsala, Sweden}

\author{Erik Sjöqvist}
\affiliation{Department of Physics and Astronomy, Uppsala University, Box 516, SE-75120 Uppsala, Sweden}

\author{Anna Delin}
\affiliation{Department of Applied Physics, School of Engineering Sciences, KTH Royal Institute of Technology, AlbaNova University Center, SE-10691 Stockholm, Sweden}
\affiliation{Swedish e-Science Research Center (SeRC), KTH Royal Institute of Technology, SE-10044 Stockholm, Sweden}

\author{Lars Nordstr\"om}
\affiliation{Department of Physics and Astronomy, Uppsala University, Box 516, SE-75120 Uppsala, Sweden}

\author{Olle Eriksson}
\affiliation{Department of Physics and Astronomy, Uppsala University, Box 516, SE-75120 Uppsala, Sweden}
\affiliation{School of Science and Technology, \"Orebro University, SE-701 82, \"Orebro, Sweden}

\date{\today}
\begin{abstract}
The Dzyaloshinskii-Moriya (DM) interaction, as well as symmetric anisotropic exchange, are important ingredients for stabilizing topologically non-trivial magnetic textures, such as, e.g., skyrmions, merons and hopfions. These types of textures are currently in focus from a fundamental science perspective and they are also discussed in the context of future spintronics information technology. While the theoretical understanding of the Heisenberg exchange interactions is well developed, it is still a challenge to access, from first principles theory, the DM interaction as well as the symmetric anisotropic exchange, which both require a fully-relativistic treatment of the electronic structure, in magnetic systems where substantial electron-electron correlations are present. Here, we present results of a theoretical framework which allows to compute these interactions in any given system and demonstrate its performance for several selected cases, for both bulk and low-dimensional systems. We address several representative cases, including the bulk systems CoPt and FePt, the B20 compounds MnSi and FeGe as well as the low-dimensional transition metal bilayers Co/Pt(111) and Mn/W(001). The effect of electron-electron correlations is analyzed using dynamical mean-field theory on the level of the spin-polarized $T$-matrix + fluctuating exchange (SPTF) approximation, as regards the strength and character of the isotropic (Heisenberg) and anisotropic (DM) interactions in relation to the underlying electronic structure.  Our method can be combined with more advanced techniques for treating correlations, e.g., quantum Monte Carlo and exact diagonalization methods for the impurity solver of dynamical mean-field theory. We find that correlation-induced changes of the DM interaction can be rather significant, with up to five-fold modifications in the most distinctive case.
\end{abstract}

\maketitle

\section{Introduction}
Magnetic systems with topologically non-trivial spin textures, in particular, with individual skyrmions, antiskyrmions or periodic skyrmionic lattices as well as merons, antimerons and hopfions, attract growing attention due to their promising application potential in the next-generation spintronics devices such as, e.g., skyrmion-based logic circuits and skyrmion-based artificial synapses for neuromorphic computing.\cite{EverschorSitte2018,Song2020} Very often the formation of such magnetic textures relies on the presence of a sizeable antisymmetric magnetic exchange, also known as the Dzyaloshinskii-Moriya (DM) interaction, which competes with the usual Heisenberg interaction. As already demonstrated in the original paper of Dzyaloshinskii,\cite{Dzyaloshinsky1958} the DM interaction requires (i) broken space-inversion symmetry and (ii) non-zero spin-orbit coupling. In view of these criteria, it is very natural that considerable DM-induced magnetic effects are observed, for example, in thin transition metal films on the surfaces of heavy metals, such as in Co/Pt(111), Mn/W(110), Ir/Pt(111), Fe/Pt(111)\cite{Boettcher2012} and Fe/Ir(111).\cite{Heinze2011} On the other hand, there is a number of bulk systems with sizeable DM interaction, including the intensively studied B20 compounds\cite{Muehlbauer2009,Muenzer2010,Yu2011} (e.g., MnSi, Fe$_{1-x}$Co$_x$Si, and FeGe) which are the first known bulk systems with the skyrmionic behavior. In these B20 compounds, the chiral crystal structure breaks the inversion symmetry and, despite the absence of heavy elements, the DM interaction is surprisingly strong, in fact strong enough to induce a variety of non-collinear magnetic orders. The high interest in materials with topological magnetic structures motivates a search to widen the class of magnetic systems with a large DM interaction.

The search for new materials can be accelerated by means of high-throughput first-principles calculations within large structural databases.\cite{Curtarolo2013,Zunger2018} This allows to efficiently filter out less interesting structures and to obtain a ``shortlist'' of candidate systems with the desired quali\-ties, as has been exemplified by recent works focusing on the search for new permanent magnets.\cite{Vishina2020,Sanvito2017} Further analysis of the relations between the materials properties, the electronic-structure features, the details of the crystal structure and the chemical composition provides valuable information on general trends, which can significantly facilitate materials design and optimization.

Such systematic first-principles studies require state-of-the-art theoretical methods and techniques for describing accurately the complex quantum-mechanical behavior of electrons in the crystal lattice. In this respect, density functional theory (DFT)\cite{Hohenberg1964,Kohn1965} is a powerful and commonly used tool for studying the electronic structure of solid state systems. The combination of this theory with modern many-body techniques, e.g., dynamical mean-field theory,\cite{Georges1996,Kotliar2006} (DMFT) allows to capture the effect of electronic correlations which can dramatically change the overall behavior of the system. Application of these theories to magnetic systems can be further augmented by efficient techniques for extracting the magnetic interactions between the individual atoms. This implies a mapping of the total energy of the original electronic system onto a suitable Hamiltonian, for example, the Heisenberg spin model
\begin{equation}
	H = -\sum\limits_{i\neq j} J_{ij} \vec{e}_i \vec{e}_j,
	\label{e:isotropic_Heisenberg_model}
\end{equation}
where $J_{ij}$ is the isotropic magnetic exchange between spins on sites $i$ and $j$, and $\vec{e}_i$ is the unit vector describing the orientation of a classical spin on site $i$.

A well-known method to extract the $J_{ij}$ parameters is the approach suggested by Lichtenstein, Katsnelson, Antropov and Gubanov (LKAG approach).\cite{LKAG1987} It considers the total energy change, in the non-relativistic case, of the magnetic system when two atomic spins are infinitesimally rotated by angles $\pm\delta \theta$. From this energy change the Heisenberg exchange can be mapped out as
\begin{equation}
	J_{ij} = \frac{1}{4\pi} \mathrm{Im}\!\! \int\limits_{-\infty}^{E_\mathrm{F}} \mathrm{Tr} [\hat\Delta_i \hat G_{ij}^{\uparrow}(\varepsilon) \hat\Delta_j \hat G_{ji}^{\downarrow}(\varepsilon) ]\,\mathrm{d}\varepsilon.
	\label{e:LKAG_nonrel}
\end{equation}

The relevant quantities in this expression are the on-site spin splitting $\hat\Delta_i$ and the spin-dependent inter-site Green function $\hat G_{ij}^{\uparrow,\downarrow}$, which are both matrices in the orbital space and can be determined from the solution of the electronic structure problem. The trace in the above expression is calculated with respect to the orbital indices. Hence, the LKAG approach determines the relation between the electronic properties and the magnetic interactions. This approach has proved to be very efficient for studying magnetic interactions in a wide range of systems. However, the original LKAG formula (\ref{e:LKAG_nonrel}) can only provide information on the isotropic Heisenberg exchange parameters. In order to obtain information on more complex magnetic interactions, one has to generalize this approach to the relativistic case, which has been done already within the Korringa-Kohn-Rostoker (KKR) Green-function method.\cite{PhysRevB.68.104436,Ebert2009} Application of the generalized method to different magnetic systems provides a detailed picture of the anisotropic Heisenberg and DM interactions. Simulations of the magnetic properties based on the calculated exchange parameters show, in most cases, a good agreement with the measured equilibrium magnetic textures and the low-temperature magnon spectra (for a review, see Ref.~\onlinecite{Etz_2015}).

Despite this clear progress, little is known about the effect of electronic correlations on the character and size of the DM interaction. Most studies in the literature\cite{Grytsiuk2019} have addressed this issue using the DFT+$U$ method\cite{ggau1991} which takes into account only the static mean-field part of correlations. Furthermore, in some systems the dynamic nature of correlations makes a non-trivial contribution to the Heisenberg exchange, which can depend on subtle details of the electronic structure.\cite{PhysRevB.61.8906,Savrasov-jijDMFT,Kvashnin2015}

In the present work, we study the effect of dynamical correlations on the Heisenberg exchange, the DM interaction and the anisotropic but symmetric exchange, using DFT+DMFT for a few representative cases. The studied systems are the B20 skyrmion-host compounds MnSi and FeGe and transition metal monolayers on surfaces, Co/Pt(111) and Mn/W(001). In addition, we consider two simple bulk systems CoPt and FePt with a hypothetical structure where the emergence of the DM interaction, as a result of the inversion-symmetry breaking, and non-trivial correlation effects can be well illustrated. These model systems are constructed with the aim to have a simple structure with only one magnetic atom in the unit cell and are used in our work as test cases for the proposed theoretical framework. Our implementation of the generalized LKAG formula combined with DMFT is done in an all-electron, full-potential fully relativistic electronic structure software that uses linear muffin-tin orbitals as basis functions, as implemented in the RSPt electronic structure code. This software was first reported in Ref.~\onlinecite{Wills1987}, with subsequent, more complete descriptions of the implementation being published in Ref.~\onlinecite{Wills2000} and Ref.~\onlinecite{Wills2010}. The DMFT implementation in this method was first published in Refs.~\onlinecite{rspt-dmft} and \onlinecite{rspt-csc}. The implementation of exchange interaction was reported in Ref.~\onlinecite{Kvashnin2015} and DM interactions in Ref.~\onlinecite{Kvashnin2020}. Here, we use this method to analyze the magnetic interactions as function of the correlation strength.

\section{Methods}

\begin{figure}[h]
\centering
 \includegraphics[width=0.95\columnwidth]{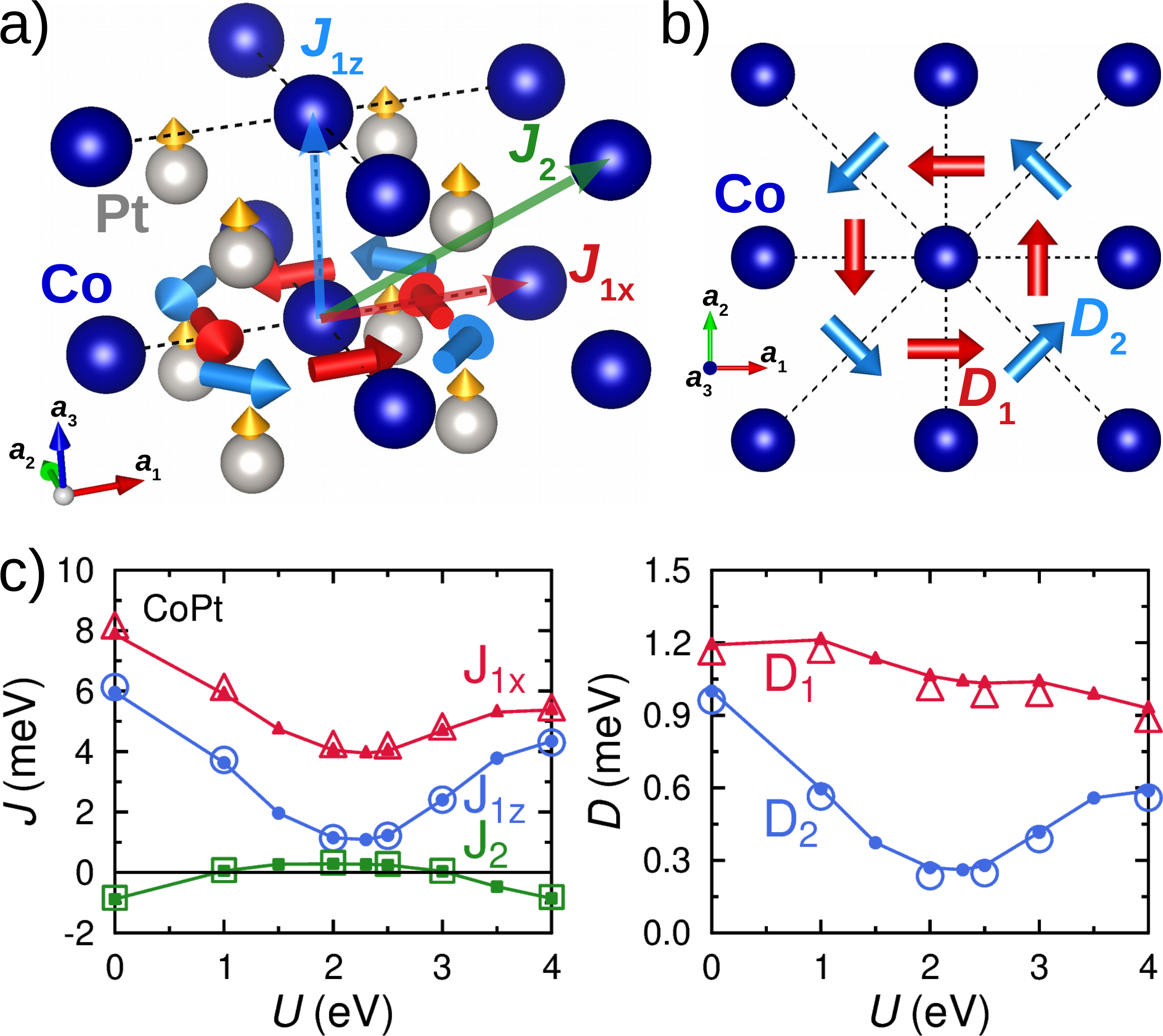}
 \caption{a) An illustration of DM interactions for the nearest ($D_1$, red arrows) and next-nearest ($D_2$, blue arrows) neighbor bonds in the CoPt compound, with a distorted CsCl structure and optimized lattice volume. The space-inversion symmetry is broken in the structure, by shifting the Pt site to the position $(\frac12, \frac12, \frac12 + \delta)$, where $\delta=0.1$ (Pt displacements are indicated by yellow arrows). b) Top view of the same structure (Pt atoms are omitted for clarity). c) The effect of electronic correlations with strength $U$ and $J_\text{H} = \unit[0.9]{eV}$ on the Heisenberg exchange and DM parameters for nearest (subscript ``1'') and next-nearest neighbors (subscript ``2''), i.e., $J_{1x}$, $J_{1z}$, $J_2$ and $D_1$ and $D_2$. Note that the calculations are performed on the DFT level of approximation (results for $U=\unit[0]{eV}$) as well as using DFT+DMFT within the SPTF approximation for a range of $U$ values (for details, see text). Results for the MTH (filled symbols) and L\"owdin projections (open symbols) are shown in comparison to each other.}
\label{f:CoPt}
\end{figure}

For all studied system, the exchange-correlation energy was approximated with the PBE parameterization\cite{PBE1996} of the generalized-gradient correction in density functional theory (DFT). In the following, we will refer to the data obtained using this setup as DFT results, which correspond to $U=\unit[0]{eV}$ in Figs.~\ref{f:CoPt}--\ref{f:B20_compounds}, \ref{f:Co_Pt111}, \ref{f:Mn_W001}. On top of this DFT approximation, correlation effects were accounted for using the relativistic version of the spin-polarized
$T$-matrix combined with fluctuating exchange (SPTF) approximation\cite{sptf-original,pourovski-sptf} approach within DMFT. This approach takes into account the frequency dependence of the self-energy, going beyond the DFT+$U$ approximation, but is computationally less demanding than the more accurate exact diagonalization and quantum Monte Carlo techniques. The details on the DMFT implementation used for the present calculations are given in Refs.~\onlinecite{rspt-dmft,rspt-csc}. Further calculation parameters, including the structural details for all systems, are given in Appendix~A.

In the next step, the energy of the electronic system is mapped onto a generalized Heisenberg model:
\begin{equation}
    H = -\sum\limits_{i\neq j} e_i^\alpha \hat J_{ij}^{\alpha\beta} e_j^\beta, \hspace{10pt} \alpha,\beta=x,y,z
    \label{e:general_Heisenberg_model}
\end{equation}
Here, the magnetic exchange parameters $\hat{J}_{ij}$ are $(3\times 3)$ tensors in the considered fully-relativistic case, which is a generalization of the $J_{ij}$ scalar parameters, that are relevant in the isotropic Heisenberg model (\ref{e:isotropic_Heisenberg_model}), derived for the non-relativistic case. In order to calculate all the components of these tensors for different site pairs $i-j$, we consider small perturbations of the initial magnetic state where all spins point along the same direction $\vec{e}_0$ defined by the common spin axis pointing along the $z$-axis. As a result, rotation of moments can be done along the $x$- and $y$-directions, and we obtain the following expressions for the relevant components of the magnetic exchange tensor\cite{PhysRevB.68.104436,Ebert2009,SECCHI201561,Kvashnin2020}:
\begin{align*}
    J_{ij}^{xx} = \frac{T}{4} \sum\limits_n \mathrm{Tr} \Big( &[\hat{H}_i + \hat{\Sigma}_i(i\omega_n), \hat\sigma^y] G_{ij}(i\omega_n) \times \\
    & [\hat{H}_j + \hat{\Sigma}_j(i\omega_n), \hat\sigma^y] G_{ji}(i\omega_n) \Big),\\
    J_{ij}^{yy} = \frac{T}{4} \sum\limits_n \mathrm{Tr} \Big( & [\hat{H}_i  + \hat{\Sigma}_i(i\omega_n), \hat\sigma^x] G_{ij}(i\omega_n) \times\\
    & [\hat{H}_j  + \hat{\Sigma}_j(i\omega_n), \hat\sigma^x] G_{ji}(i\omega_n) \Big),
\end{align*}
and
\begin{align*}
    J_{ij}^{xy} = -\frac{T}{4} \sum\limits_n \mathrm{Tr} \Big( & [\hat{H}_i + \hat{\Sigma}_i(i\omega_n), \hat\sigma^y] G_{ij}(i\omega_n) \times \\
    & [\hat{H}_j + \hat{\Sigma}_j(i\omega_n), \hat\sigma^x] G_{ji}(i\omega_n) \Big).
\end{align*}
In the equations above, $T$ stands for the electronic temperature and the summation runs over different Matsubara frequencies $\omega_n = 2\pi T(n + 1)$ (following Ref.~\onlinecite{PhysRevB.61.8906}), for which inter-site Green function, $G_{ij}(i\omega_n)$, is evaluated using DMFT. 
Square brackets denote commutators, $\sigma$ are the Pauli matrices, and $\hat H_{i}$ and $\hat \Sigma_{i}(i\omega)$ are the site-projected local Hamiltonian and self-energy, respectively.
Note that the trace in the expressions above is performed over spin-orbitals.

For a crystal structure without symmetry operations that bring the $x$-, $y$- or $z$-axis into each other, the $J_{ij}$ tensor in Eq.~\eqref{e:general_Heisenberg_model} may lack symmetry also. In this case one needs to evaluate all components, $J_{ij}^{\alpha\beta}$ for $\alpha,\beta=x,y,z$. This is done similarly to the description presented above, but with a rotated axis of the original spin configuration\cite{PhysRevB.68.104436}. For instance,  with the spin axis pointing in $x$ direction, we obtain $J_{ij}^{yy}$,$J_{ij}^{zz}$, $J_{ij}^{yz}$ and $J_{ij}^{zy}$ components.

Based on the calculated $J_{ij}$ tensor, the symmetric and antisymmetric anisotropic interactions can be evaluated. Since the latter can be identified with DM interactions represented by vectors $\vec{D}$, it is common to specify its individual vector component with a sub- or superscript, and we write:
\begin{equation}
    D_{ij}^z = \frac12 (J_{ij}^{xy} - J_{ij}^{yx}). \label{e:DM_definition}
\end{equation}
The symmetric anisotropic interaction is commonly represented by:
\begin{equation}
	C_{ij}^z = \frac12 (J_{ij}^{xy} + J_{ij}^{yx}). 
\label{e:C_definition}
\end{equation}
and similar expressions for the other components. In contrast to the DM interaction, the symmetric anisotropic interaction in general does not transform as a vector. The exchange interactions in the studied systems are obtained for many neighbor shells of interacting spins and show, in general, a long-range character (more details in Appendix~B). For the discussion of correlations effects on these exchange parameters, we provide also the information on the electronic structure calculated on the DFT and DFT+DMFT levels (Appendix~C). Also, the discussion of correlation-induced changes of the magnetic moments in the studied systems is provided in Appendix D. While in the main text we will focus on the Heisenberg and DM interactions, the details of the less significant symmetric anisotropic exchange $C_{ij}$ defined by Eq.~(\ref{e:C_definition}) are given in Appendix~E.

The calculation of the exchange parameters involves the evaluation of local quantities, such as inter-site Green functions and spin splittings, which requires a proper definition of the local site-specific orbital projection. In this work, we consider the muffin-tin-head (MTH) and the L\"owdin orthonormalized (ORT) projections\cite{rspt-dmft}. Following the procedure outlined in Ref.~\onlinecite{Kvashnin2015}, we employ the MTH projection scheme to define the effective impurity problem, but investigate the impact of using either the MTH or the multiple-$\kappa$ ORT projection scheme in the calculations of the magnetic exchange tensor. In particular, Figs.~\ref{f:CoPt}--\ref{f:B20_compounds}, \ref{f:Co_Pt111}, \ref{f:Mn_W001} show the exchange parameters for the MTH (filled symbols) and L\"owdin projections (open symbols), and it is apparent that, in most cases, the two projection schemes produce very similar results and general trends for the magnetic exchange interactions.

In the non-relativistic case, each magnetic exchange interaction (\ref{e:LKAG_nonrel}) can be decomposed into different orbital components. For this, the on-site spin-splitting matrix $\Delta_i$ must be diagonalized in the orbital space which requires the definition of a new orbital basis (for details, see, e.g., Ref.~\onlinecite{PhysRevLett.116.217202,Cardias2017}). The latter reflects the symmetry of the crystal field. In that new basis, the magnetic exchange reads:
\begin{equation}
J_{ij} = \sum_{m_1,m_2} J^{m_1 m_2}_{ij}
\end{equation}
where
\begin{equation}
	J_{ij}^{m_1 m_2} = \frac{T}{4}\sum_n \Delta_i^{m_1}  G_{ij}^{\uparrow m_1 m_2} (i\omega_n) \Delta_j^{m_2} G_{ji}^{\downarrow m_1 m_2}(i\omega_n).
	\label{e:LKAG_nonrel_orbital}
\end{equation}

This decomposition makes it possible to understand the impact of different electronic states on the magnetic interactions and provides a more complete physical picture, as discussed, e.g., in recent works.\cite{Cardias2017,Nomoto2020} Due to the origin of the expression in Eq.~(\ref{e:LKAG_nonrel_orbital}), the orbital decomposition can be done only in the non-relativistic case, i.e., for the Heisenberg interactions but not for the DM interactions. In this work, the orbital-resolved Heisenberg exchange will be briefly described for selected system on the non-relativistic DFT and DFT+DMFT level, in order to show a general relation to the electronic structure, and a detailed analysis is left for a future study. Summing up all the different orbital contributions $J_{ij}^{m_1 m_2}$ in Eq.~(\ref{e:LKAG_nonrel_orbital}) gives the total value of the magnetic exchange between the considered sites $i$ and $j$ that can be also calculated from (\ref{e:LKAG_nonrel}). We note that, in case of correlated systems, the spin splitting, $\hat{\Delta}_i $, in Eq.~(\ref{e:LKAG_nonrel_orbital}) includes also the self-energy contribution.

Using the calculated exchange parameters, both isotropic and anisotropic, and  magnetic moments we apply the Monte Carlo technique with the Metropolis algorithm, as implemented in the \textsc{UppASD} code,\cite{uppasd,Eriksson2017} in order to determine the magnetic ground state for selected systems. These simulations are based on the Hamiltonian (\ref{e:general_Heisenberg_model}) with an additional term $ \sum\limits_{i} K_i (\vec{e}_i \cdot \vec{n}_i)^2 $ describing the on-site anisotropy, where the summation is done over different spins oriented along unit vectors $\vec{e}_i$, $K_i$ is the magnetic anisotropy energy per atom and $\vec{n}_i$ is the easy axis direction. The anisotropy is estimated based on the one-shot relativistic calculations for magnetization pointing along the $x$, $y$ and $z$ directions, which are run on top of the self-consistent non-relativistic electronic structure. The anisotropy is characterized then by the difference in the sum of the electronic eigenvalues for the three aforementioned cases. It should be noted that this approach, which is based on the force theorem, often is a more accurate way of determining the magnetic anisotropy, compared to relativistic total energies. An annealing protocol is used for the calculation of the magnetic ground state. Comparison with the measured magnetic properties or theoretical works for some of the studied systems would indicate, whether the theoretical results for the magnetic interactions are reliable.

\section{Bulk systems}
In this section, we present numerical results for a selection of bulk materials: CoPt, FePt and the B20 compounds MnSi and FeGe. We focus both on the Heisenberg exchange and on the DM interaction, where the latter is commonly discussed to be finite only in systems with a substantial spin-orbit coupling in combination with a broken space-inversion symmetry. Although more general considerations that involve non-collinear magnetic configurations have been discussed,\cite{Cardias2020} we study here bulk materials that are primarily collinear, have a large spin-orbit coupling and a crystal structure which breaks space-inversion symmetry. We find that the symmetric anisotropic exchange in general appears to be  considerably smaller than the DM interactions. The only exception is the CoPt system, where this exchange is comparable to the DM interaction.

\subsection{CoPt} The first example we discuss is a tetragonal CoPt ordered compound where the initial CsCl-type structure is distorted by shifting Pt away from the high-symmetry position (see Fig.~\ref{f:CoPt} and Appendix~A). This structural distortion breaks the space-inversion symmetry, thereby allowing a non-zero DM interaction. The magnetism in this system is dominated by Co with a moment of $\unit[1.91]{\mu_\text{B}}$, while the induced Pt moment is $\unit[0.34]{\mu_\text{B}}$. In contrast to Co, the Pt sites also show a sizeable orbital moment of $\unit[0.10]{\mu_\text{B}}$. These values are obtained for the magnetic moments pointing along the $z$ direction. When the moments are oriented along the $x$ direction, their magnitude is almost the same: $m(\text{Co}) = \unit[1.91]{\mu_\text{B}}$, $m(\text{Pt}) = \unit[0.35]{\mu_\text{B}}$ and the Pt orbital moment becomes $\unit[0.12]{\mu_\text{B}}$. When electronic correlations are included using DFT+DMFT, the spin and orbital Co moments mostly increase and the induced Pt moments decrease (see Appendix~D). We note that the total energies of the ferromagnetic configurations with Co moments oriented along the $x$ and $z$ directions differ by an unusually large value $E_x - E_z \sim\!\unit[-3.3]{meV}$ (for calculations with $U=\unit[0]{eV}$). Interestingly, the magnetic anisotropy shows significant non-monotonic variations as a function of the correlation strength $U$ (data not shown) and can even change its sign for $U\geq\unit[3.5]{eV}$, indicating a transition from an easy-plane to an easy-axis anisotropy, which may be of paramount importance for permanent magnet applications.

Using the full relativistic generalization of the LKAG formula, the exchange parameters for different atomic neighbors can be calculated and the results indicate a long-range character of the Heisenberg and DM interactions in this metallic CoPt system. More information on the distance-dependence of the exchange parameters in this and other studied systems can be found in Appendix~B. Here, we concentrate on the largest interactions that show interesting trends. We start our analysis by a discussion of results on the DFT level of approximation, i.e., data obtained without Hubbard-$U$ corrections ($U=\unit[0]{eV}$ in Fig.~\ref{f:CoPt}c). We start by a technical comment, that the two projection schemes give consistent values, clearly a rewarding finding. Secondly, we note that the nearest-neighbor (NN) exchange $J_1$ is ferromagnetic (FM) while there is a weaker antiferromagnetic (AFM) exchange $J_2$ between the next-nearest neighbors. Furthermore, the structural distortion in this system creates a large anisotropy $J_{1x}/J_{1z}=1.3$ in the nearest-neighbor Heisenberg exchange parameters within the $xy$-plane ($J_{1x}$) and along the $z$ direction ($J_{1z}$) (Fig.~\ref{f:CoPt}). Further neighbors can show an even larger anisotropy ratio.

An orbital-resolved analysis (see Appendix F) indicates that all orbitals, except for the $d_{z^2}$, contribute significantly to the in-plane exchange interaction $J_{1x}$, while the out-of-plane exchange $J_{1z}$ is due to the $d_{xz}/d_{yz}$ and $d_{z^2}$ orbitals. In orbital space, $J_{1z}$ is fully diagonal and, for the $J_{1x}$, the off-diagonal contributions, such as $d_{x^2-y^2} \to d_{z^2}$, are relatively small. Furthermore, the DM interaction is of the order of $10\%$ of the Heisenberg exchange. This is an unusually large value that is due to the very strong structural distortion and the presence of a heavy element Pt, that induces a large spin-orbit coupling. The DM vectors lie fully in the $xy$-plane (Fig.~\ref{f:CoPt}b) and in the directions expected, based on the Moriya symmetry rules.\cite{Moriya1960}

Next, we study the influence of dynamical electronic correlations on the strength of the exchange interactions discussed above. As shown in Fig.~\ref{f:CoPt}c, the anisotropy between the Heisenberg exchange along the $xy$-plane and $z$ directions is maintained, but both interactions show a pronounced minimum around $U = \unit[2.3]{eV}$, which enhances significantly the anisotropy ratio $J_{1x}/J_{1z}$ from 1.3 (for $U=\unit[0]{eV}$) to 3.6 (for $U=\unit[2.3]{eV}$). The overall variations of the Heisenberg exchange are rather large, as $U$ is modified: $\sim\!50\%$ for $J_{1x}$ and $\sim\!82\%$ for $J_{1z}$. In case of $J_{1x}$, the $d_{xy}$ orbital contribution shows the largest decrease due to correlations, while the suppression of $J_{1z}$ is mostly due to the $d_{z^2}$ contribution. In the aforementioned region of $U$ values, the next-nearest neighbor (NNN) interaction $J_2$ even changes the sign. Similar anomalies are observed for the DM interactions in this system, especially for the NNN DM parameter $D_2$, which decreases by 74\% for $U = \unit[2.3]{eV}$ compared to the case $U=\unit[0]{eV}$, while $D_1$ is reduced only by 13\%. Further increase of correlations up to $U = \unit[4.0]{eV}$ reduces the anisotropy $J_{1x}/J_{1z}$ down to 1.2 and enhances considerably the NNN DM interaction $D_2$ making it comparable with the NN $D_1$ interaction, similarly to the case $U=\unit[0]{eV}$.

These results indicate that the competition between different kinds of magnetic interactions in this system is strongly affected by electronic correlations. Considerable frustration of the Heisenberg interactions and the presence of sizable DM interaction can produce interesting non-trivial magnetic states in the studied CoPt system. For that reason, it is worth investigating the possibility of synthesizing alloys with similar structure as we have considered here, either in the bulk or thin-film form. The modification of the exchange parameters with increasing value of $U$ is accompanied by a modification of the band dispersion, as discussed in Appendix~C. In Appendix F we provide a more detailed account of the trends exhibited by $J$ and $D$ as functions of strength of the Coulomb repulsion, $U$ (Fig.~\ref{f:CoPt}c).

\begin{figure}[h]
\centering
 \includegraphics[width=0.95\columnwidth]{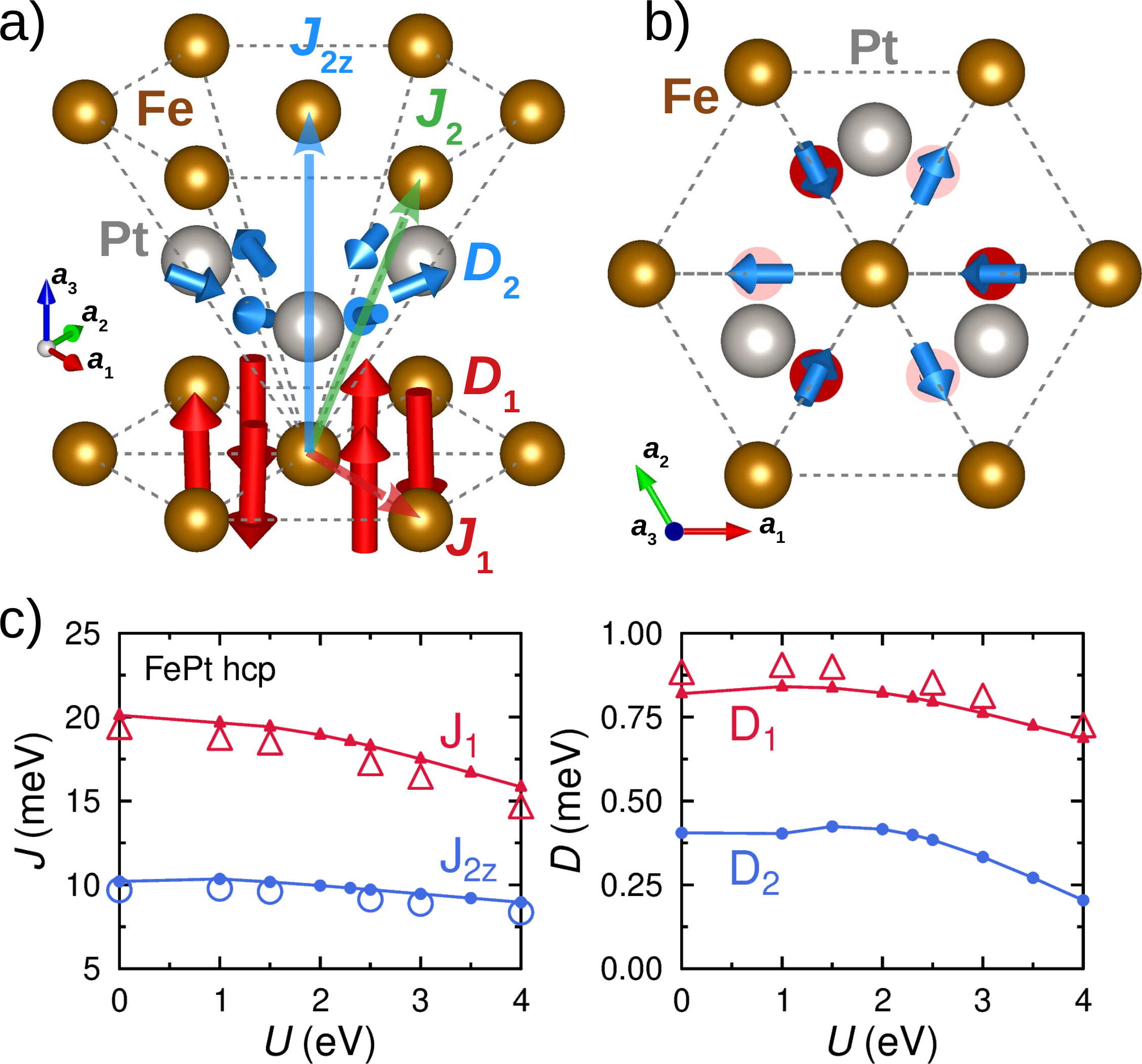} \hspace{5pt}
 \caption{a) Illustration of DM interactions for the nearest ($D_1$, red arrows) and next-nearest ($D_2$, blue arrows) neighbor bonds in the FePt compound with broken space-inversion symmetry and optimized lattice volume (see text for details).
 b) Top view of the same structure with DM vectors in each Fe plane shown by shaded circles and DM vectors between the Fe planes shown by arrows.
 c) The influence of electronic correlations with strength $U$ and $J_\text{H} = \unit[0.9]{eV}$ on the Heisenberg and DM exchange parameters obtained from DFT+DMFT (with SPTF).
 Results for the MTH (filled symbols) and L\"owdin projections (open symbols) are shown in comparison to each other.}
\label{f:FePt}
\end{figure}

\subsection{FePt} Another bulk system that we consider here is FePt. We have chosen to investigate this system as an ordered compound in a hypothetical hexagonal crystal structure (see Appendix~A), where the space-inversion symmetry is broken by the hexagonal crystal structure, with alternating Fe and Pt layers. In the experimentally observed L1$_0$ structure, the space-inversion is not broken, and this system is less interesting for studies of DM interactions. While magnetism resides mostly in the Fe layers ($m_s(\text{Fe})=\unit[2.87]{\mu_\text{B}}$), Pt has a large spin-orbit coupling, which is crucial for the DM interaction, and a smaller induced spin moment $\sim\unit[0.30]{\mu_\text{B}}$. The spin moments of both atoms are only slightly affected by the magnetization direction, similarly to the case of CoPt discussed above. The orbital moments, however, show larger variations, with the maximal value $\unit[0.09]{\mu_\text{B}}$ observed for the Fe moments along the $z$ direction. Similarly to CoPt, electronic correlations enhance the Fe spin moments and suppress the Pt induced moments (see Appendix~D). We note here, however, that the Fe orbital moments reveal non-monotonic variations as functions of $U$ with a maximum around $U=\unit[2.3]{eV}$. Another observation is the effect of correlations on the magnetic anisotropy. Our results for the total energies along the $x$-, $y$- and $z$-directions indicate that small correlations with $U$ around 1\,eV can switch the anisotropy from the easy-plane to the easy-axis type and significantly enhance the anisotropy energy scale.

The calculated nearest-neighbor Heisenberg exchange $J_1$ is shown in Fig.~\ref{f:FePt}c, and it may be noted that pure DFT calculations ($U=\unit[0]{eV}$) suggest a FM coupling between the atoms in each Fe layer and between the neighboring Fe layers, i.e., both $J_{1}$ and $J_{2z}$ are positive. Due to the metallic character of this system, the magnetic interactions have a long-range character, as discussed in Appendix~B. The orbital decomposition (data not shown) of the nearest-neighbor in-plane exchange interaction, $J_1$ along [100]-direction, reveals dominant contributions of the $d_{x^2-y^2}$ and $d_{xz}$ orbitals. The ferromagnetic exchange, $J_{2z}$, between the layers (see Fig.~\ref{f:FePt}a) involves the $d_{yz} \to d_{x^2-y^2}$ and $d_{xz} \to d_{xy}$ couplings, while the exchange between orbitals of the same kind, e.g., $d_{xy} \to d_{xy}$, is rather weak. For the in-plane nearest-neighbor bonds, the calculated DM vectors are almost fully out-of-plane, while the deviation from the $z$ direction is around $1/30$ which can be related to symmetry breaking effects due to the chosen magnetization direction. For the interlayer DM coupling $D_2$, the DM vectors are almost lying in-plane but have a non-vanishing $z$ component, which is not restricted by the crystal symmetry.

\begin{figure*}
\centering
 \includegraphics[width=0.95\columnwidth]{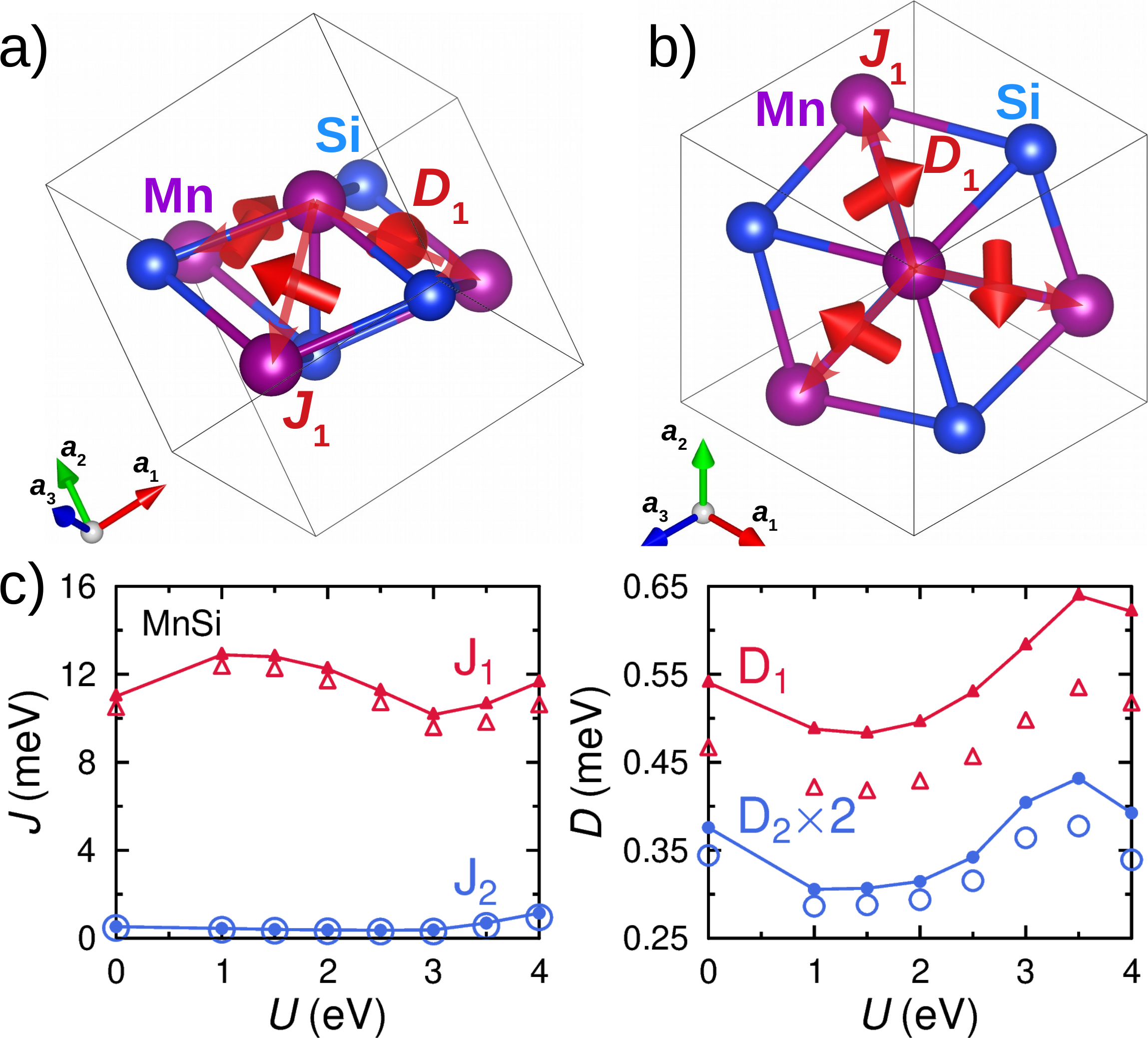} \hspace{10pt}
 \includegraphics[width=0.95\columnwidth]{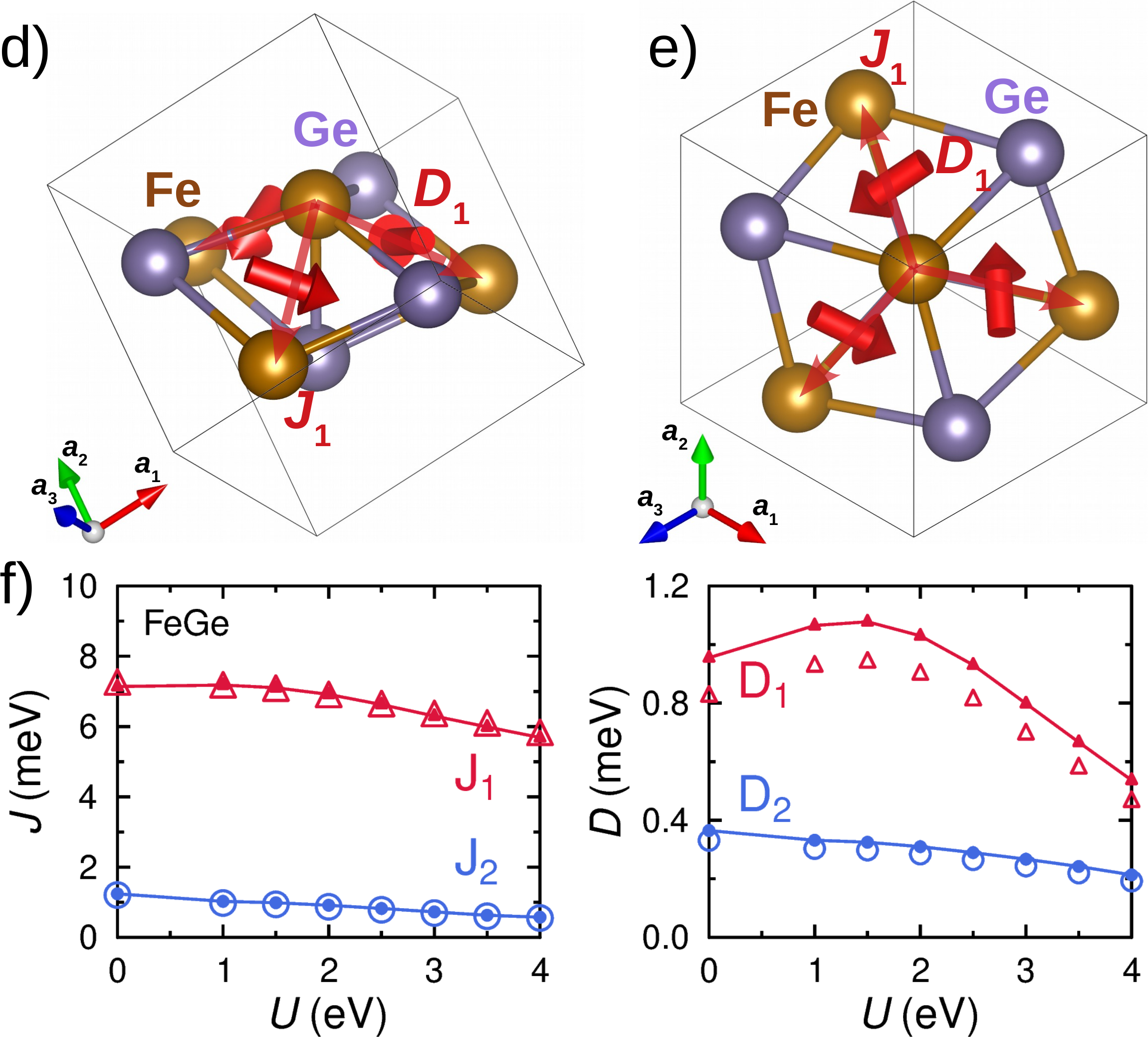}
 \caption{a,d) Illustration of DM interactions ($D_1$, red arrows) for the nearest-neighbor bonds in the B20 compounds MnSi and FeGe.
 b,e) View of the same B20 structures showing the $C_{3v}$ symmetry of the DM vectors. For the sake of better presentation, the crystal structure is shifted by $(\frac12, \frac12, \frac12)$ with respect to the unit cell boundary in plots a, b, d and e.
 c,f) The effect of electronic correlations with strength $U$ and $J_\text{H} = 0.9\:\text{eV}$ on the Heisenberg nearest neighbor ($J_1$) and next-nearest neighbor ($J_2$) and corresponding DM exchange parameters $D_1$ and $D_2$, obtained from DFT+DMFT.
 Results for the MTH (filled symbols) and L\"owdin projections (open symbols) are shown in comparison to each other.
 }
\label{f:B20_compounds}
\end{figure*}
The exchange parameters $J_1$ and $J_{2z}$ decrease basically monotonically by $\sim\!20\%$ and $\sim\!15\%$, with increasing correlation strength $U$ in the DFT+DMFT treatment, where the $U$ parameter is varied from $0$ to $\unit[4]{eV}$ (Fig.~\ref{f:FePt}c). We note that there is a small maximum for $J_{2z}$ at $U=\unit[1]{eV}$. The reduction of $J_1$ is mostly due to the dominating $d_{x^2-y^2}$ and $d_{xz}$ orbital contributions, and, in case of $J_{2z}$, all off-diagonal orbital transitions are suppressed by the same amount. Similar correlation induced effects are observed for the nearest-neighbor DM interactions, $D_1$ and $D_2$, where the dominant $D_1$ parameter decreases by $\sim\!10\%$ while the smaller $D_2$-parameter decreases by $\sim\!40\%$. Here, the $D_1$ parameter shows a small maximum at $\unit[1]{eV}$ and $D_2$ reveals a more pronounced maximum at $\unit[1.5]{eV}$. The $D_1$, in-plane interaction, amounts in this compound to $3-4\%$ of the Heisenberg exchange $J_1$, while $D_2$ is almost $40\%$ of the exchange $J_2$ across the layers. This is partially due to the small value of $J_2 \sim \!\unit[0.8]{meV}$. Also, the calculated DM vectors show the correct $D_{3h}$ symmetry of the crystal lattice (Fig.~\ref{f:FePt}a,b), which is characterized by the existence of a six-fold roto-inversion axis.

We note that the heat of formation for the two bulk systems CoPt and FePt discussed above is positive, even though we have analyzed hypothetical crystal structures. A positive heat of formation is one of the necessary conditions for structural stability. Another condition~-- positive phonon frequencies~-- is, however, not analyzed in this work. Although the possibility of experimental realization of these materials is unknown, the two examples illustrate the emergence of the DM interaction in different scenarios of broken space-inversion symmetry and represent useful test cases for the developed technique of extracting the exchange parameters from first principles.

\subsection{B20 compounds} 
The final example of bulk materials that we describe are the B20 compounds MnSi and FeGe. These systems are well-known materials hosting magnetic skyrmions.\cite{Muehlbauer2009,Yu2011} They have a chiral crystal structure (see Fig.~\ref{f:B20_compounds}a,d), which has, in general, no space-inversion symmetry and hence allows, in principle, a non-zero DM interaction.

Our calculations for the nearest and next-nearest interactions are presented in Fig.~\ref{f:B20_compounds} and again we note that the calculated parameters depend only weakly on the projection scheme. The biggest sensitivity is found for the $D_1$ parameter of MnSi. Furthermore, our results of the long-ranged interactions are shown in Appendix~B and they, together with the data in Fig.~\ref{f:B20_compounds}, suggest that both B20 compounds investigated here are characterized by rather short-ranged Heisenberg interactions, where the nearest neighbors are coupled ferromagnetically. Despite the small value of the exchange interactions for further neighbors, they have to be included, for example, in the simulations of the Curie temperature or for calculations of the spin stiffness constant. While the Heisenberg exchange is stronger in MnSi, larger DM interaction is obtained for the FeGe system. Overall, this translates to $\nicefrac{D_1}{J_1}$ ratios of $4.9\%$ for MnSi and $13.4\%$ for FeGe (for $U=\unit[0]{eV}$), suggesting different length scales of the helical order in these two systems. Indeed, the spatial period of the magnetic helical order observed in experiments is found to be larger in FeGe ($\unit[700]{\AA}$) than in MnSi ($\unit[190]{\AA}$). From the theory presented in Fig.~\ref{f:B20_compounds}, FeGe has a larger $\nicefrac{D_1}{J_1}$ ratio than MnSi. This indicates that the simple model of the helical state where the spatial period is inversely proportional to the $\nicefrac{D_1}{J_1}$ ratio is inaccurate for the B20 compounds, where interactions between further atomic shells, i.e., $J_2$ and $D_2$ etc., are important.
 Also, one may expect significant biquadratic exchange $\sim\!(\vec{S}_i \cdot \vec{S}_j)^2$ for these systems, as our preliminary data suggest (not shown here).

We note that our values for the Heisenberg exchange in FeGe (at $U=\unit[0]{eV}$) are similar to the recently reported theoretical values,\cite{Grytsiuk2019} while the nearest-neighbor DM exchange parameter ($D_1$) in our calculations is approximately three times bigger than that reported in Ref.~\onlinecite{Grytsiuk2019}.
From the calculated interatomic interactions we can also evaluate the micromagnetic parameters $A$ and $D$, i.e., the exchange ($A$) and spiralization ($D$) related to the Heisenberg and DM interactions, respectively, using the following expressions:\cite{Grytsiuk2019}
\begin{eqnarray}
A=\frac{1}{2}\sum_{i\neq j}J_{ij}R_{ij}\otimes R_{ij} \label{microA}\\
D=\sum_{i\neq j}D_{ij}\otimes R_{ij}\label{microD}
\end{eqnarray}

From Eqs.~(\ref{microA}--\ref{microD}), we obtain the parameters $A=\unit[7.9]{pJ/m}$ and $D=\unit[0.18]{mJ/m^2}$, which can be used to construct a continuum model for FeGe. The exchange stiffness agrees well with previous theoretical works based on density functional theory (DFT). However, all currently available theoretical estimates based on DFT of the parameter $A$ exceed the measured values (which are below $\unit[1.37]{pJ/m}$), at least, by a factor of $\sim$six. As Fig.~\ref{f:B20_compounds}c,f shows, dynamical correlations in form of DMFT do not significantly improve on the situation, and for $U=\unit[2.3]{eV}$ we obtain a slightly reduced value $A=\unit[7.1]{pJ/m}$ due to weaker interatomic interactions. On the other hand, our DFT estimate of the spiralization parameter $D=\unit[0.18]{mJ/m^2}$ is in a decent agreement with the experimental value of $\unit[0.11]{mJ/m^2}$, compared to other theoretical studies which report larger values. When dynamical correlations are included with $U=\unit[2.3]{eV}$, the theoretical value becomes smaller $D=\unit[0.11]{mJ/m^2}$ and the agreement with experiment is good.

Electronic correlations show unexpected effects in these chiral systems, according to our DFT+DMFT calculations. Many of the dominant exchange parameters reach a local maximum or minimum for correlation strength $U_0 = 1.5\:\text{eV}$. Above this value ($U > U_0$), the Heisenberg interaction $J_1$ (Fig.~\ref{f:B20_compounds}c,f) decreases monotonically as expected. On the other hand, the DM interactions $D_1$ and $D_2$ follow opposite trends in MnSi and FeGe, being enhanced by correlations with $U>1.5\:\text{eV}$ in MnSi and weakened in FeGe. The overall variations of the magnetic exchange interactions due to correlations with $U=(0-4)\:\text{eV}$ are around 20--40\% for the nearest-neighbor bonds ($J_1$ and $D_1$) and around 40-55\% for the next-nearest neighbors ($J_2$ and $D_2$).

Electronic correlations are shown in Appendix~C to influence the energy dispersion differently, where, in particular, the life-time effects and the imaginary part of the self-energy are probably more significant for MnSi, which can be related to the fact that MnSi is closer to the half-filled limit than FeGe. Also, the $3d$-orbital occupation matrices of Mn and Fe are different, which is reflected in the spin magnetic moments $m(\text{Mn})=\unit[1.05]{\mu_\text{B}}$ and $m(\text{Fe})=\unit[1.19]{\mu_\text{B}}$, that react only slightly but differently to electronic correlations, except for the case of MnSi with $U\leq \unit[3]{eV}$. We note that the obtained value of the total moment of FeGe, $~\unit[1.1]{\mu_\mathrm{B}/\mathrm{f.u.}}$, including the negative interstitial contribution, is somewhat larger than the measured value of $\unit[0.98]{\mu_\mathrm{B}/\mathrm{f.u.}}$\cite{Spencer2018} Similarly to previous theory studies,\cite{Jeong2004,Hortamani2008} our calculations overestimate the total moment of MnSi compared to the experimental value of $\unit[0.4]{\mu_\mathrm{B}/\mathrm{f.u.}}$,\cite{Ziebeck1982} possibly suggesting that spin fluctuations, as discussed in Ref.~\onlinecite{Ziebeck1982}, are more important in this system than in FeGe, which is also noted in Ref.~\onlinecite{Grytsiuk2019}). In Ref.~\onlinecite{Ziebeck1982}, neutron scattering measurements suggest a rather complicated picture for MnSi, in which the diffuse scattering suggests a Mn moment of $\unit[0.83]{\mu_\mathrm{B}}$, which is close to the calculated values obtained in this work.

\section{Low-dimensional systems}
In this section, we present numerical results for a selection of transition metal overlayers supported on a substrate. The choice is motivated by systems that have been realized experimentally in the past and where both Heisenberg and DM interactions are of interest. These types of systems have, in general, no inversion symmetry and the DM interaction is expected to be relevant almost for any system of this geometry. We focus here on two substrates from the 5$d$ transition metal series, Pt and W. These metals produce a considerable spin-orbit coupling due to their large atomic numbers. From experiments,\cite{Corredor2017,Ferriani2008} it is known that Co can grow as a single monolayer on the Pt(111) surface, while monolayer Mn can be deposited on a W(001) surface. Hence, the magnetic interactions in these two systems are studied in detail in this work.

\subsection{Monolayer of Co on Pt(111)} 
\begin{figure}
\centering
 \includegraphics[width=0.99\columnwidth]{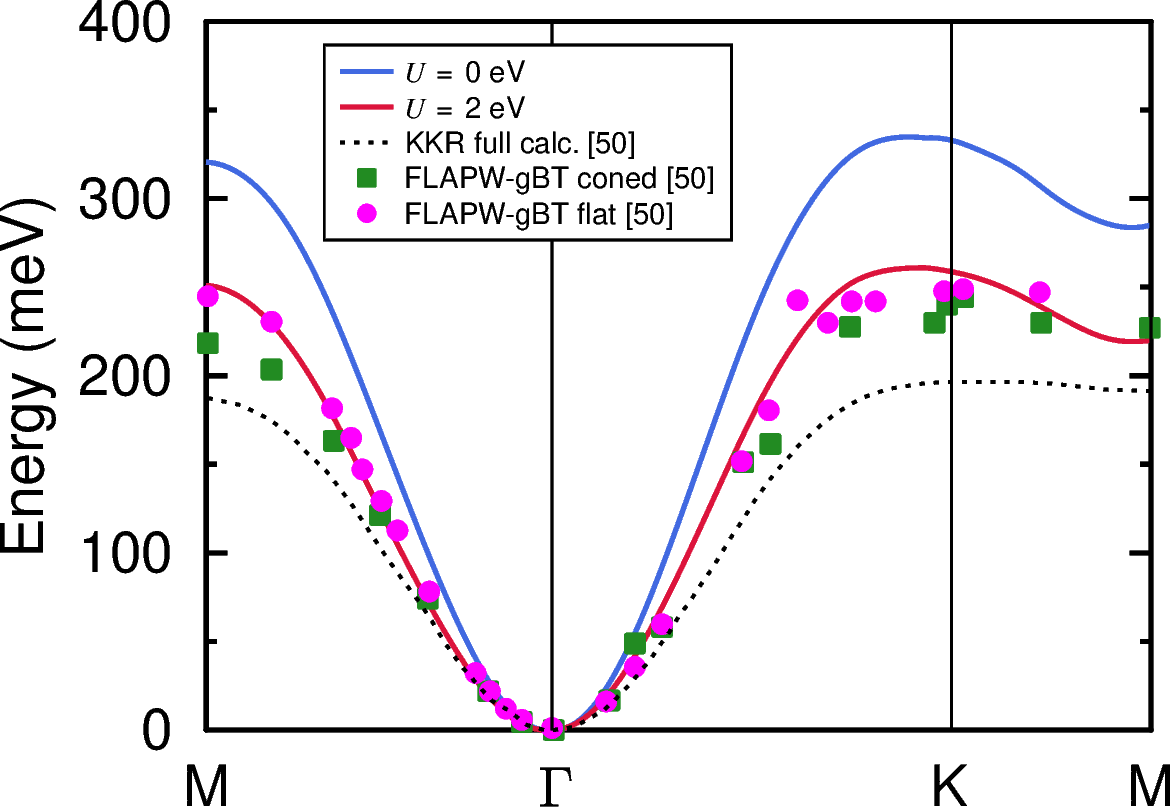}
 \caption{Adiabatic magnon spectra for the Co monolayer on the Pt(111) surface. Solid lines refer to the interactions calculated in this work for $U=\unit[0]{eV}$ (blue) and $U=\unit[2]{eV}$ (red). For comparison we also show the calculated spectra taken from Ref.~\onlinecite{Zimmermann2019}.
 }
\label{f:Co_Pt_magnon_spectra} 
\end{figure}

For Co/Pt(111), our DFT calculations reveal large Co moments ($\unit[2.04]{\mu_\text{B}}$) and induced moments on Pt in the topmost surface layer ($\unit[0.33]{\mu_\text{B}}$). The Co moments are slightly enhanced by dynamical correlations up to $\unit[2.09]{\mu_\text{B}}$ for $U=\unit[4]{eV}$ while the Pt moments are reduced down to $\unit[0.26]{\mu_\text{B}}$. The calculated total energies suggest an easy-plane anisotropy with Co moments lying within the $xy$-plane parallel to the surface, while the angular shape and the magnitude of the in-plane anisotropy strongly depend on the correlation strength $U$ (data not shown). For the eigenvalue sums $E'$ from the one-shot relativistic calculations for the magnetic configurations with $\vec{M}\parallel x$, $\vec{M}\parallel y$ and $\vec{M}\parallel z$ we find that $E'_x - E'_y = \unit[0.00238]{meV/Co}$ and $E'_z - E'_y = \unit[0.102]{meV/Co}$ for $U=0$ and $E'_x - E'_y = \unit[0.00278]{meV/Co}$ and $E'_z - E'_y = \unit[0.244]{meV/Co}$ for $U=\unit[2]{eV}$. These values are used in the following for calculating the magnon spectra (Fig.~\ref{f:Co_Pt_magnon_spectra}) which will be discussed later.

\begin{figure}
\centering
 \includegraphics[width=0.95\columnwidth]{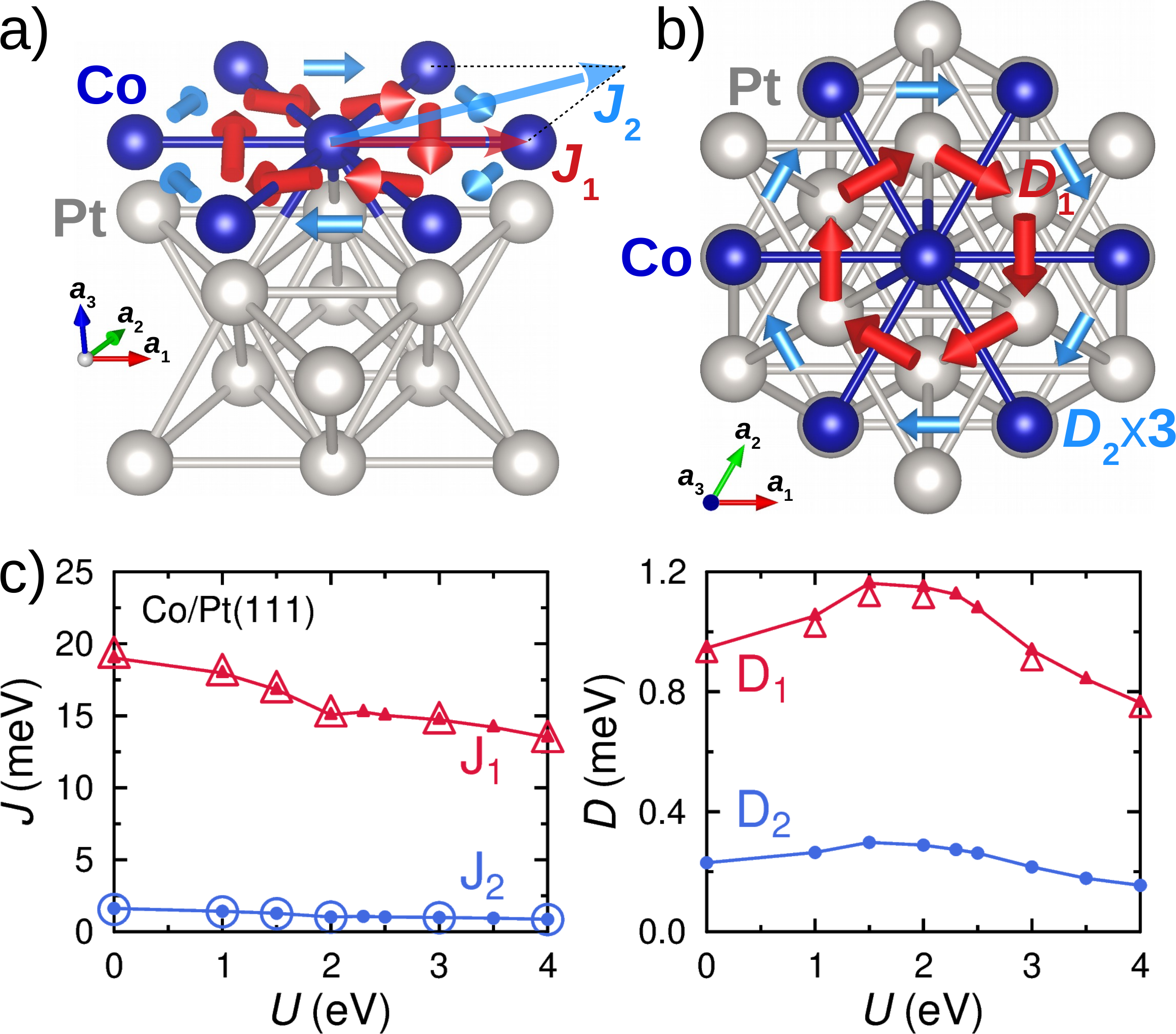}
 \caption{a) Illustration of DM interactions for the nearest ($D_1$, red arrows) and next-nearest ($D_2$, blue arrows) neighbor bonds in the Co monolayer on the Pt(111) surface. Overall $C_6$ symmetry of the in-plane component of the DM vectors is consistent with the lattice symmetry.
 b) Top view of the surface and substrate. The length of the DM vectors are drawn to indicate the size of the interaction, although in plots a) and b), the $D_2$ vectors are increased by the factor of three for clarity.
 c) The effect of electronic correlations with strength $U$ and $J_\text{H} = 0.9\:\text{eV}$ on the Heisenberg $J_1$ and $J_2$ and DM exchange parameters $D_1$ and $D_2$ obtained from DFT+DMFT.
 Results for the MTH (filled symbols) and L\"owdin projections (open symbols) are shown in comparison to each other.
 }
\label{f:Co_Pt111} 
\end{figure}

On the DFT level, we find strong ferromagnetic interactions between the nearest neighbors ($J_1\sim\unit[18.9]{meV}$) within the Co layer and much weaker interactions for more distant neighbors (see Figs.~\ref{f:Co_Pt111} and \ref{f:3d_5d_vs_R}). Orbital-resolved analysis (data not shown) suggests that $J_1$, e.g., along [100], is dominated by the exchange between the same orbitals from the manifold $d_{xz}$, $d_{xy}$ and $d_{x^2-y^2}$. The exchange, $J_1$, along other directions would also contain the admixture of the $d_{yz}$ orbitals which have the same spin splitting as the $d_{xz}$ orbitals. Relatively large DM interaction, $D_1\sim\unit[0.9]{meV}$, is obtained for the nearest-neighbor bonds (see Fig.~\ref{f:Co_Pt111}b,c) and it follows the symmetry of the underlying crystal structure. Also, the DM vectors $\vec{D}_1$ are canted away from the (001)-plane, which is a common behavior for transition metal films on (111) surfaces, since there is no horizontal mirror plane in (001), in contrast to the CoPt bulk case. Our DFT values for the exchange parameters, $J_1$ and $D_1$, are somewhat smaller than the previously reported values obtained using spin-spiral FLAPW calculations\cite{Dupe2014,Zimmermann2019}, PAW supercell approach\cite{Yang2015} and KKR method.\cite{Simon2018,Zimmermann2019} Interestingly, the micromagnetic DM interaction has been calculated for this system within the Berry phase approach.\cite{Freimuth2014,Freimuth2017} Scanning electron microscopy studies\cite{Corredor2017} suggest that the existing theoretical estimates of the DM interaction are consistent with the measured domain wall angle and width, and the values reported from weakly correlated electronic structure calculations ($U=\unit[0]{eV}$) are consistent with these experiments.

The electronic correlations, as revealed by our DMFT calculations, significantly increase the out-of-plane component of the nearest-neighbor DM interaction ($D_1$), leading to a sizeable rotation of the corresponding DM vectors. The canting angle $\alpha_\text{DM}$ describing the orientation of these vectors with respect to the Co plane equals $21^\circ$ at $U=\unit[0]{eV}$, a value that increases at first with the Hubbard $U$, reaching the maximum value of $49^\circ$ for $U=\unit[2]{eV}$. For larger $U$-values the angle decreases monotonously down to $29^\circ$ for $U=\unit[4]{eV}$. Note that, for $U=\unit[(2.0-2.3)]{eV}$, the out-of-plane component of $D_1$ becomes bigger than the in-plane one ($\alpha_\text{DM}>45^\circ$), while the $\nicefrac{D_1}{J_1}$ ratio increases visibly from 5.0\% ($U=\unit[0]{eV}$) to 7.9\% ($U=\unit[2]{eV}$), demonstrating a strong influence of correlations on the magnetic interactions. At the same time, the DM vectors for the next-nearest neighbors ($D_2$) have an almost vanishing out-of-plane component ($\sim\!\unit[10^{-5}]{meV}$) due to symmetry restrictions, independently of the correlation strength. Both interactions $D_1$ and $D_2$ reach a maximum at $U =\unit[1.5]{eV}$ (Fig.~\ref{f:Co_Pt111}c), similarly to the B20 compounds (Fig.~\ref{f:B20_compounds}c,\,f) discussed in the previous section. The difference between the two cases, however, is that the Heisenberg exchange parameters in the Co/Pt(111) system show a monotonous decrease as a function of $U$, where the calculated variations of the $J_1$ and $J_2$ parameters equal $29\%$ and $46\%$ for $U=\unit[(0-4)]{eV}$. For all values of $U$, we find calculated $J$- and $D$-parameters that are rather insensitive to the projection scheme.

In order to validate our calculations, we also calculate the adiabatic magnon spectra\cite{Halilov_1997} for two different values of U ($U=\unit[0]{eV}$, $U=\unit[2]{eV}$). The calculated spectra are shown in Fig.~\ref{f:Co_Pt_magnon_spectra}, and can be compared to previous theoretical  calculations, obtained from different methods, taken from Ref.~\onlinecite{Zimmermann2019}. We note that by introducing correlations, i.e., $U\neq0$, the magnon dispersion becomes softer, with a weaker $k$-dependence. This result is in a better agreement with other reported methods\cite{Zimmermann2019}, where in particular the agreement with the FLAPW results is quite good. These FLAPW results did not include dynamic correlations, but were based on the Vosko-Wilk-Nusair parametrisation\cite{Vosko1980} of the local density approximation. The $U=0$ results from the present set of calculations were based on the generalized gradient approximation of the energy functional, and it is possible that the difference observed between the $U=0$ results presented here and the data of Ref.~\onlinecite{Zimmermann2019} are due to the difference in energy functional, since the structural parameters for Co/Pt(111) in our work and Ref.~\onlinecite{Zimmermann2019} are very similar (see Appendix~A).

In Appendix~C, we discuss the influence of correlations on the electronic structure. As discussed, the influence is rather strong for the Co/Pt(111) system, with correlation-induced features emerging at the Fermi level, and with significant life-time broadening at higher binding energies.

\begin{figure}
\centering
 \includegraphics[width=0.99\columnwidth]{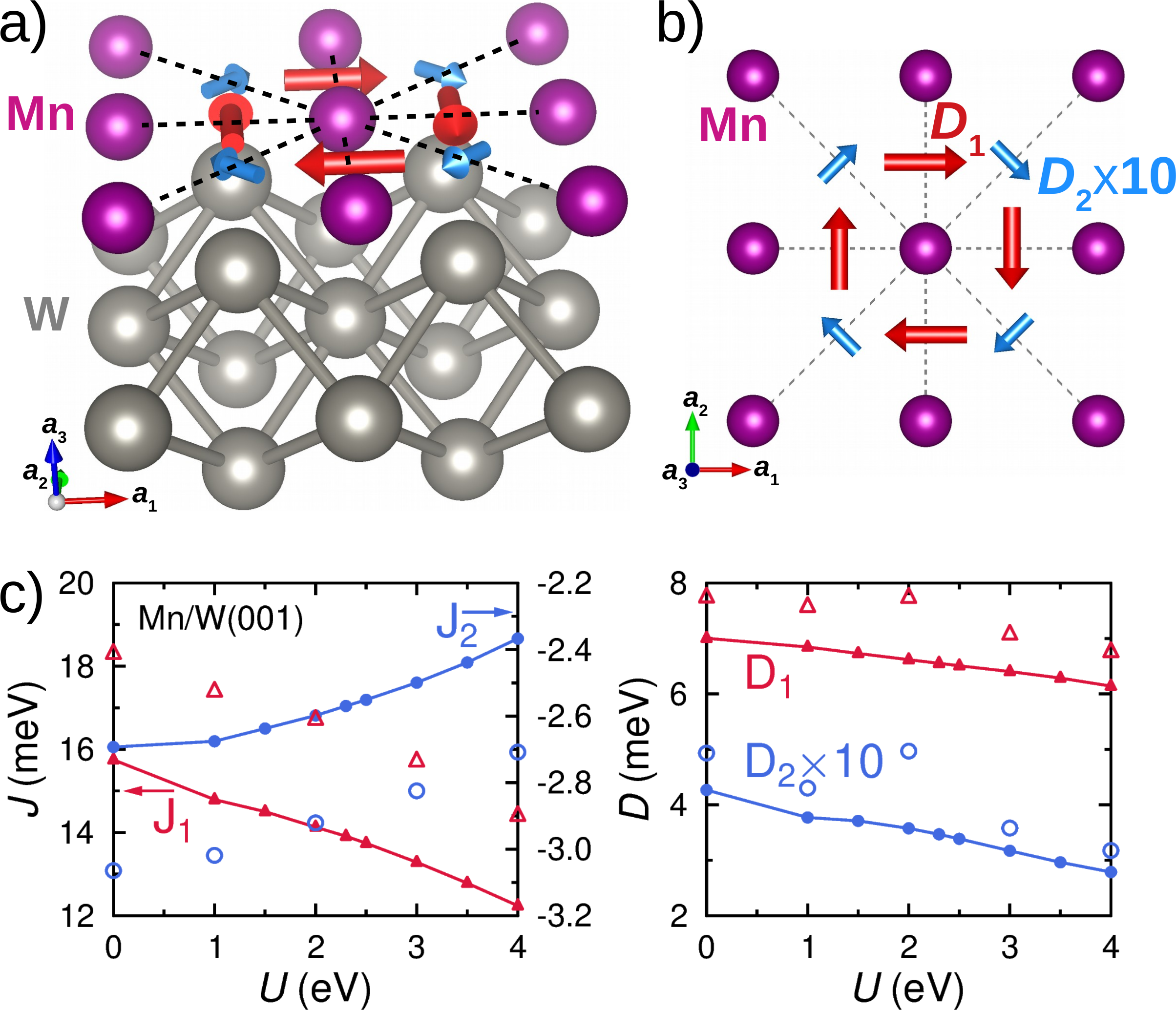}
 \caption{a) Illustration of DM interactions for the nearest ($D_1$, red arrows) and next-nearest ($D_2$, blue arrows) neighbor bonds in the Mn monolayer on the W(001) surface. Overall $C_4$ symmetry of the DM vectors is consistent with the lattice symmetry.
 b) Top view of the surface and substrate structure (W atoms are omitted for clarity).
 c) The effect of electronic correlations with strength $U$ and $J_\text{H} = \unit[0.9]{eV}$ on the Heisenberg $J_1$ and $J_2$ and DM exchange parameters $D_1$ and $D_2$ obtained from DFT+DMFT. In all plots, the length of the $D_2$ vectors are increased by the factor of ten for clarity.
 Results for the MTH (filled symbols) and L\"owdin projections (open symbols) are shown in comparison to each other.}
\label{f:Mn_W001}
\end{figure}

\subsection{Monolayer of Mn on W(001)} 
The other low-dimensional system investigated here, the Mn monolayer on the W(001) surface, is characterized by large Mn moments around $\unit[3.43-3.45]{\mu_\text{B}}$, which can increase or decrease in response to electronic correlations (see Appendix~D), and smaller induced moments on W which are the largest in the topmost surface layer ($m(\text{W})=\unit[0.38]{\mu_\text{B}}$) and antiparallel to the Mn moments. These W moments are reduced down to $\unit[0.34]{\mu_\text{B}}$ when the correlation strength is tuned up to $\unit[4]{eV}$. The calculated eigenvalue sums for $U=0$ suggest that the out-of-plane magnetic configuration $\vec{M}\parallel z$ is lower in energy by $\unit[2.51]{meV/Mn}$ than the in-plane configurations $\vec{M}\parallel x$ and $\vec{M}\parallel y$, which is opposite to the trend shown by the relativistic total energies. Further below, the modelling of the magnetic ground state will be based on the more accurate estimate of anisotropy based on the eigenvalue sums for $U=0$, while the exchange parameters for $U=0$ and $U=\unit[2]{eV}$ will be used for comparison.

The Mn/W(001) bilayer is the only case out of all examples discussed here where the difference between the results based on the two projection schemes is the largest, in particular for the $J$ parameters (Fig.~\ref{f:Mn_W001}). The differences are seen to be of order 10--15\,\% for the NN and NNN $J$ parameters, something which does not change the results qualitatively but could be of importance for detailed quantitative results. In the analysis that follows we focus specifically on the results from the MTH projection.

\begin{figure}
\centering
 \includegraphics[width=0.99\columnwidth]{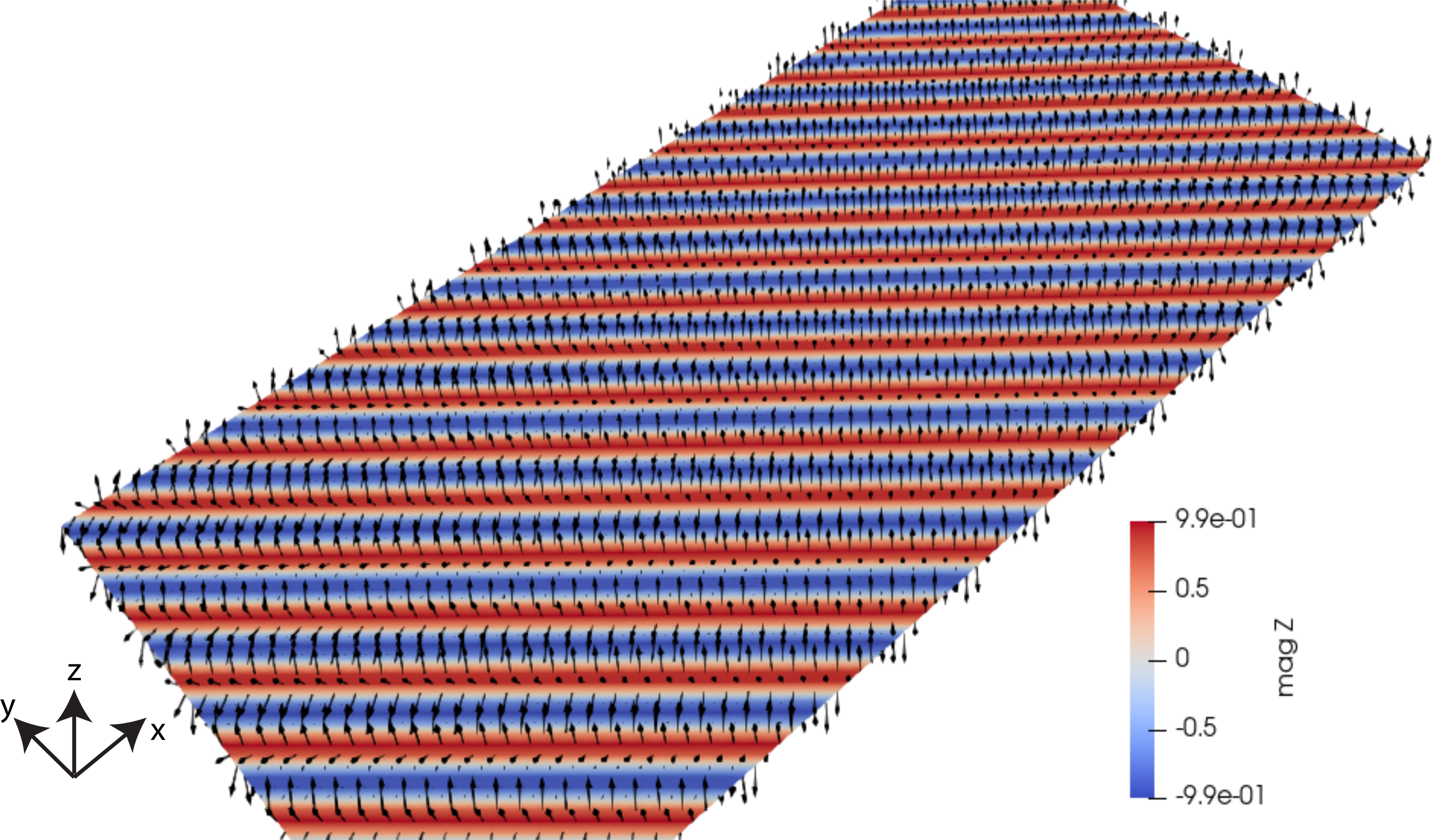}
 \caption{Magnetic ground state of Mn monolayer on W(001) surface with correlation strength $U = \unit[0]{eV}$. The arrows represent the direction of the Mn magnetic moments and the color code shows the normalized out-of-plane component of these magnetic moments.}
\label{f:Mn_W001_ground_state}
\end{figure}

The magnetic interactions in this system are dominated by a strong ferromagnetic nearest-neighbor exchange $J_1$ and a considerably smaller antiferromagnetic NNN exchange $J_2$ (Fig.~\ref{f:Mn_W001}c). Exceptionally strong DM interaction is obtained in our DFT calculations for the same NN ($\unit[7.0]{meV}$) and NNN ($\unit[0.4]{meV}$) bonds, in agreement with previous first-principles studies of this low-dimensional system.\cite{Ferriani2008,Udvardi2008,Freimuth2014} Large value of the DM interaction leads to a helical magnetic ground state with a relatively small period of $\unit[2.9]{nm}$ ($\unit[2.2]{nm}$ in experiment\cite{Ferriani2008}) and the propagation vector along the [110] in-plane direction, revealed by Monte Carlo simulations (Fig.~\ref{f:Mn_W001_ground_state}). The NN exchange in this system is mediated mostly by the $d_{xy}$, $d_{xz}$ and $d_{yz}$ orbitals (data not shown), which is natural considering the geometry of the Mn layer. However, we also find large off-diagonal contributions to this exchange process, in particular, $d_{xy} \to d_{xz}/d_{yz}$, $d_{x^2-y^2} \to d_{xz}/d_{yz}$ and $d_{x^2-y^2} \to d_{z^2}$ interactions.

When dynamical correlations are included, we observe a monotonous decrease of the Heisenberg exchange as a function of the Hubbard $U$ describing the correlation strength (Fig.~\ref{f:Mn_W001}c). The relative change is $-22\%$ for $J_1$ and $-12\%$ for $J_2$, in the studied range $U = \unit[(0-4)]{eV}$. The calculated DM interaction parameters $D_1$ and $D_2$ both decrease with increasing $U$, where the variations are $-12\%$ and $-35\%$ when $U$ changes from $0$ to $\unit[4]{eV}$. In contrast to the Co/Pt(111), the DM vectors for the NN and NNN bonds in Mn/W(001) lie fully in-plane and do not rotate out-of-plane even when the electronic correlations are tuned. This is a consequence of the specific crystal symmetry and the Moriya rules.\cite{Moriya1960} For the case of $U = \unit[2]{eV}$, the  magnetic ground state of this system has a spatial period of $\unit[2.7]{nm}$, which is closer to the experimental findings, compared to the case of $U = \unit[0]{eV}$.

\section*{Conclusions}
We demonstrate the performance of a theoretical method to determine the exchange interactions, including the DM interaction and symmetric anisotropic exchange, in bulk and nanoscale magnetic systems. The method takes into account dynamical electronic correlations, using DFT+DMFT, as well as the relativistic spin-orbit coupling. Several representative cases are used to illustrate numerical results, including both bulk and low-dimensional systems. The effect of electron-electron interactions on the magnitude of the Heisenberg and DM exchange is quantified within the SPTF approximation to DMFT. It is demonstrated here that different projection schemes (orthogonal and muffin-tin heads) give very similar results for all interatomic exchange interactions, irrespective of the strength of the Hubbard $U$. In addition, we find that the DFT+DMFT results with $U,J_\text{H}>0$ extrapolate smoothly to the DFT results ($U=\unit[0]{eV}$, $J_\text{H} = \unit[0]{eV}$), both shown in Figs.~\ref{f:CoPt}--\ref{f:B20_compounds}, \ref{f:Co_Pt111}, \ref{f:Mn_W001}.

We find in this study that for some of the systems investigated, e.g., the CoPt distorted alloy and the B20 skyrmion-host compounds MnSi and FeGe, correlations can produce large non-trivial variations of the magnetic interactions, leading to local minima and maxima (Figs.~\ref{f:CoPt} and~\ref{f:B20_compounds}). Also, the low-dimensional Co/Pt(111) system shows an interesting effect, namely a rotation of the DM vectors in response to the inclusion of correlations to the electronic structure. On the other hand, some systems, e.g., the Mn/W(001) bilayer, are characterized by more monotonous changes of the magnetic interactions vs the $U$ parameter. We do note however that the measured period of the spin-spiral structure of Mn/W(001) is reproduced with slightly better accuracy by calculations based on DMFT, as compared to DFT-based theory.

In general, we observe that correlations can produce up to 30\% variations of the leading Heisenberg and DM exchange interactions, while weaker interactions between further neighbors can show even larger relative variations. Whether monotonic or non-monotonic, the observed correlation-induced changes in the magnetic exchange parameters can be crucial for a quantitative description of the material properties, such as the Curie temperature, spin stiffness (see Fig.~\ref{f:Co_Pt_magnon_spectra} for Co/Pt(111)) and the spatial period of possible non-collinear magnetic orders. For the B20 compound FeGe, dynamical correlations do not change the Heisenberg exchange interactions significantly, which also implies that the evaluated micromagnetic exchange parameter, $A$, is not modified substantially. This means that the results of previous theory\cite{Grytsiuk2019} and the calculations presented here (both with and without dynamical correlations) agree, but that theory grossly fails to reproduce the experimental value of the exchange parameter, $A$, for this system. Curiously, the DM parameter is reproduced much better by theory. In our opinion, the inclusion of the symmetric anisotropic interactions represented by Eq.~(\ref{e:C_definition}), which we show here to be significant, may result in different values of $A$ for the analysis of the magnon spectra. Thus, it would contribute to resolve this problem.

It is interesting to compare MnSi and FeGe, which have similar crystal structures whereas the electronic occupation of the $3d$ states differs with one electron. Accordingly, MnSi is closer to the half-filled case than FeGe, which explains larger correlations effects for the Heisenberg interactions (Fig.~\ref{f:B20_compounds}c,f). However, the DM interaction in FeGe is more sensitive to correlations (Fig.~\ref{f:B20_compounds}f) which is a more subtle issue. In this respect, the two B20 systems investigated here still remain challenging in terms of an accurate first-principles description of the magnetic properties.

In the proposed \textit{ab initio} framework, it is also possible to obtain a complete picture of magnetic interactions between different atomic neighbors in terms of various orbital contributions. It appears that the magnetic exchange does not always involve interactions between orbitals of the same kind on the two considered atomic sites and, in certain cases (e.g., for \textit{hcp} FePt), can be dominated by cross-interactions between different orbitals. In this case, the electronic correlations can have a different effect compared to the interactions between similar orbitals. This information would be useful for finding ways of tuning the interactions in such systems, e.g., by mechanical strain or charge doping, and for interpreting their magnetic properties in terms of the underlying electronic structure.

\section*{Acknowledgments}

This work was financially supported by the Knut and Alice Wallenberg Foundation through Grant No. 2018.0060. O.E. also acknowledges support by the Swedish Research Council (VR), the Foundation for Strategic Research (SSF), the Swedish Energy Agency (Energimyndigheten), the European Research Council (854843-FASTCORR), eSSENCE and STandUP. D.T., Y.K., A.D. and L.N. acknowledge support from the Swedish Research Council (VR). The computations/data handling were enabled by resources provided by the Swedish National Infrastructure for Computing (SNIC) at the National Supercomputing Centre (NSC, Tetralith cluster) and the Chalmers Centre for Computational Science and Engineering (C3SE, Hebbe cluster), partially funded by the Swedish Research Council through grant agreement No.\,2016-07213. We would like to thank Dr. Erna Delczeg for doing test calculations using the SPR-KKR code for comparison with our DFT results for FePt at an early stage of this work. Structural sketches in plots (a,b) of Figs.~\ref{f:CoPt}--\ref{f:B20_compounds}, \ref{f:Co_Pt111}, \ref{f:Mn_W001} have been produced by the \textsc{VESTA3} software.\cite{vesta}

\newpage
\appendix

\section{Calculation parameters}

The structural parameters of the studied systems are listed below:
\begin{enumerate}
    \item[CoPt:] Cubic cell with the optimized lattice parameter $a=3.013\,\text{\AA}$ is used. This lattice parameter was obtained for Co and Pt in the high-symmetry positions $(0,0,0)$ and $(\frac12, \frac12, \frac12)$. Afterwards, Pt was shifted to the position $(\frac12, \frac12, \frac12 + \delta)$ with $\delta = 0.1$ while keeping the new lattice parameter fixed.
    
    Since the magnetization direction along $x$ is usually preferred compared to $z$ direction, as we find for this system, the exchange parameters are determined from the $\vec{M}\parallel x$ configuration and Fig.~\ref{f:CoPt}c shows the $J_{zz}$ component of the Heisenberg exchange. Due to the significant magnetic anisotropy, other components $J_{xx}$ and $J_{yy}$ may be different. However, we focus here on the general magnetic trends influenced by electronic correlations and do not intend to analyze in detail the magnetic interactions.

    \item[FePt:] Hexagonal cell is used with the optimized lattice parameters $a=2.730\,\text{\AA}$ and $c/a=1.599$ and the lattice vectors:
    \begin{equation}
        \begin{array}{rccc}
             \vec{a}_1 &= (1,&0,&0) \\[5pt]
             \vec{a}_2 &= (-\frac12,& \frac{\sqrt{3}}{2},& 0) \\[5pt]
             \vec{a}_3 &= (0,& 0,& c/a)
        \end{array}
    \end{equation}
    and the internal positions $(0,0,0)$ for Fe and $(\frac13, \frac23, \frac12)$ for Pt.
    
    Here, the preferred magnetization direction is, for most values of $U$, the $z$ direction. From the $\vec{M}\parallel z$ configuration, the $J_{xx}$ and $J_{yy}$ components of the Heisenberg interaction can be determined and Fig.~\ref{f:FePt}c shows the behavior of their average $(J_{xx}+J_{yy})/2$.

    \item[MnSi:] Cubic cell with the lattice parameter $a = 4.556\,\text{\AA}$ and four formula units is used. The Wyckoff positions are $(0.1395, 0.1395, 0.1395)$ for Mn and $(0.8474, 0.8474, 0.8474)$ for Si.
    
    \item[FeGe:] Cubic cell with the lattice parameter $a = 4.700\,\text{\AA}$ and four formula units is used. The Wyckoff positions are $(0.1352, 0.1352, 0.1352)$ for Fe and $(0.8419, 0.8419, 0.8419)$ for Ge.
    
    For both B20 systems (MnSi and FeGe), the $x$, $y$ and $z$ directions are equivalent due to the cubic symmetry and, in Fig.~\ref{f:B20_compounds}(c,f), we plot the average $(J_{xx}+J_{yy})/2$ determined from the $\vec{M}\parallel z$ configuration.
    
    \item[Co/Pt:] This system is modelled by a supercell with 5 Pt layers with the \textit{fcc} lattice parameter $a=3.900\,\text{\AA}$ and a nearest-neighbor Pt-Pt distance of 2.758\,\AA. For the surface layer, the optimized Pt-Pt distance is set to 2.879\,\AA{} and the Co-Pt distance is 2.570\,\AA. Co atoms are in the hole positions of the Pt surface. The corresponding Pt-Pt and Co-Pt interlayer distances are $\unit[2.40]{\AA}$ and $\unit[2.02]{\AA}$ which are similar to the values reported in Table~I of Ref.~\onlinecite{Zimmermann2019}.
    
    We find that the total energies for the $\vec{M}\parallel x$ and $\vec{M}\parallel y$ configurations are lower than for $\vec{M}\parallel z$. For most values of $U$, the $\vec{M}\parallel y$ state has the lowest energy among the three studied configurations and, for that reason, the Heisenberg exchange component $J_{zz}$ was calculated for $\vec{M}\parallel y$ and plotted in Fig.~\ref{f:Co_Pt111}c.
    
    \item[Mn/W:] This system is modelled by a supercell with 5 W layers where the ideal \textit{bcc} structure is assumed with a nearest-neighbor W-W distance of 3.165\,\AA. On the W surface, Mn forms a square lattice with the same lattice parameter and a Mn-W nearest-neighbor distance of 2.741\,\AA. The vacuum region has a thickness of 26.9\,\AA.
    
    For this system, we find that the in-plane orientation of the Mn magnetic moments is preferred, while the $\vec{M}\parallel x$ and $\vec{M}\parallel y$ configurations are equivalent by symmetry. The Heisenberg interaction $J_{zz}$ was determined for $\vec{M}\parallel x$ and plotted in Fig.~\ref{f:Mn_W001}c.
\end{enumerate}

For the aforementioned CoPt and FePt bulk alloys and Co/Pt(111) bilayer system, the structural optimization was performed on the DFT level using the projector-augmented wave method\cite{Bloechl1994} as implemented in the Vienna \textit{Ab initio} Simulation Package (VASP).\cite{Kresse1996} The ferromagnetic order and the energy cutoff of $\unit[800]{eV}$ for the plane-wave expansion of the wavefunction was set for all systems. The $\Gamma$-centered $(10\times 10\times 10)$ and $(20\times 20\times 20)$ $k$-meshes were used for CoPt and FePt, respectively. For the CoPt bulk alloy, the internal coordinates were fixed, while for the FePt alloy the atoms remained in the high-symmetry positions. For the Co/Pt(111) system, the two top Pt layers and the Co monolayer were fully relaxed and the rest of the Pt subsystem was fixed in the ideal bulk structure.

The Brillouin-zone (BZ) sampling for the studied systems was done using the following $k$-meshes with the total number $n_k$ of $k$-points in the BZ:
\begin{itemize}
  \item CoPt: $(25 \times 25 \times 25)$, $n_k = 15625$.
  Denser $(40 \times 40 \times 40)$ $k$-mesh changes the DFT estimates of the $J_{ij}$ parameters, that are bigger than $\unit[0.01]{meV}$, by less than $\unit[1.2]{\%}$. The $D_x > \unit[0.01]{meV}$ components of the DM vectors are changed by less than $\unit[3.5]{\%}$.
  \item FePt: $(80 \times 80 \times 40)$, $n_k = 256000$.
  In order to calculate the exchange parameters using the L\"owdin projection (Fig.~\ref{f:FePt}c), we had to reduce the dimensions of the $k$-mesh for this system down to $(40 \times 40 \times 20)$, which corresponds to the total of $n_k = 32000$ $k$-points. For a few selected $U$ values, we have verified for only the nearest neighbor exchange parameters that the denser $(80 \times 80 \times 40)$ $k$-mesh changes the results by $\unit[+(0.1-0.2)]{\%}$ for $J_1$, $\unit[-(1-2)]{\%}$ for $J_2$ and by around $\unit[-6]{\%}$ for $D_1$.
  \item MnSi and FeGe: $(10 \times 10 \times 10)$, $n_k = 1000$.
  Denser $(20 \times 20 \times 20)$ $k$-mesh changes the DFT estimates of the dominating $J_{ij}$ parameters, that are bigger than $\unit[0.7]{meV}$, by less than $\unit[0.4]{\%}$ (MnSi) and $\unit[3.4]{\%}$ (FeGe). For the DM interactions above $\unit[0.05]{meV}$, the differences are $\unit[6.4]{\%}$ (MnSi) and $\unit[5.9]{\%}$ (FeGe).
  \item Co/Pt(111): $(30 \times 30 \times 2)$, $n_k = 1800$. Denser $(45 \times 45 \times 3)$ $k$-mesh changes the DFT estimates of the $J_{ij}$ parameters, that are bigger than $\unit[0.7]{meV}$, by less than $\unit[1.0]{\%}$, and, for example, the DM components $D_x$ above $\unit[0.05]{meV}$ change by less than $\unit[7.3]{\%}$.
  \item Mn/W(001): $(30 \times 30 \times 3)$, $n_k = 2700$. Denser $(40 \times 40 \times 4)$ $k$-mesh changes the DFT estimates of the $J_{ij}$ parameters, that are bigger than $\unit[0.01]{meV}$, by less than $\unit[2.2]{\%}$ and the $D_x > \unit[0.01]{meV}$ components of the DM vectors~-- by less than $\unit[1.6]{\%}$.
\end{itemize}

For all bulk systems, the $k$-mesh was shifted around the $\Gamma$ point by half the grid step. For the 3$d$/5$d$ bilayers, the $\Gamma$-centered $k$-mesh was used, in order to be consistent with the lattice symmetry. The Fermi smearing for the electronic occupations was used for all systems, with the smearing temperature set to $\unit[158]{K}$ for FePt and $\unit[395]{K}$ for the other systems.

In addition, we used symmetry arguments when calculating the DM interaction in various systems, in order to reduce the amount of calculations. For example, the $\vec{D}_1$ and $\vec{D}_2$ vectors in the CoPt alloy (Fig.~\ref{f:CoPt}a,b) only have in-plane components and are perpendicular to the radius vectors connecting the corresponding atoms. This allows to determine the vectors from a single calculation with the magnetization along the $x$-direction ($\vec{M}\parallel x$), which was done for different $U$-values in Fig.~\ref{f:CoPt}c. Another example are the B20 compounds where the $\vec{D}_1$ vectors for the three nearest-neighbor bonds can be transformed into each other by means of a 120$^\circ$-rotation, which is equivalent to interchanging the $x$, $y$ and $z$ components (disregarding the sign). Therefore, the absolute value $|\vec{D}_1|$ can be obtained from a single calculation with $\vec{M}\parallel z$ for different $U$-values (result in Fig.~\ref{f:B20_compounds}). In both aforementioned cases, the amount of calculations is reduced by the factor of three, compared to a case where symmetry arguments cannot be used and one has to perform three calculations with $\vec{M}\parallel x$, $\vec{M}\parallel y$ and $\vec{M}\parallel z$ for each value of the $U$ parameter.

\section{Long-range character of interactions}

\begin{figure}
\centering
 \includegraphics[width=0.49\columnwidth]{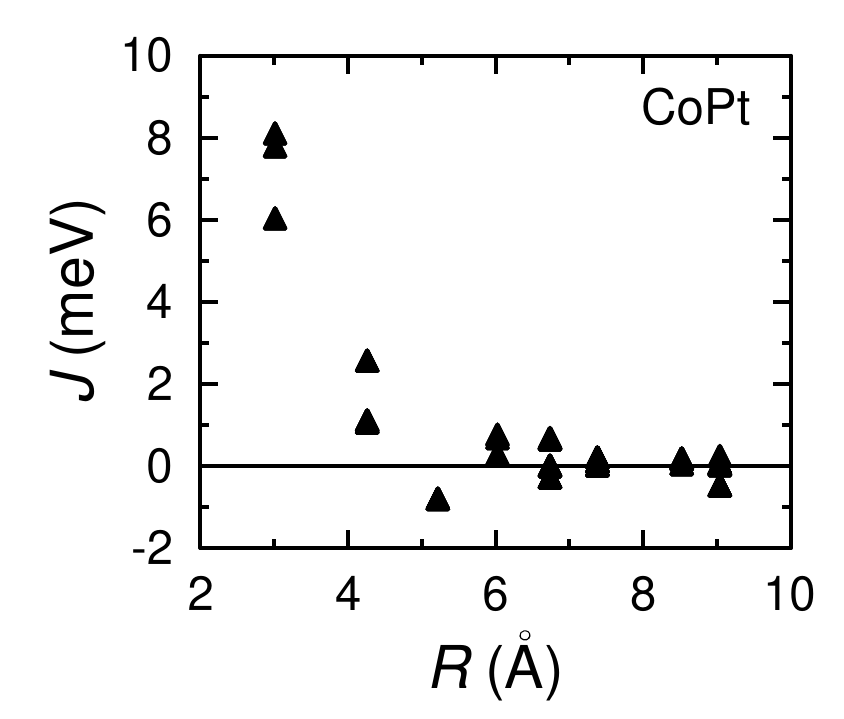}
 \includegraphics[width=0.49\columnwidth]{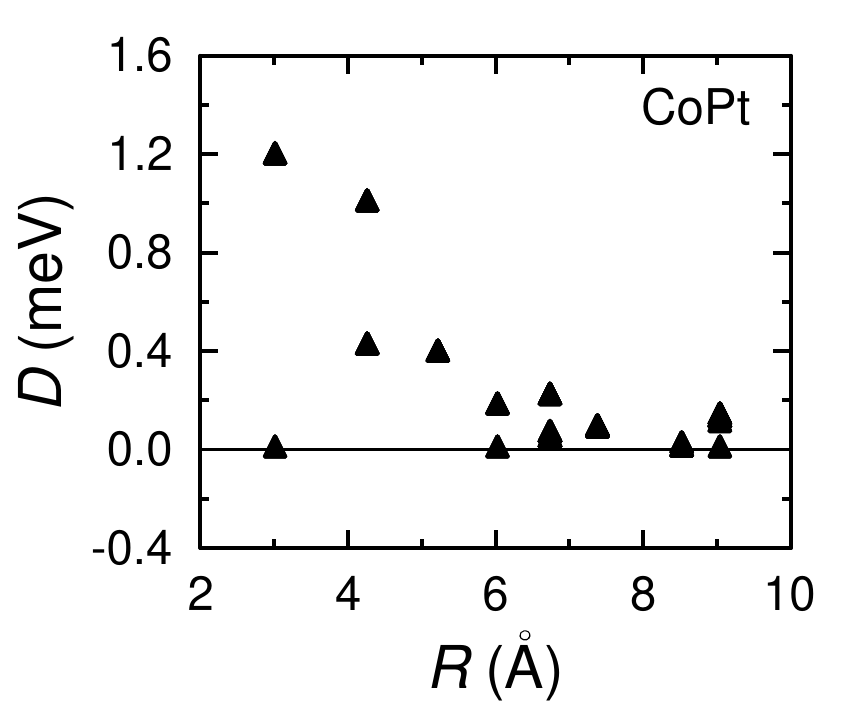}
 \includegraphics[width=0.49\columnwidth]{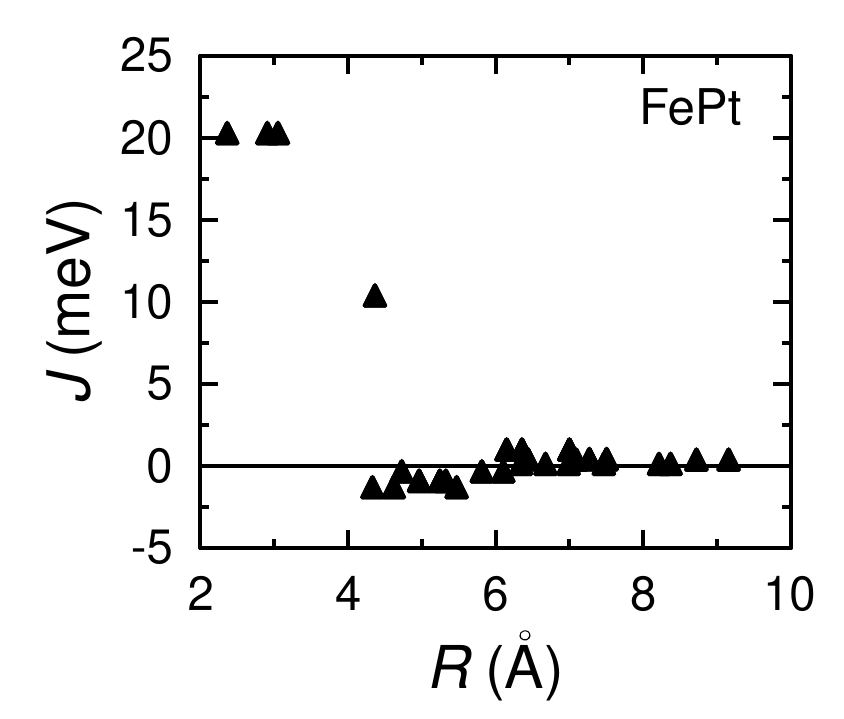}
 \includegraphics[width=0.49\columnwidth]{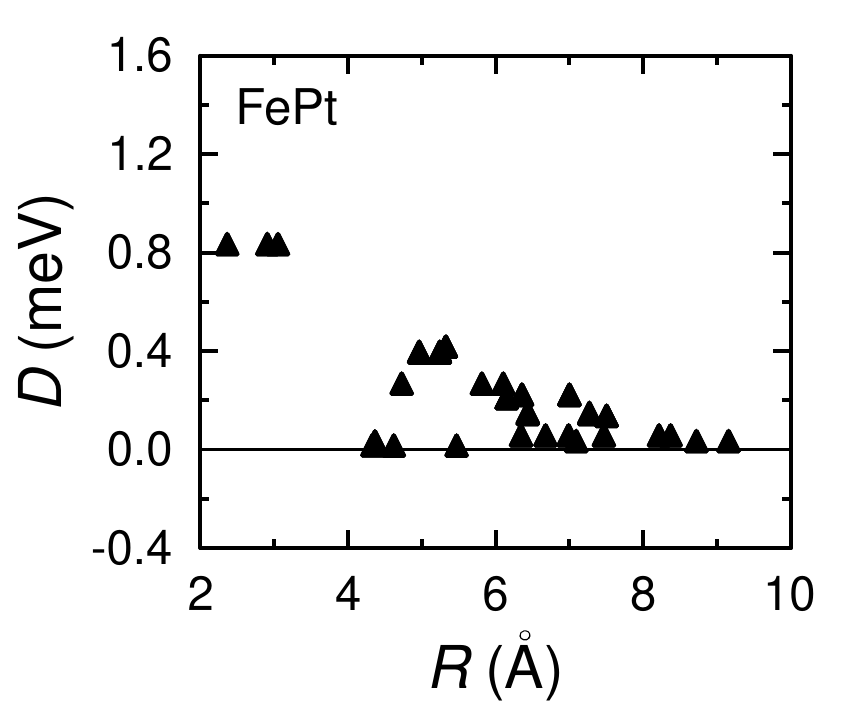}
 \caption{The Heisenberg and DM exchange parameters of the CoPt (upper two panels, cf. Fig.~\ref{f:CoPt}a) and FePt (lower two panels, cf. Fig.~\ref{f:FePt}a) alloys with optimized structure vs. the distance between the interacting Co and Fe magnetic moments. Calculations are done with the GGA functional ($U=\unit[0]{eV}$).}
\label{f:CoPt_and_FePt_vs_R}
\end{figure}

\begin{figure}
\centering
 \includegraphics[width=0.49\columnwidth]{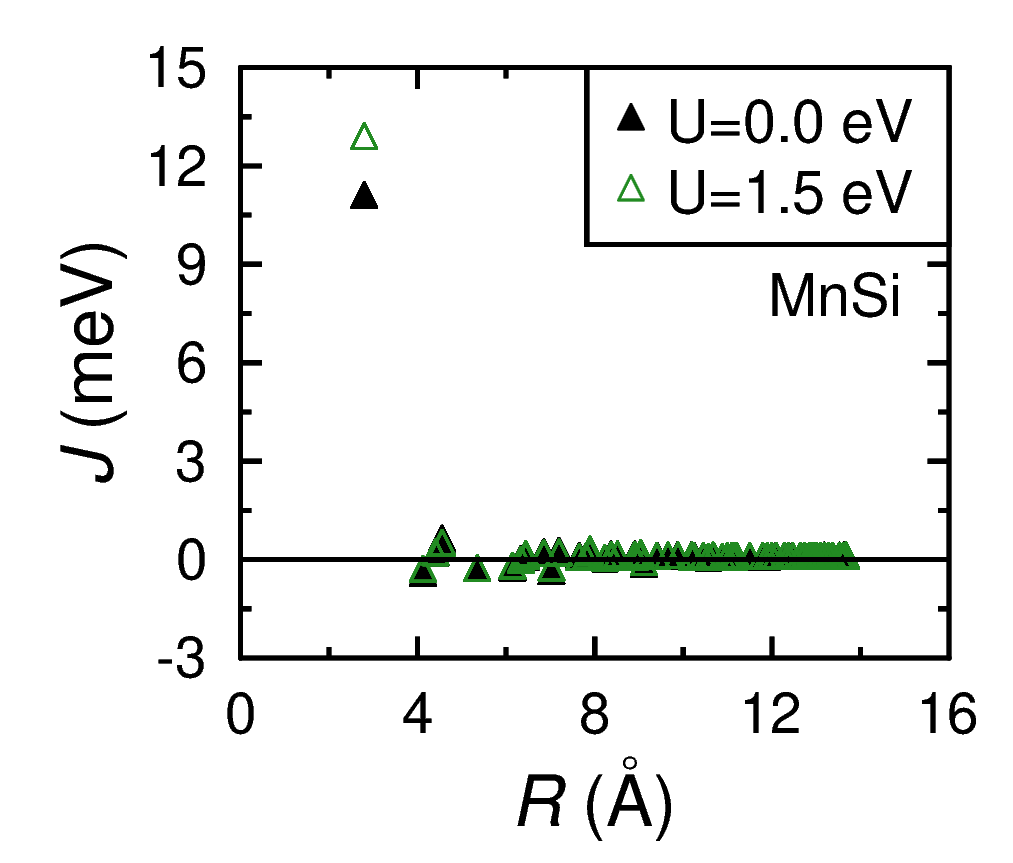}
 \includegraphics[width=0.49\columnwidth]{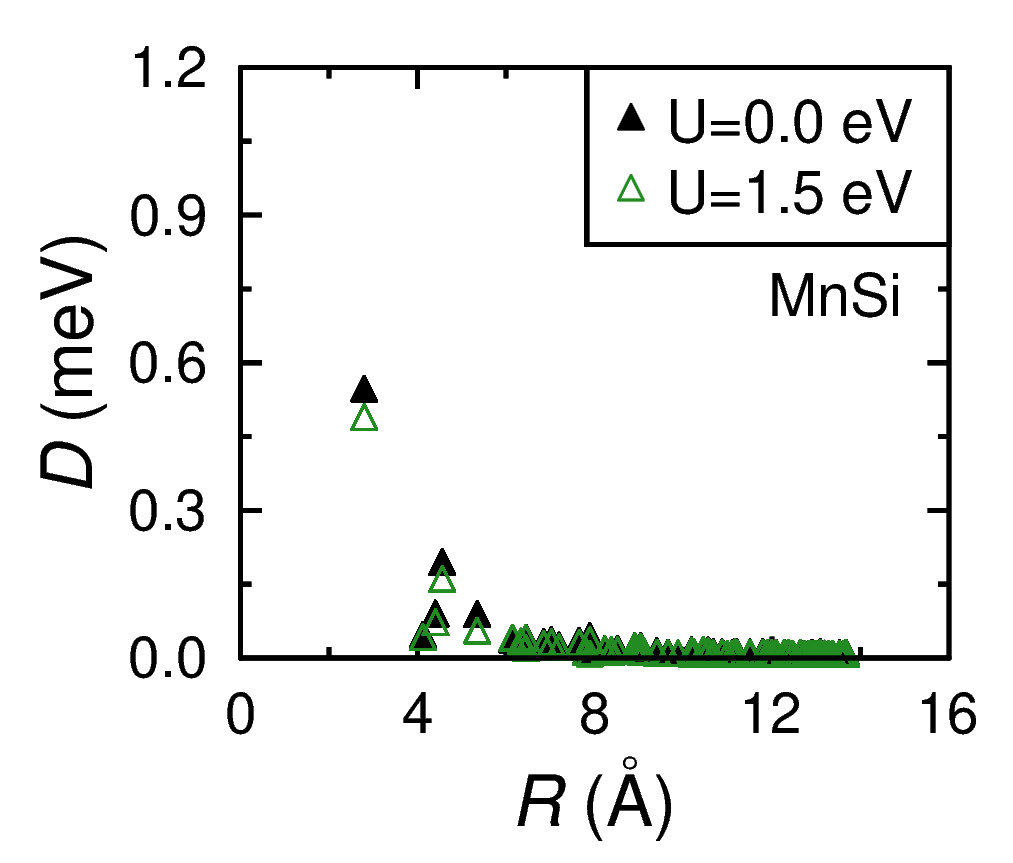}
 \includegraphics[width=0.49\columnwidth]{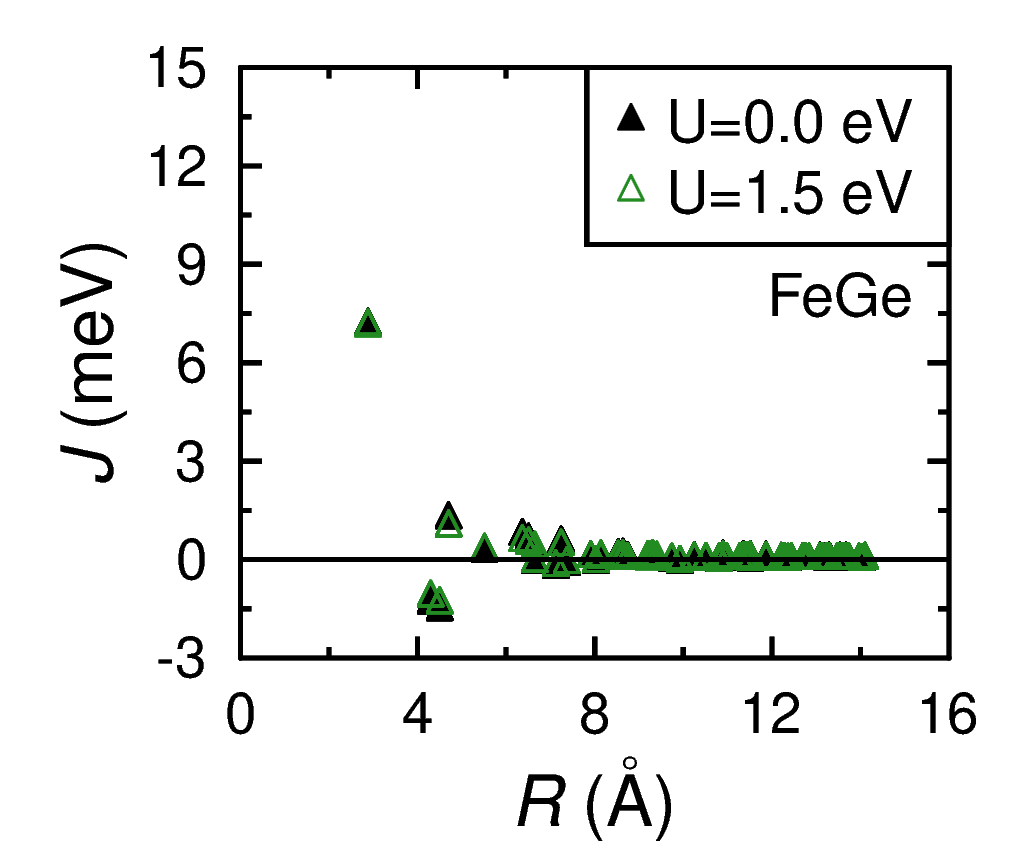}
 \includegraphics[width=0.49\columnwidth]{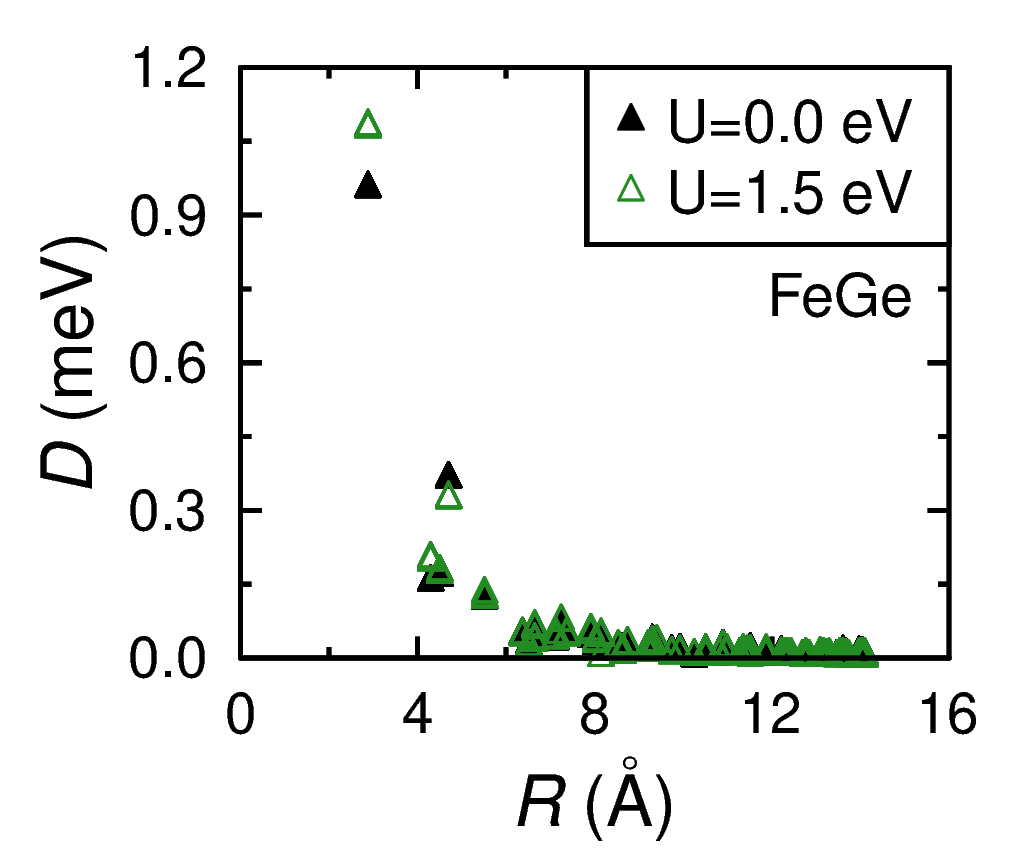}
 \caption{The Heisenberg exchange parameters of the B20 compounds MnSi (upper two panels) and FeGe (lower two panels) vs. the distance between the interacting transition metal magnetic moments. Calculations are done with DFT ($U=\unit[0]{eV}$, filled symbols) and DFT+DMFT ($U=\unit[1.5]{eV}$, open symbols).
 }
\label{f:B20_vs_R}
\end{figure}

\begin{figure}
\centering
 \includegraphics[width=0.49\columnwidth]{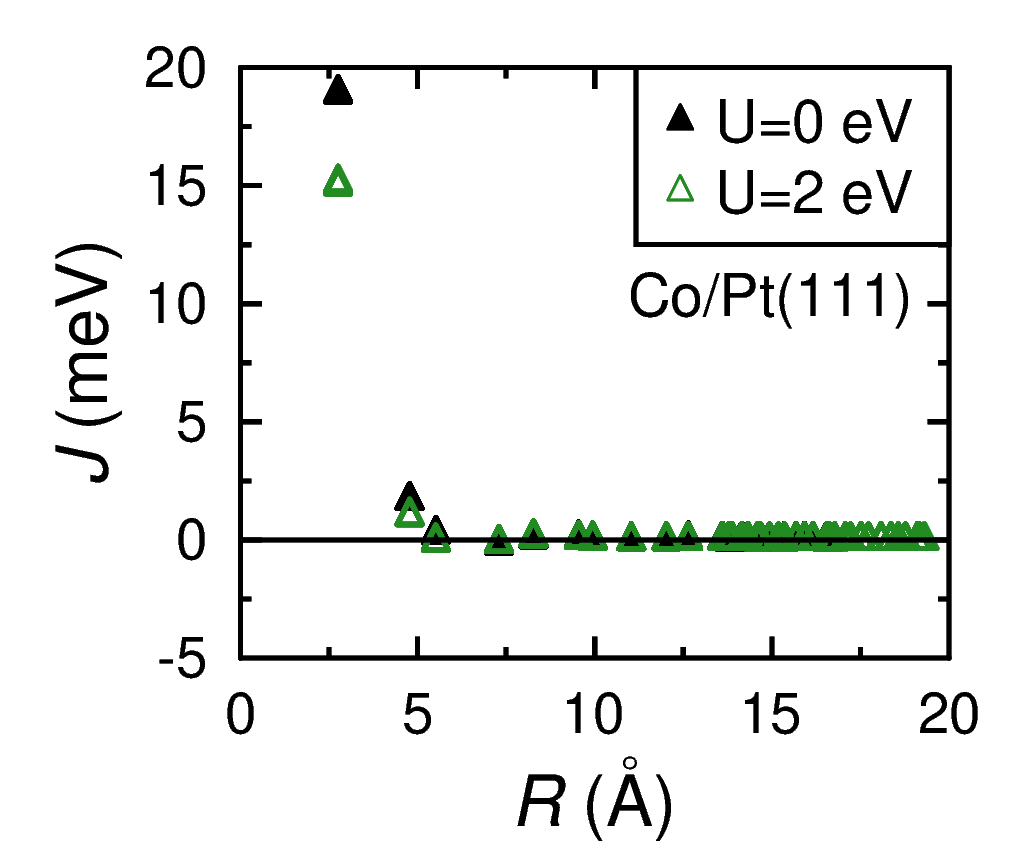}
 \includegraphics[width=0.49\columnwidth]{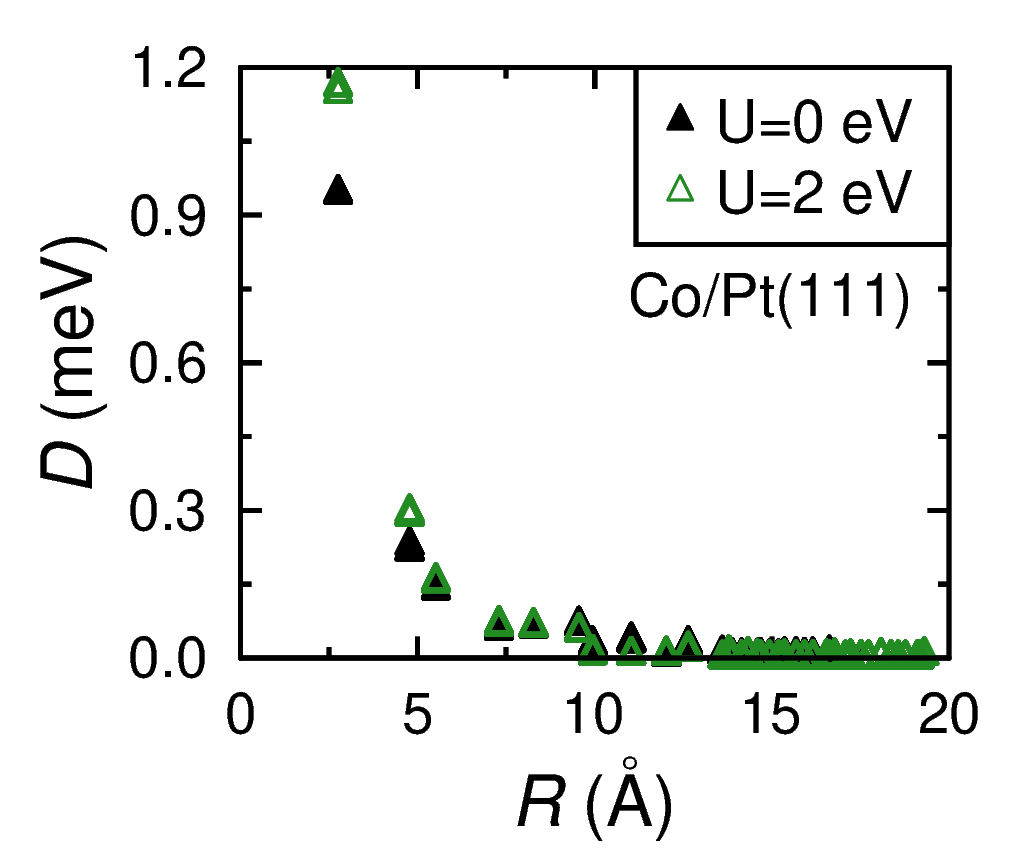}
 \includegraphics[width=0.49\columnwidth]{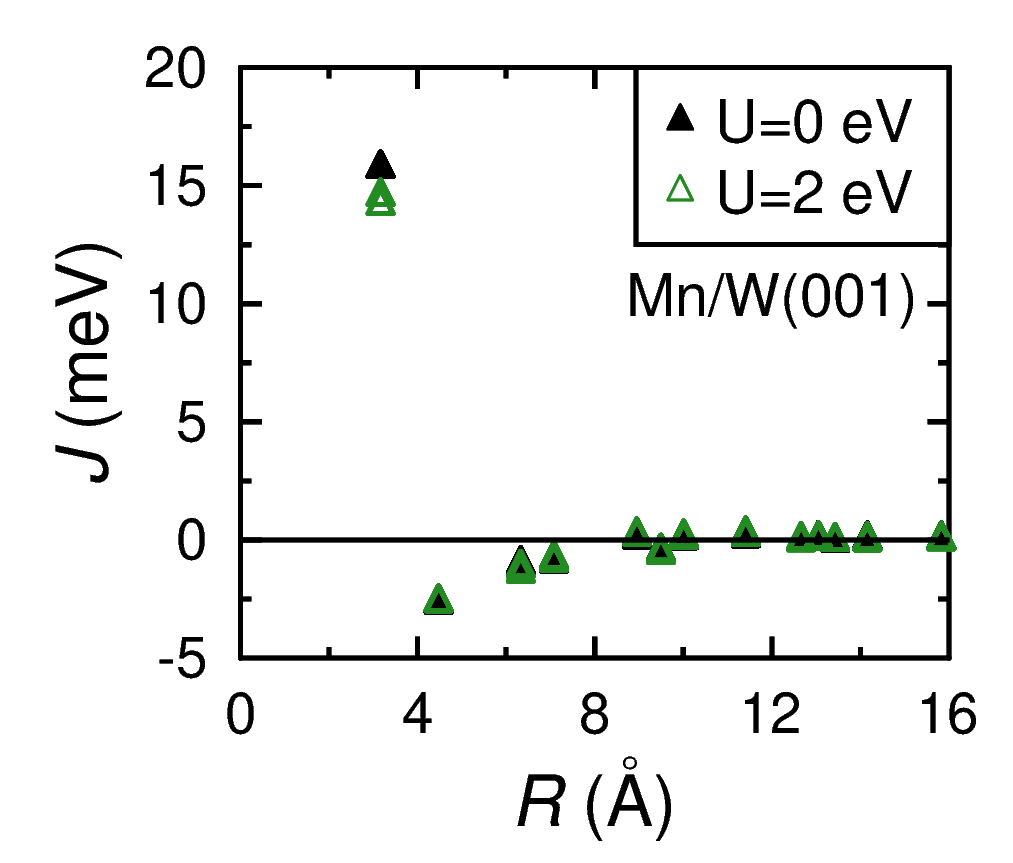}
 \includegraphics[width=0.49\columnwidth]{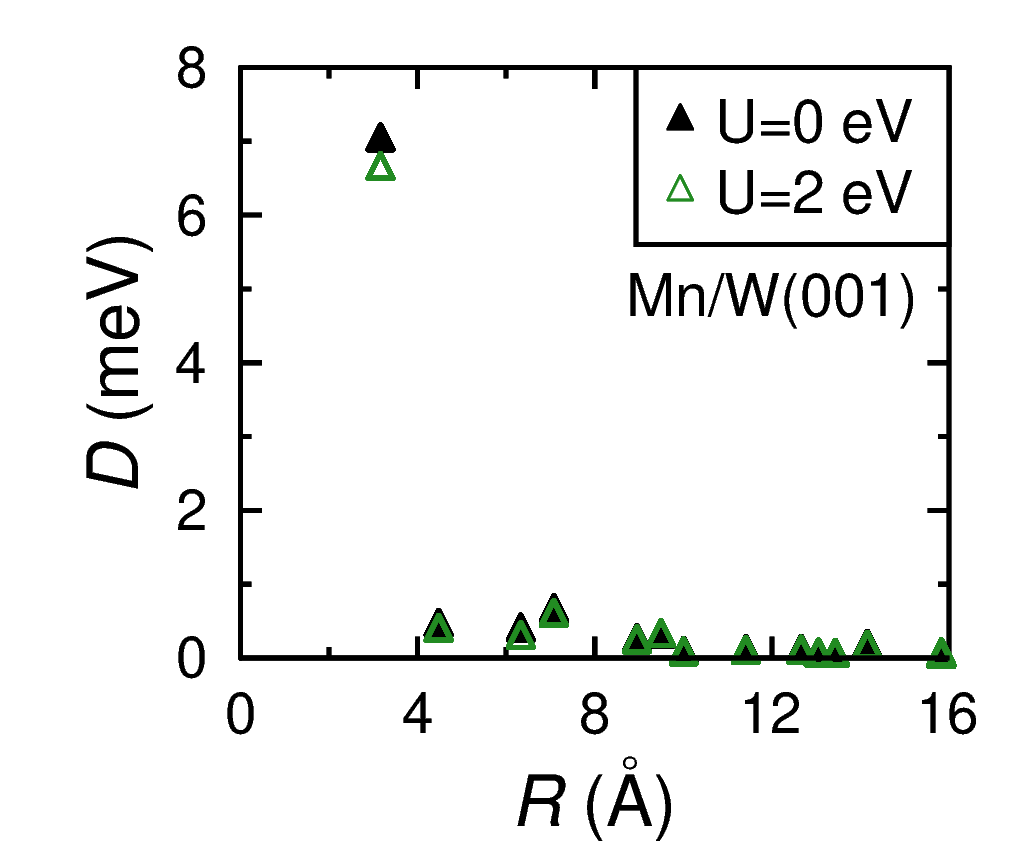}
 \caption{The Heisenberg exchange parameters of the low-dimensional Co/Pt(111) (upper two panels, cf. Fig.~\ref{f:Co_Pt111}a) and Mn/W(001) (lower two panels, cf. Fig.~\ref{f:Mn_W001}a) systems vs. the distance between the interacting 3$d$ transition metal magnetic moments. Calculations are done with DFT ($U=\unit[0]{eV}$, filled symbols) and DFT+DMFT ($U=\unit[2]{eV}$, open symbols).}
\label{f:3d_5d_vs_R}
\end{figure}

Since the studied systems are metallic, the magnetic exchange interactions have a long-range character which is resolved by our calculations. Figures~\ref{f:CoPt_and_FePt_vs_R}--\ref{f:3d_5d_vs_R} show the distance dependence of the Heisenberg and DM exchange interactions between the interacting atoms, calculated on the DFT and DFT+DMFT ($U=\unit[2]{eV}$) levels. While interactions in several neighboring shells up to a distance of around 5\,\AA{} seem to dominate in each system, smaller interactions at larger distances also need to be included, for example, in the analysis of the magnon spectra and the simulation of the magnetic order. Especially, the spin stiffness is known to converge slowly with respect to the maximal considered interaction distance, which has been pointed out, for example, for FeGe already in previous studies.\cite{Grytsiuk2019} Furthermore, we notice that the DM interactions are more long-ranged than the Heisenberg exchange for the studied systems. The electronic correlations influence the magnitude of magnetic interactions at all distances, while the average interaction range is the same.

Detailed information on some of the nearest-neighbor interactions for all systems is provided also in Tables~I-VI where the DFT estimates of the Heisenberg exchange $J$ and the components $D_x$, $D_y$ and $D_z$ of the DM interaction are given in meV (with two decimals), while the components $n_x$, $n_y$ and $n_z$ of the radius vector between the interacting spins are given in units of the lattice vectors (for structural details, see Appendix~A). In these tables, small exchange parameters below $\unit[10^{-4}]{meV}$ are shown as zeros. For the neighbors with all $n'_i$ components inversed ($n'_i = -n_i$), the $J$ parameters are the same and the DM vectors are inversed $\vec{D}' = -\vec{D}$. In case of the B20 compounds, which have four magnetic sites in the unit cell, the exchange parameters are shown only between sites~1 and 2, while interactions between other site pairs can be obtained using symmetry operations.

\begin{table}[h!]
 \caption{Magnetic interactions in the ordered CoPt alloy with the optimized structure, obtained from DFT calculations ($U=\unit[0]{eV}$).}
 \vspace{5pt}
 \centering
 \begin{tabular}{|c|c|c|c|r|r|r|}
    \hline
    \hspace{5pt}$n_x$\hspace{5pt} & \hspace{5pt}$n_y$\hspace{5pt} & \hspace{5pt}$n_z$\hspace{5pt} & \hspace{10pt}$J$\hspace{10pt} & \hspace{10pt}$D_x$\hspace{10pt} & \hspace{10pt}$D_y$\hspace{10pt} & \hspace{10pt}$D_z$\hspace{10pt} \\ \hline
    1 & 0 & 0   &  8.03 &  0.00 &  1.19 &  0.00 \\
    0 & 1 & 0   &  7.72 & -1.19 &  0.00 &  0.00 \\
    0 & 0 & 1   &  5.96 &  0.00 &  0.00 &  0.00 \\
    1 &  1 & 0  &  2.49 &  0.71 & -0.71 &  0.00 \\
    1 & -1 & 0  &  2.49 & -0.71 & -0.71 &  0.00 \\
    1 &  0 &  1 &  1.03 &  0.00 &  0.42 &  0.00 \\
    1 &  0 & -1 &  1.03 &  0.00 &  0.42 &  0.00 \\
    0 &  1 &  1 &  0.98 & -0.42 &  0.00 &  0.00 \\
    0 &  1 & -1 &  0.98 & -0.42 &  0.00 &  0.00 \\
    1 &  1 &  1 & -0.88 & -0.28 &  0.28 &  0.00 \\
    1 &  1 & -1 & -0.88 & -0.28 &  0.28 &  0.00 \\
    1 & -1 &  1 & -0.88 &  0.28 &  0.28 &  0.00 \\
    1 & -1 & -1 & -0.88 &  0.28 &  0.28 &  0.00 \\
    2 &  0 &  0 &  0.69 &  0.00 & -0.17 &  0.00 \\
    0 &  2 &  0 &  0.60 &  0.17 &  0.00 &  0.00 \\
    0 &  0 &  2 &  0.20 &  0.00 &  0.00 &  0.00 \\ \hline
 \end{tabular}
    \label{tab:CoPt_exchange_parameters}
\end{table}

\begin{table}[h!]
 \caption{Magnetic interactions in the ordered FePt alloy with the optimized structure, obtained from DFT calculations ($U=\unit[0]{eV}$).}
 \vspace{5pt}
 \centering
 \begin{tabular}{|c|c|c|c|r|r|r|}
    \hline
    \hspace{5pt}$n_x$\hspace{5pt} & \hspace{5pt}$n_y$\hspace{5pt} & \hspace{5pt}$n_z$\hspace{5pt} & \hspace{10pt}$J$\hspace{10pt} & \hspace{10pt}$D_x$\hspace{10pt} & \hspace{10pt}$D_y$\hspace{10pt} & \hspace{10pt}$D_z$\hspace{10pt} \\ \hline
    1 & 0 &  0 &  20.1 &  0.00 &  0.00 &  0.82 \\
    0 & 1 &  0 &  20.1 &  0.00 &  0.03 &  0.82 \\
    1 & 1 &  0 &  20.1 &  0.00 &  0.03 & -0.82 \\
    0 & 0 &  1 &  10.2 &  0.02 &  0.00 &  0.00 \\
    1 & 2 &  0 & -1.48 &  0.00 &  0.00 &  0.00 \\
    2 & 1 &  0 & -1.48 &  0.00 &  0.00 &  0.00 \\
    1 & -1 & 0 & -1.48 &  0.00 &  0.00 &  0.00 \\
    1 & 0 &  1 & -1.10 & -0.38 &  0.00 & -0.15 \\
    1 & 0 & -1 & -1.10 &  0.38 &  0.00 & -0.15 \\
    0 & 1 &  1 & -1.10 &  0.16 & -0.32 & -0.15 \\
    0 & 1 & -1 & -1.10 & -0.16 &  0.32 & -0.15 \\
    1 & 1 &  1 & -1.10 &  0.16 &  0.32 &  0.15 \\
    1 & 1 & -1 & -1.10 & -0.16 & -0.32 &  0.15 \\
    2 & 0 &  0 & -0.51 &  0.00 &  0.00 &  0.26 \\
    0 & 2 &  0 & -0.51 &  0.00 &  0.00 &  0.26 \\
    2 & 2 &  0 & -0.51 &  0.00 &  0.00 & -0.26 \\
 \hline
 \end{tabular}
    \label{tab:FePt_exchange_parameters}
\end{table}

\begin{table}[h!]
 \caption{Magnetic interactions in the B20 compound MnSi, obtained from DFT calculations ($U=\unit[0]{eV}$).}
 \vspace{5pt}
 \centering
 \begin{tabular}{|c|c|c|c|c|r|r|r|}
    \hline
    site 2 & \hspace{5pt}$n_x$\hspace{5pt} & \hspace{5pt}$n_y$\hspace{5pt} & \hspace{5pt}$n_z$\hspace{5pt} & \hspace{10pt}$J$\hspace{10pt} & \hspace{10pt}$D_x$\hspace{10pt} & \hspace{10pt}$D_y$\hspace{10pt} & \hspace{10pt}$D_z$\hspace{10pt} \\ \hline
    2 & 0 & 0 & 0 &   11.00  &  0.34 & -0.27 &  0.32 \\
    2 & 0 & 0 & 1 &   11.00  &  0.34 & -0.27 & -0.32 \\
    2 & 0 & 1 & 0 &   -0.52  &  0.01 &  0.03 & -0.02 \\
    2 & 0 & 1 & 1 &   -0.52  &  0.01 &  0.03 &  0.02 \\
    2 & -1 & 0 & 0 &   0.10  & -0.08 & -0.01 &  0.01 \\
    2 & -1 & 0 & 1 &   0.10  & -0.08 & -0.01 & -0.01 \\
    1 & 1 & 0 & 0 &    0.53  & -0.01 &  0.09 &  0.16 \\
    1 & 0 & 1 & 0 &    0.53  &  0.16 & -0.01 &  0.09 \\
    1 & 0 & 0 & 1 &    0.53  &  0.09 &  0.16 & -0.01 \\
    2 & 1 & 0 & 0 &   -0.37  & -0.02 & -0.02 &  0.02 \\
    2 & 1 & 0 & 1 &   -0.37  & -0.02 & -0.02 & -0.02 \\
    2 & 0 & -1 & 0 &  -0.09  & -0.02 &  0.01 &  0.00 \\
    2 & 0 & -1 & 1 &  -0.09  & -0.02 &  0.01 &  0.00 \\
    2 & -1 & 1 & 0 &  -0.37  & -0.01 &  0.04 &  0.07 \\
    2 & -1 & 1 & 1 &  -0.37  & -0.01 &  0.04 & -0.07 \\ \hline
 \end{tabular}
    \label{tab:MnSi_exchange_parameters}
\end{table}

\begin{table}[h!]
 \caption{Magnetic interactions in the B20 compound FeGe, obtained from DFT calculations ($U=\unit[0]{eV}$).}
 \vspace{5pt}
 \centering
 \begin{tabular}{|c|c|c|c|c|r|r|r|}
    \hline
    site 2 & \hspace{5pt}$n_x$\hspace{5pt} & \hspace{5pt}$n_y$\hspace{5pt} & \hspace{5pt}$n_z$\hspace{5pt} & \hspace{7pt}$\left\langle J\right\rangle$\hspace{7pt} & \hspace{10pt}$D_x$\hspace{10pt} & \hspace{10pt}$D_y$\hspace{10pt} & \hspace{10pt}$D_z$\hspace{10pt} \\ \hline
    2 & 0 & 0 & 0 &   7.15  & -0.61 &  0.52 & -0.52 \\
    2 & 0 & 0 & 1 &   7.15  & -0.61 &  0.52 &  0.52 \\
    2 & 0 & 1 & 0 &  -1.36  & -0.15 &  0.02 &  0.04 \\
    2 & 0 & 1 & 1 &  -1.36  & -0.15 &  0.02 & -0.04 \\
    2 & -1 & 0 & 0 & -1.57  &  0.04 &  0.16 & -0.04 \\
    2 & -1 & 0 & 1 & -1.57  &  0.04 &  0.16 &  0.04 \\
    1 & 1 & 0 & 0 &   1.23  &  0.08 & -0.21 & -0.29 \\
    1 & 0 & 1 & 0 &   1.24  & -0.29 &  0.08 & -0.21 \\
    1 & 0 & 0 & 1 &   1.26  & -0.21 & -0.29 &  0.08 \\
    2 & 1 & 0 & 0 &   0.74  & -0.01 & -0.02 & -0.04 \\
    2 & 1 & 0 & 1 &   0.74  & -0.01 & -0.02 &  0.04 \\
    2 & 0 & -1 & 0 &   0.60  & 0.01 & -0.01 & -0.02 \\
    2 & 0 & -1 & 1 &   0.60  & 0.01 & -0.01 &  0.02 \\
    2 & -1 & 1 & 0 &   0.21  & 0.11 & -0.04 &  0.01 \\
    2 & -1 & 1 & 1 &   0.21  & 0.11 & -0.04 & -0.01 \\ \hline
 \end{tabular}
    \label{tab:FeGe_exchange_parameters}
\end{table}

\begin{table}[h!]
 \caption{Magnetic interactions in the Co/Pt(111) system, obtained from DFT calculations ($U=\unit[0]{eV}$).}
 \vspace{5pt}
 \centering
 \begin{tabular}{|c|c|c|c|r|r|r|}
    \hline
    \hspace{5pt}$n_x$\hspace{5pt} & \hspace{5pt}$n_y$\hspace{5pt} & \hspace{5pt}$n_z$\hspace{5pt} & \hspace{10pt}$J$\hspace{10pt} & \hspace{10pt}$D_x$\hspace{10pt} & \hspace{10pt}$D_y$\hspace{10pt} & \hspace{10pt}$D_z$\hspace{10pt} \\ \hline
     1 &  0 & 0  & 18.9 &  0.00 & -0.88 & -0.34 \\
     0 &  1 & 0  & 18.9 &  0.77 & -0.43 &  0.34  \\
     1 & -1 & 0  & 18.9 & -0.77 & -0.43 &  0.34  \\
     1 &  1 & 0  & 1.70 &  0.11 & -0.19 &  0.00 \\
     1 & -2 & 0  & 1.70 & -0.23 &  0.00 &  0.00 \\
     2 & -1 & 0  & 1.70 & -0.11 & -0.19 &  0.00 \\
     2 &  0 & 0  & 0.26 &  0.00 &  0.12 &  0.08 \\
     0 &  2 & 0  & 0.26 & -0.10 &  0.06 & -0.08 \\
     2 & -2 & 0  & 0.26 &  0.10 &  0.06 & -0.08 \\ \hline
 \end{tabular}
    \label{tab:CoPt111_exchange_parameters}
\end{table}

\begin{table}[h!]
 \caption{Magnetic interactions in the Mn/W(001) system, obtained from DFT calculations ($U=\unit[0]{eV}$).}
 \vspace{5pt}
 \centering
 \begin{tabular}{|c|c|c|c|r|r|r|}
    \hline
    \hspace{5pt}$n_x$\hspace{5pt} & \hspace{5pt}$n_y$\hspace{5pt} & \hspace{5pt}$n_z$\hspace{5pt} & \hspace{10pt}$J$\hspace{10pt} & \hspace{10pt}$D_x$\hspace{10pt} & \hspace{10pt}$D_y$\hspace{10pt} & \hspace{10pt}$D_z$\hspace{10pt} \\ \hline
     1 &  0 &  0  &  15.8 &  0.00 & -7.00 & 0.00 \\
     0 &  1 &  0  &  15.8 &  7.00 &  0.00 & 0.00 \\
     1 &  1 &  0  & -2.69 &  0.30 & -0.30 & 0.00 \\
     1 & -1 &  0  & -2.69 & -0.30 & -0.30 & 0.00 \\
     2 &  0 &  0  & -1.02 &  0.00 & -0.36 & 0.00 \\
     0 &  2 &  0  & -1.02 &  0.36 &  0.00 & 0.00 \\
     1 &  2 &  0  & -0.94 & -0.62 &  0.11 & 0.00 \\
     1 & -2 &  0  & -0.94 &  0.62 &  0.11 & 0.00 \\
     2 &  1 &  0  & -0.94 & -0.11 &  0.62 & 0.00 \\
     2 & -1 &  0  & -0.94 &  0.11 &  0.62 & 0.00 \\
     \hline
 \end{tabular}
    \label{tab:MnW001_exchange_parameters}
\end{table}

\section{Correlated electronic structure}

\begin{figure}
\centering
 \includegraphics[width=0.95\columnwidth]{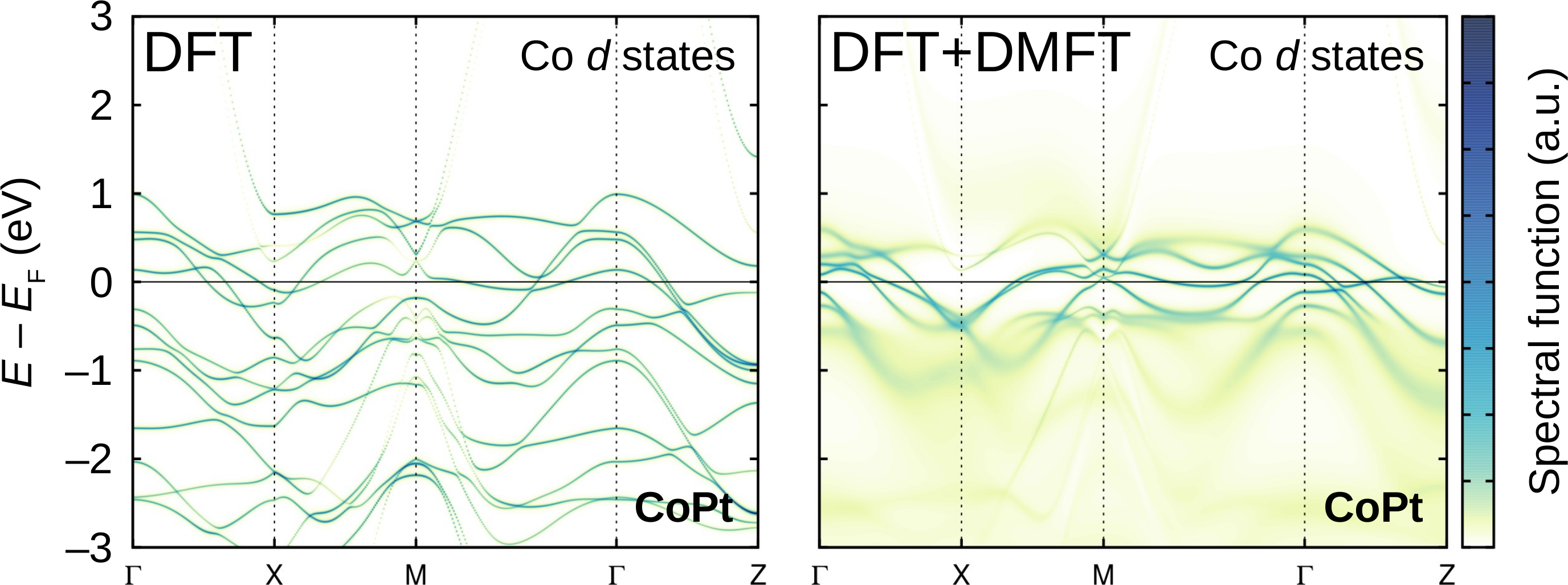}\\ \vspace{10pt}
 \includegraphics[width=0.95\columnwidth]{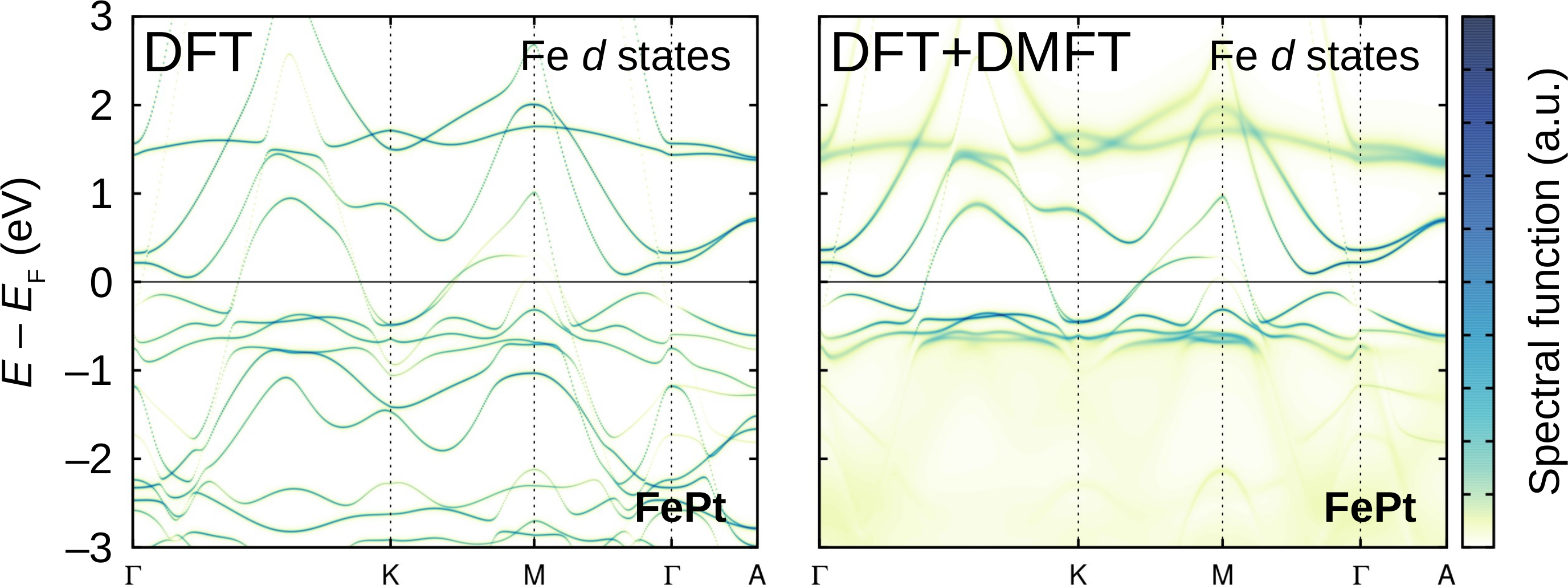}
 \caption{The electronic band structure/spectral function of the CoPt (upper plots) and FePt (lower plots) compounds, obtained from DFT and DFT+DMFT with the SPTF solver ($U=\unit[2.3]{eV}$ for Co \textit{d} states, $U=\unit[1.0]{eV}$ for Fe \textit{d} states and $J_\text{H} = 0.9\:\text{eV}$ in both cases).}
\label{f:CoPt_and_FePt_bands}
\end{figure}

\begin{figure}
\centering
 \includegraphics[width=0.95\columnwidth]{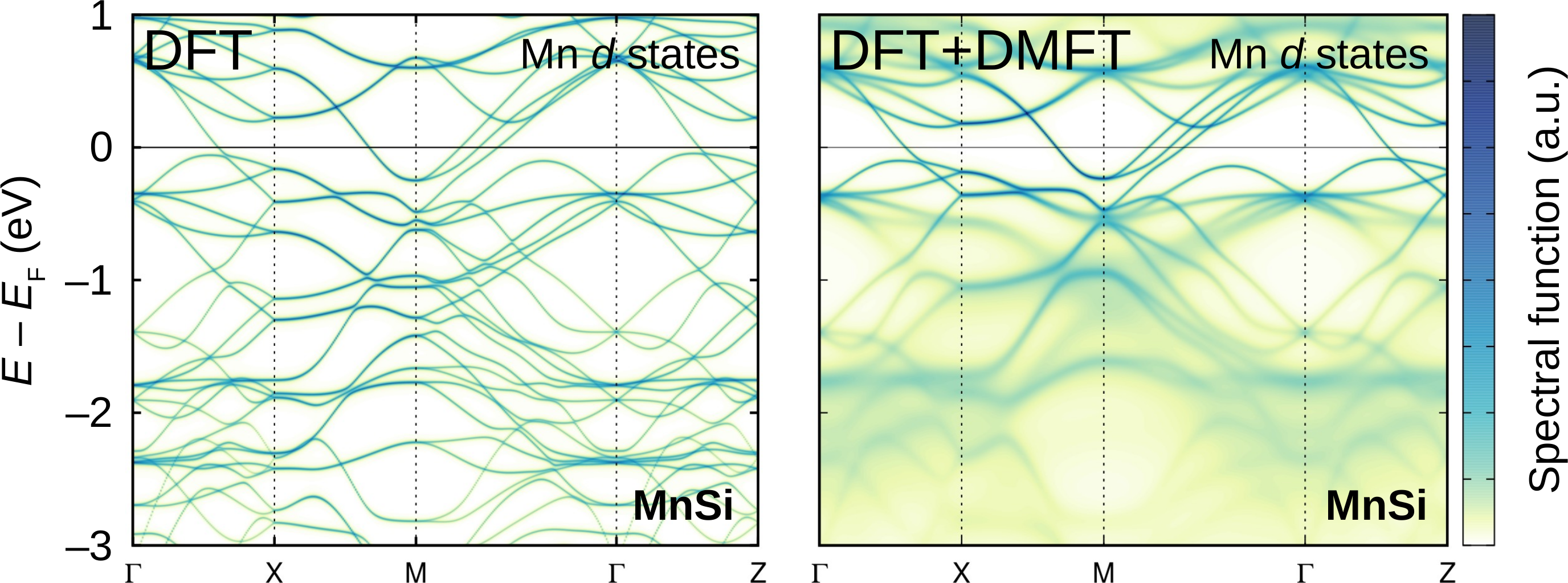}\\ \vspace{10pt}
 \includegraphics[width=0.95\columnwidth]{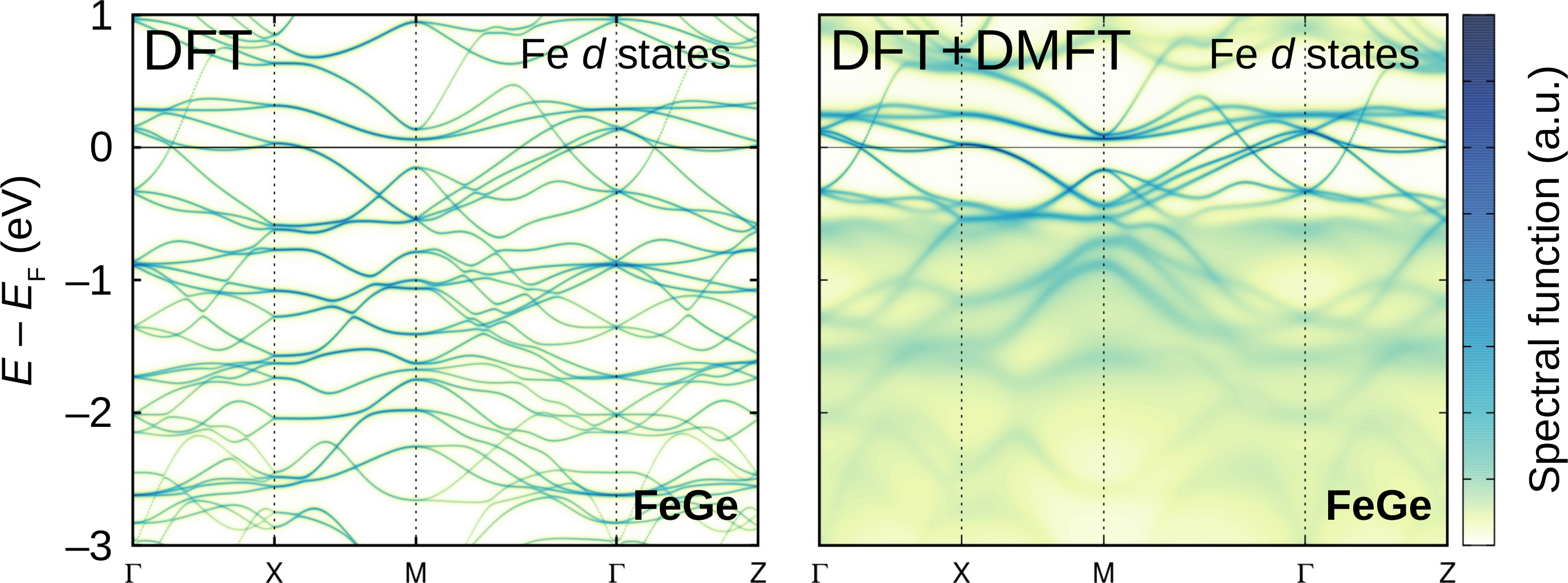}
 \caption{The electronic band structure/spectral function of the B20 compounds MnSi (upper plots) and FeGe (lower plots) obtained from DFT and DFT+DMFT with the SPTF solver ($U=1.5\:\text{eV}$ and $J_\text{H} = 0.9\:\text{eV}$ for Mn/Fe \textit{d} states).}
\label{f:B20_bands}
\end{figure}

\begin{figure}
\centering
 \includegraphics[width=0.95\columnwidth]{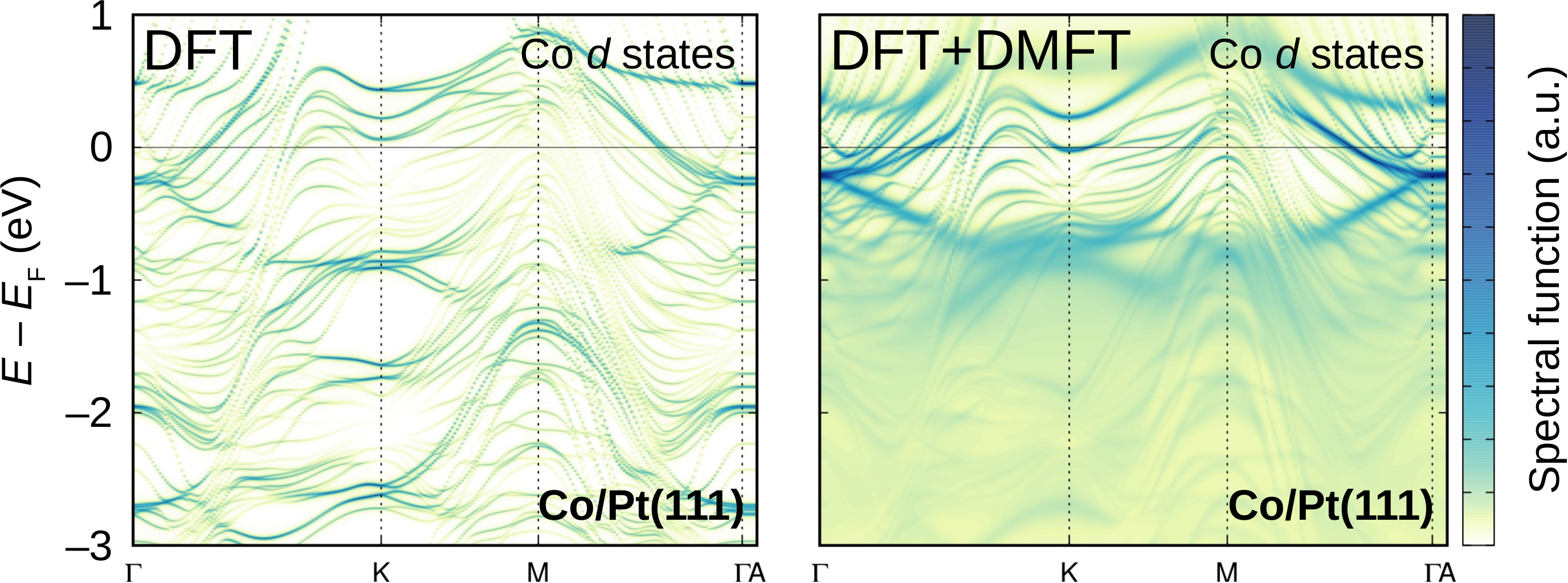}\\ \vspace{10pt}
 \includegraphics[width=0.95\columnwidth]{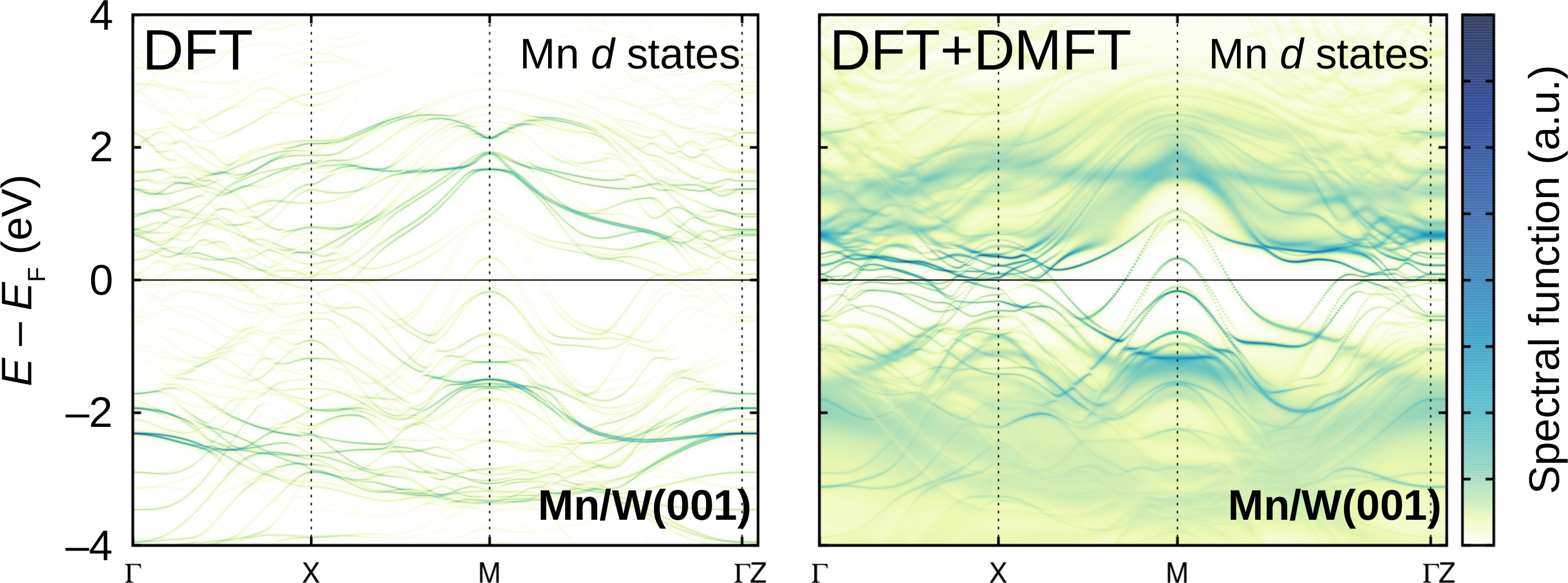}
 \caption{The electronic band structure/spectral function of the Co monolayer on the Pt(111) surface (upper plots) and Mn monolayer on the W(001) surface (lower plots) obtained from DFT and DFT+DMFT with the SPTF solver ($U=2.0\:\text{eV}$ and $J_\text{H} = 0.9\:\text{eV}$ for Co/Mn \textit{d} states).}
\label{f:3d_5d_bilayer_bands}
\end{figure}

For some of the studied systems, e.g., CoPt alloy and B20 compounds, we observe non-trivial variations of the magnetic exchange parameters as functions of the correlation strength $U$. More specifically, for moderate values of $U$ the exchange parameters reach a local maximum or minimum, which is in contrast to simpler trends of, e.g., magnetic insulators that are governed by superexchange, with the well known $1/U$-behavior. For metallic systems, the exchange interaction is typically more complicated,\cite{PhysRevLett.116.217202} but if possible, it is desirable to understand the origin of the aforementioned trends of how the exchange interactions depend on the correlation strength. We therefore analyzed the effect of correlations on the electronic states near the Fermi level, and for this reason we display the spectral functions in Figs.~\ref{f:CoPt_and_FePt_bands}-\ref{f:3d_5d_bilayer_bands} for each relevant system for $U=0$ and $U=U_0$. Here $U_0$ is chosen to give the local maximum or minimum in the exchange interactions, $J$ and/or $D$. For example, $U_0=1.5\:\text{eV}$ for the B20 compounds, as apparent from Fig.~\ref{f:B20_compounds}. Overall, reduction of the bandwidth and broadening of the transition metal $d$ states further away from the Fermi level are observed due to the electronic correlations, especially for the Co-based systems, CoPt and Co/Pt(111). For Co/Pt(111) (top plots in (Fig.~\ref{f:3d_5d_bilayer_bands})), there are even new features that emerge at the Fermi level, when $U$ is increased, specifically at the K-point. This is significant, since this is the energy region relevant for at least the long-ranged magnetic exchange interactions. Comparably large band renormalizations and broadening effects induced by correlations are observed for another low-dimensional system of Mn/W(001) (bottom plots in Fig.~\ref{f:3d_5d_bilayer_bands}). For the compounds with the B20 structure, we find that MnSi and FeGe react strongly to electron correlations, which is apparent from the correlation-induced broadening of the electronic states, e.g., $\unit[1-3]{eV}$ below the Fermi level. The significant impact of the dynamical correlations on these two systems is also reflected in the calculated magnetic exchange parameters (Fig.~\ref{f:B20_compounds}c,f).

\section{Magnetic moments vs correlations}

\begin{figure}[h]
\centering
 \includegraphics[width=0.49\columnwidth]{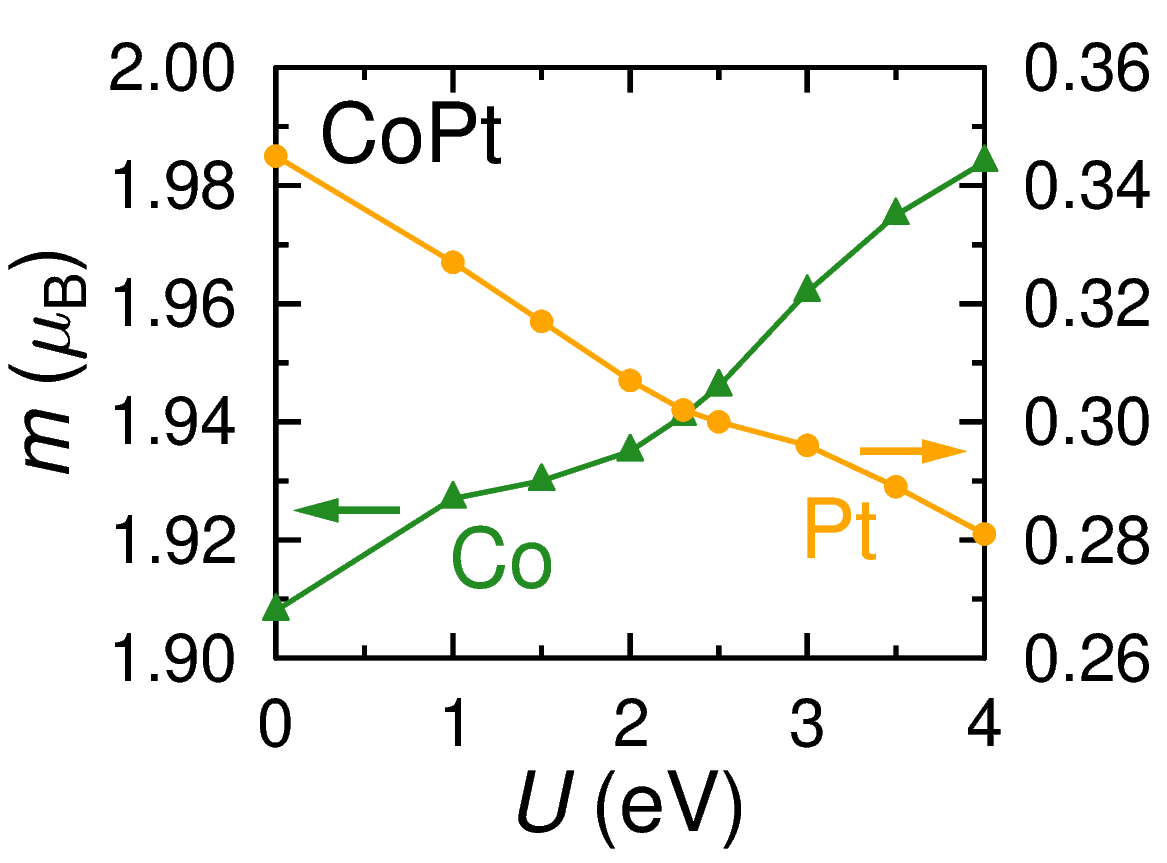}
 \includegraphics[width=0.49\columnwidth]{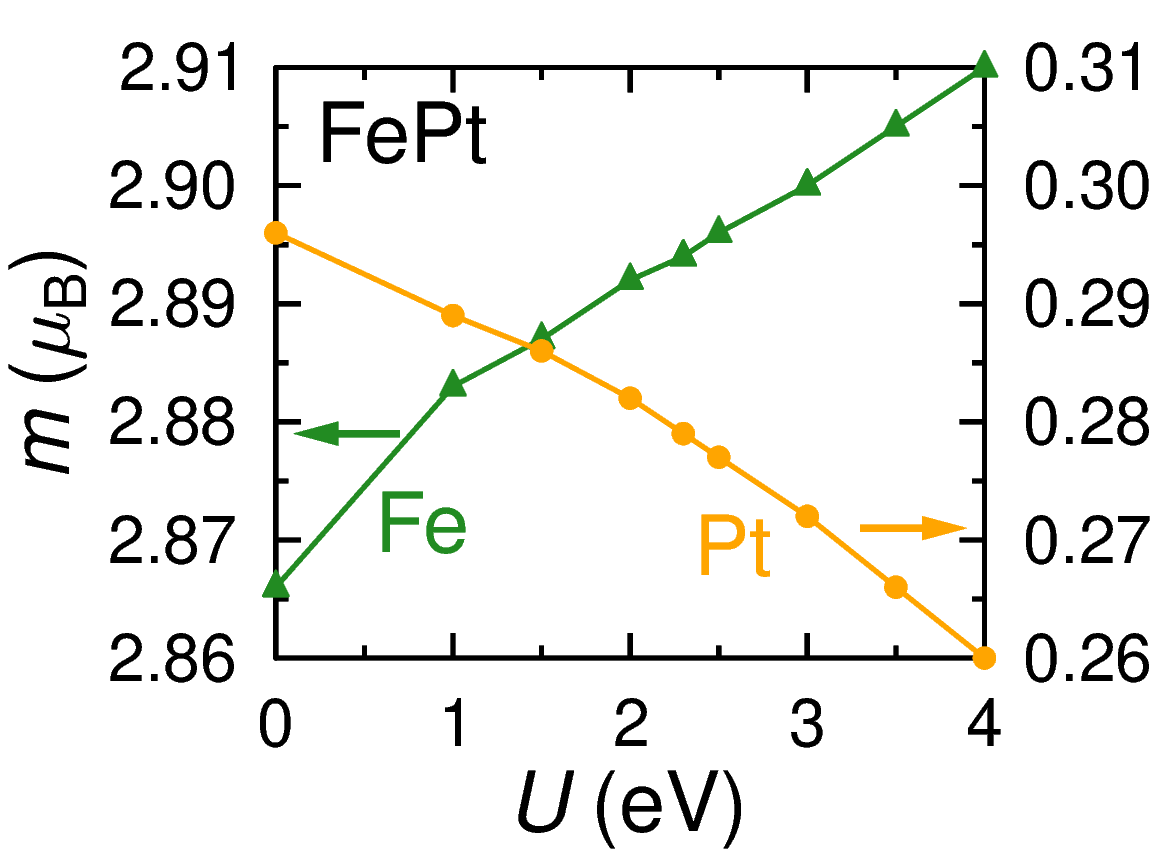}
 \vspace{-20pt}
 \caption{The site-resolved spin magnetic moments vs the correlation strength $U$ for the CoPt ($M\!\parallel\!x$) and FePt ($M\!\parallel\!z$) alloys, with the optimized structure, obtained from DFT+DMFT ($J_\text{H} = \unit[0.9]{eV}$).}
\label{f:CoPt_and_FePt_moments_vs_U}
\end{figure}

\begin{figure}[h]
\centering
 \includegraphics[width=0.49\columnwidth]{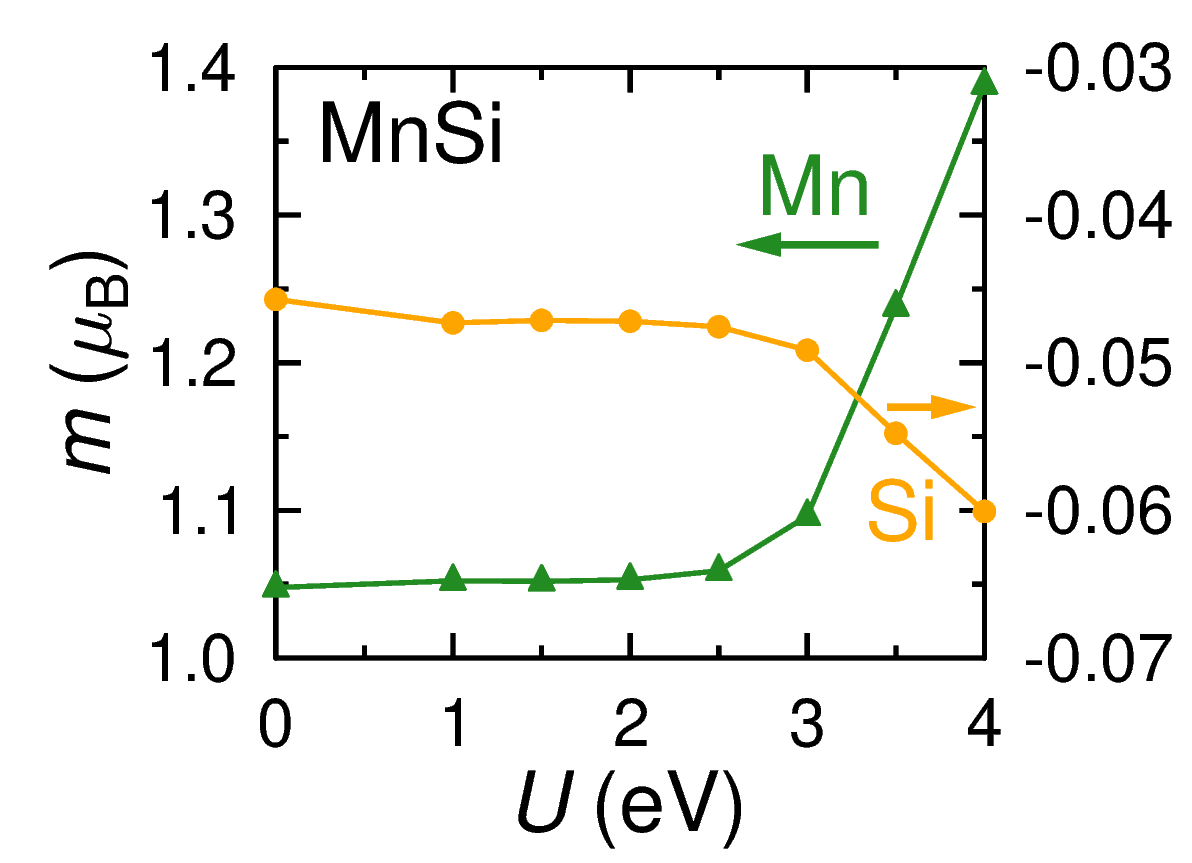}
 \includegraphics[width=0.49\columnwidth]{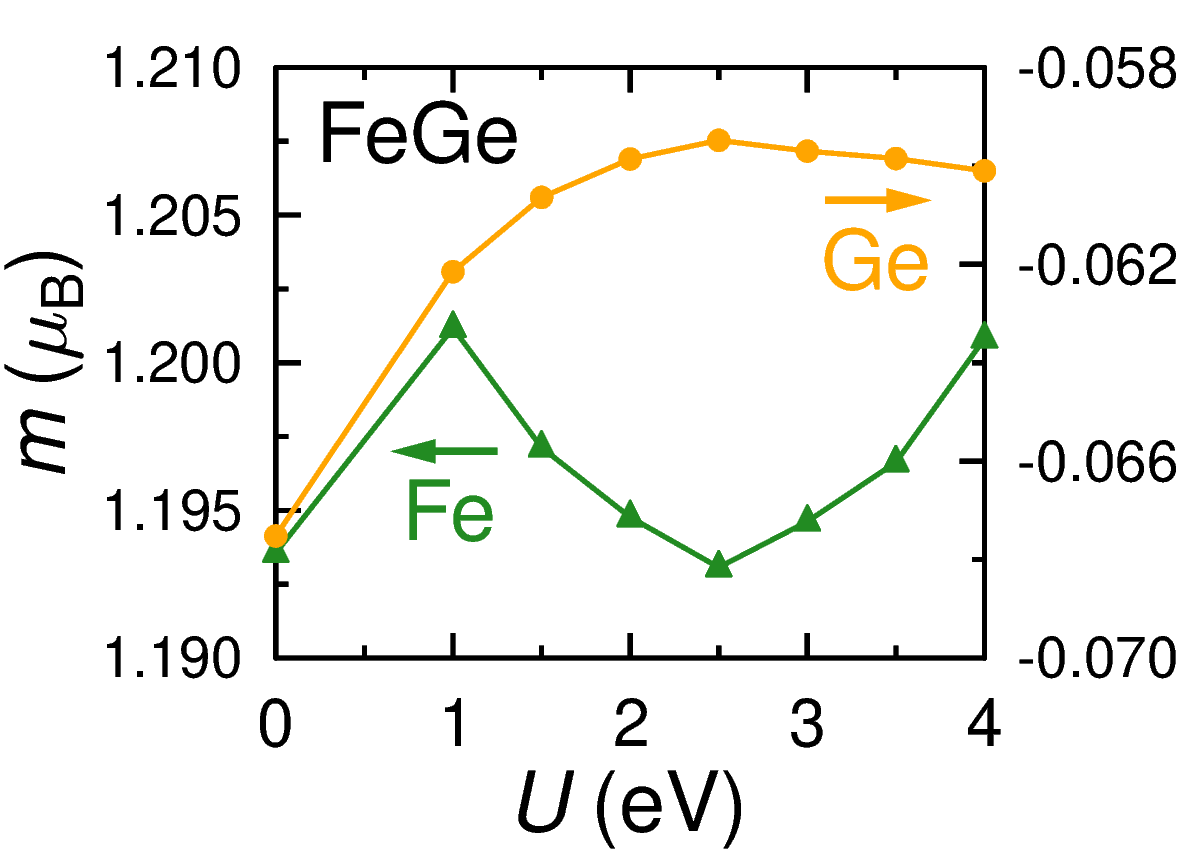}
 \vspace{-20pt}
 \caption{The site-resolved spin magnetic moments vs the correlation strength $U$ for the B20 compounds MnSi and FeGe ($M\!\parallel\!z$) obtained from DFT+DMFT ($J_\text{H} = \unit[0.9]{eV}$).}
\label{f:MnSi_and_FeGe_moments_vs_U}
\end{figure}

\begin{figure}[h]
\centering
 \includegraphics[width=0.49\columnwidth]{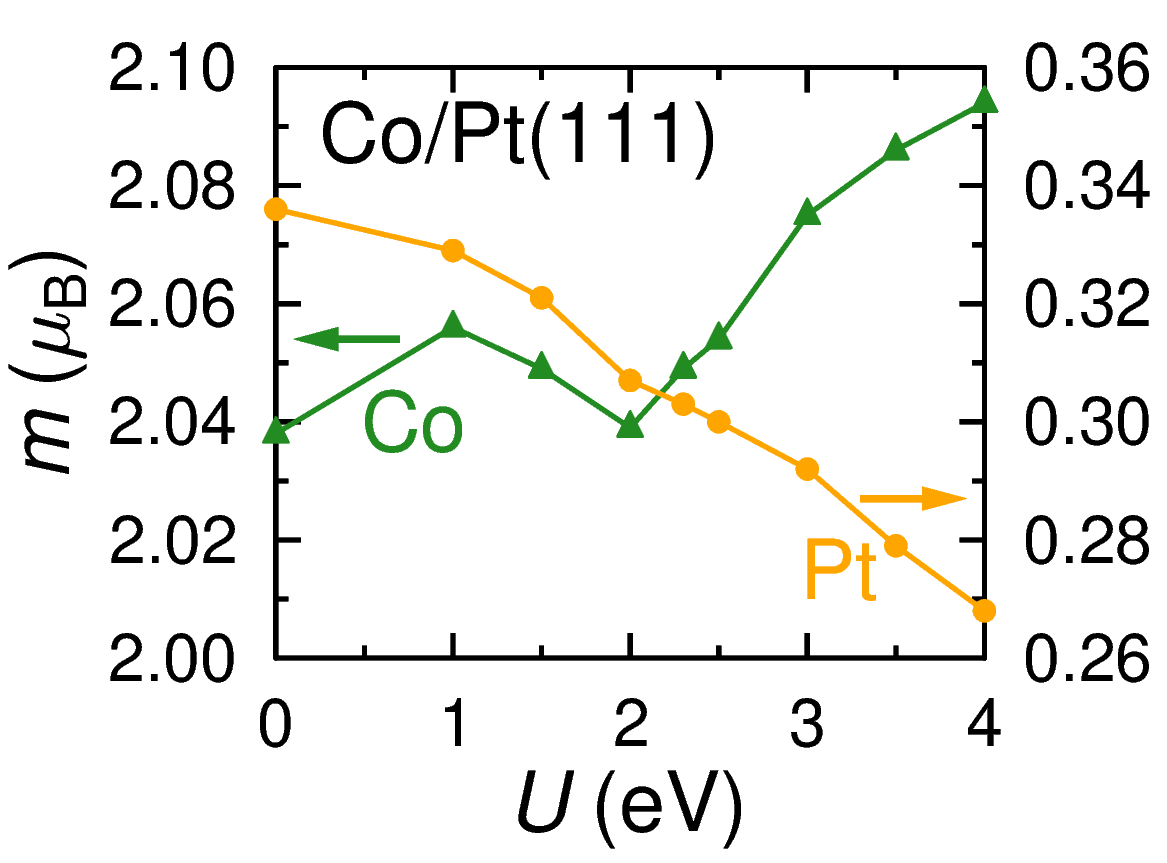}
 \includegraphics[width=0.49\columnwidth]{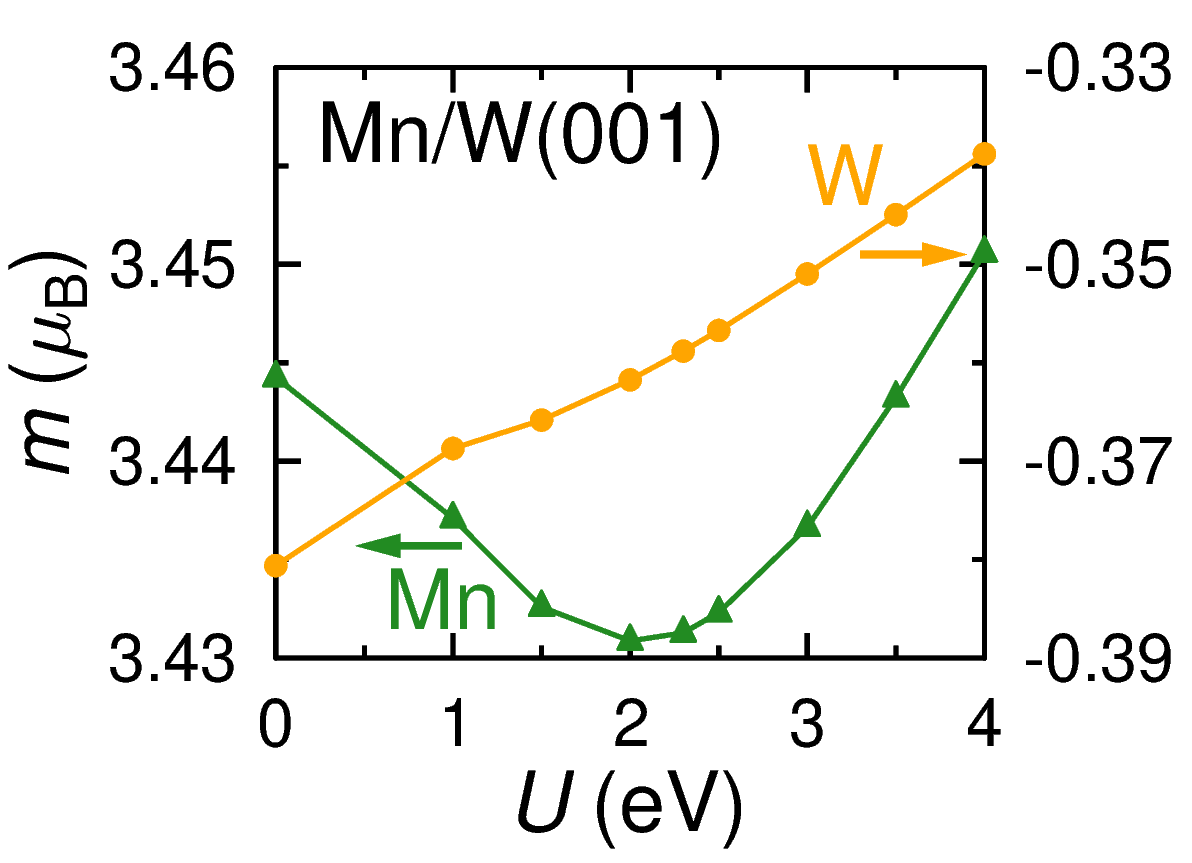}
 \vspace{-20pt}
 \caption{The site-resolved spin magnetic moments vs the correlation strength $U$ for the Co monolayer on the Pt(111) surface ($M\!\parallel\!z$) and the Mn monolayer on the W(001) surface ($M\!\parallel\!x$) obtained from DFT+DMFT ($J_\text{H} = \unit[0.9]{eV}$).}
\label{f:CoPt111_MnW001_moments_vs_U}
\end{figure}

The influence of electronic correlations on the magnitude of the spin magnetic moments of the constituting atoms in the studied bulk and low-dimensional systems is depicted in Figs.~\ref{f:CoPt_and_FePt_moments_vs_U}--\ref{f:CoPt111_MnW001_moments_vs_U}. In most cases, the dominating spin moment is enhanced by correlations and the induced moment on the neighboring sites is reduced, although some deviations from this trend are observed, for example, for the B20 FeGe system and 3$d$/5$d$ transition metal bilayers. In the later systems, the moment variations are, however, quite small.

The magnetic moment in DFT based calculations is the result of a balance between kinematic effects, where band formation favours spin-degenerate states, and the exchange-correlation energy, that favours spin-pairing, e.g., as discussed in Ref.~\onlinecite{Eriksson2017}. The effect of correlations is typically to reduce the electronic bandwidth, which would result in correlation-enhanced moments. However, a concomitant effect is to reduce the exchange splitting, which reduces the moment. Hence, correlation-driven trends of the magnetic moments are less clear-cut. As an example we note that in Ref.~\onlinecite{rspt-dmft}, using similar methods as used in the present paper, correlation effects in bulk \textit{bcc} Fe and \textit{hcp} Co were found to only marginally influence the saturation moment. In the calculations presented here, the atomic site with dominating spin moment (Mn, Fe or Co) has an increased value of the moment, when $U$ is increased, suggesting that the correlation-induced bandwidth reduction dominates.

\section{Symmetric anisotropic exchange}

\begin{figure}[h]
\centering
 \includegraphics[width=0.49\columnwidth]{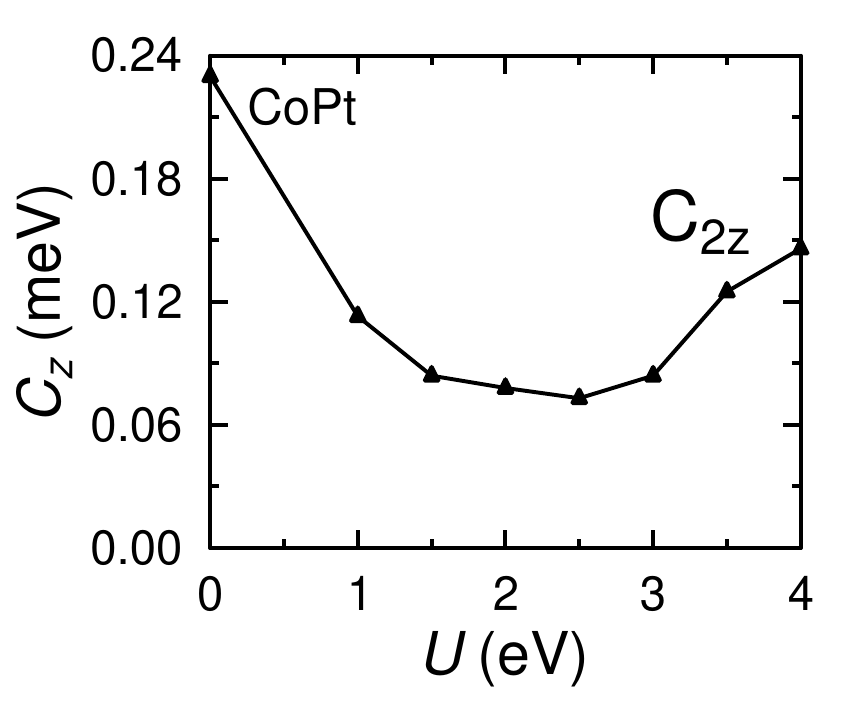}
 \includegraphics[width=0.49\columnwidth]{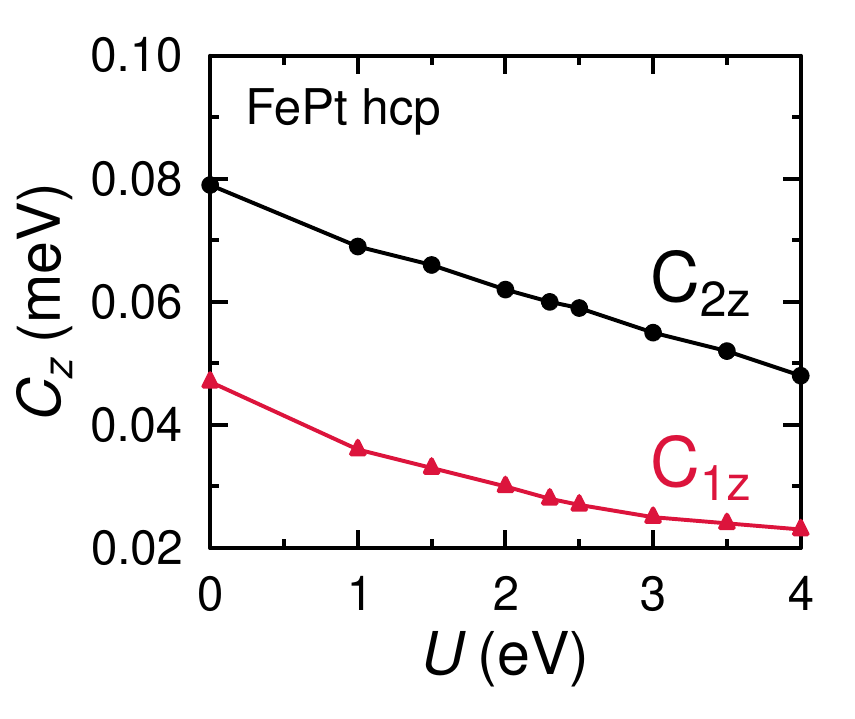}
 \vspace{-20pt}
 \caption{The $z$-component of the symmetric anisotropic exchange vs the correlation strength $U$ for the CoPt and FePt alloys, with the optimized structure, obtained from DFT+DMFT ($J_\text{H} = \unit[0.9]{eV}$). For CoPt, $C_{2z}$ corresponds to the NNN bonds within the Co $xy$-planes. For FePt, $C_{1z}$ and $C_{2z}$ correspond to the NN and NNN bonds within the Fe layers.}
\label{f:CoPt_FePt_Cz_optimized}
\end{figure}

\begin{figure}[h]
\centering
 \includegraphics[width=0.49\columnwidth]{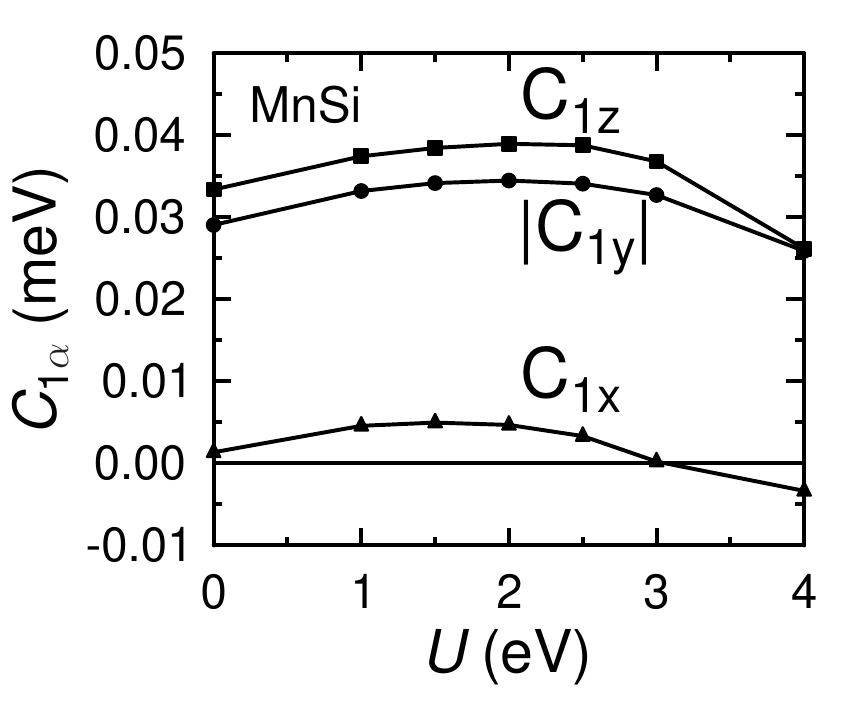}
 \includegraphics[width=0.49\columnwidth]{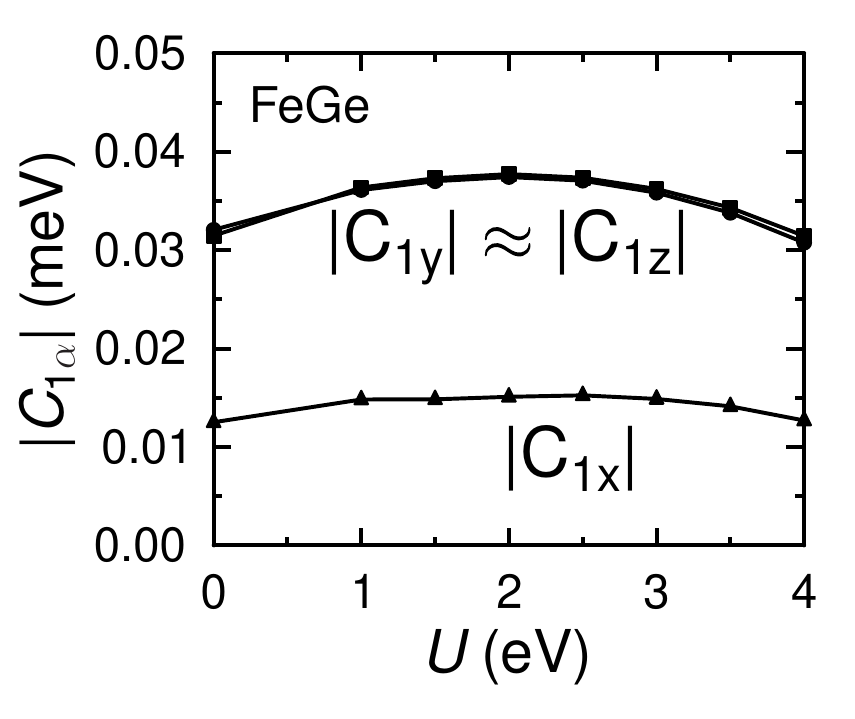}
 \vspace{-20pt}
 \caption{The $x$-, $y$- and $z$-components of the symmetric anisotropic exchange vs the correlation strength $U$ for one of the nearest-neighbor (NN) bonds in MnSi and FeGe obtained from DFT+DMFT ($J_\text{H} = \unit[0.9]{eV}$). The components of this exchange for other NN bonds are obtained by a 120$^o$-rotation around the [111] axis.}
\label{f:MnSi_FeGe_C1}
\end{figure}

\begin{figure}[h]
\centering
 \includegraphics[width=0.49\columnwidth]{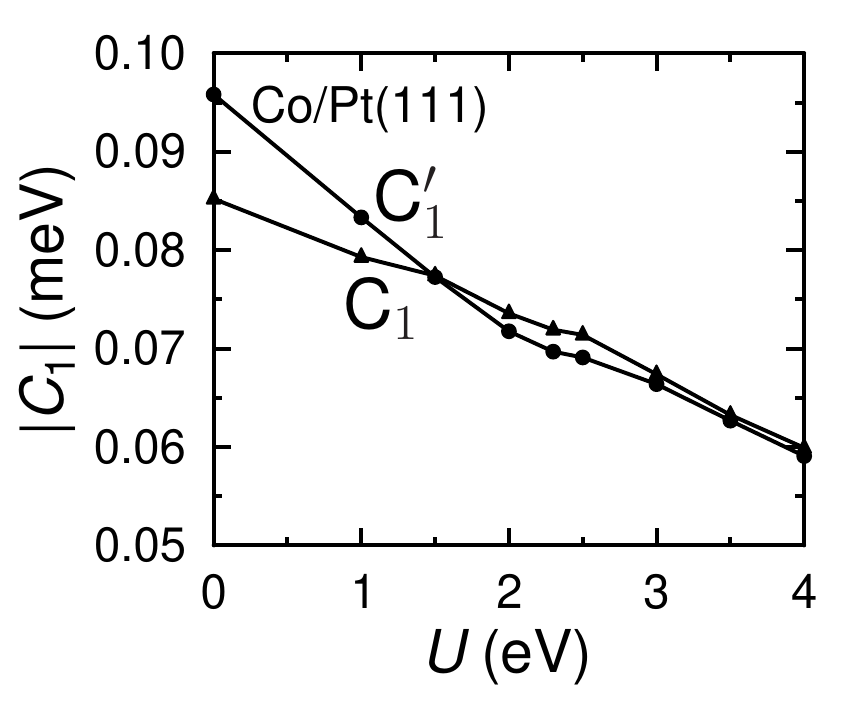}
 \includegraphics[width=0.49\columnwidth]{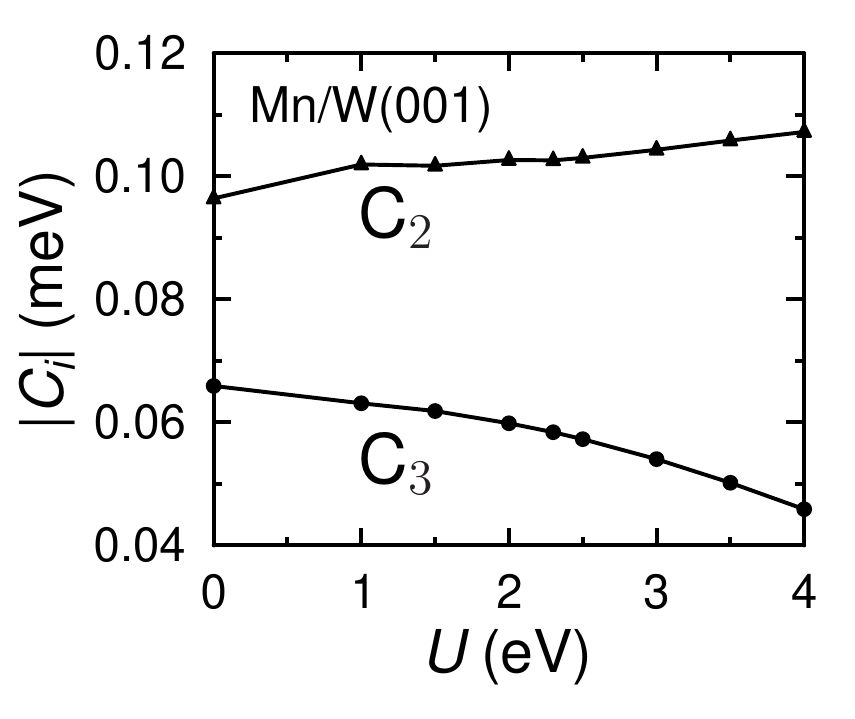}
 \vspace{-20pt}
 \caption{The magnitude of the symmetric anisotropic exchange vs the correlation strength $U$ for selected bonds in Co/Pt(111) and Mn/W(001) obtained from DFT+DMFT ($J_\text{H} = \unit[0.9]{eV}$). For Co/Pt(111), the $C_1$ parameter is for the NN bonds with $\Delta\vec{r} = \pm \vec{a}_1$ and $C'_1$ is for the other four NN bonds. For Mn/W(001), $C_2$ interaction for the in-plane NNN bonds and $C_3$ for further neighbors with $\Delta\vec{r}=\pm\vec{a}_1\pm2\vec{a}_2$ and $\Delta\vec{r}=\pm2\vec{a}_1\pm\vec{a}_2$ are shown. These interactions only have a $z$-component.}
\label{f:CoPt111_MnW001_C}
\end{figure}

Apart from the antisymmetric anisotropic exchange (DM interaction), we have also calculated the symmetric anisotropic interactions $(C_x, C_y, C_z)$, defined by Eq.~(\ref{e:C_definition}), for all here investigated systems and analyzed their sensitivity to the electronic correlations (Figs.~\ref{f:CoPt_FePt_Cz_optimized}--\ref{f:CoPt111_MnW001_C}). These are the anisotropic exchange interactions that may result in bond-dependent interactions of the Heisenberg-Kitaev Hamiltonian,\cite{Jackeli2009} that has been frequently discussed for materials holding quantum spin-liquids. Figure~\ref{f:CoPt_FePt_Cz_optimized} shows the $C_z$ for a few nearest-neighbor bonds for the artificial CoPt and FePt alloys, respectively. The considered $C_z$ in CoPt is five times smaller than the DM interaction but follows the same $U$-dependence (Fig.~\ref{f:CoPt}c). For FePt, on the other hand, we find a monotonically decreasing $C_z$ which is, on average, one order of magnitude weaker than the DM interaction. In case of the B20 compounds MnSi and FeGe (Fig.~\ref{f:MnSi_FeGe_C1}), the NN symmetric anisotropic exchange $C_1$ is the only non-negligible interaction of that kind but is still one order of magnitude smaller than the DM interaction. Interestingly, the components of $C_1$ in these B20 compounds are transformed in the same way, as the components of the DM interaction. The transformation corresponds to a 120$^\text{o}$-rotation which is equivalent to a cyclic permutation of the $x$, $y$ and $z$ components. The calculated $C_1$ interaction shows only small variations as a function of correlation strength $U$, revealing a maximum around $U=\unit[2]{eV}$. For the low-dimensional systems Co/Pt(111) and Mn/W(001), the symmetric anisotropic exchange changes basically monotonically due to electronic correlations (Fig.~\ref{f:CoPt111_MnW001_C}) and more pronounced changes are observed for the Co-based system.

\section{Orbital-resolved analysis of CoPt}

\begin{figure}
\centering
 \includegraphics[width=0.99\columnwidth]{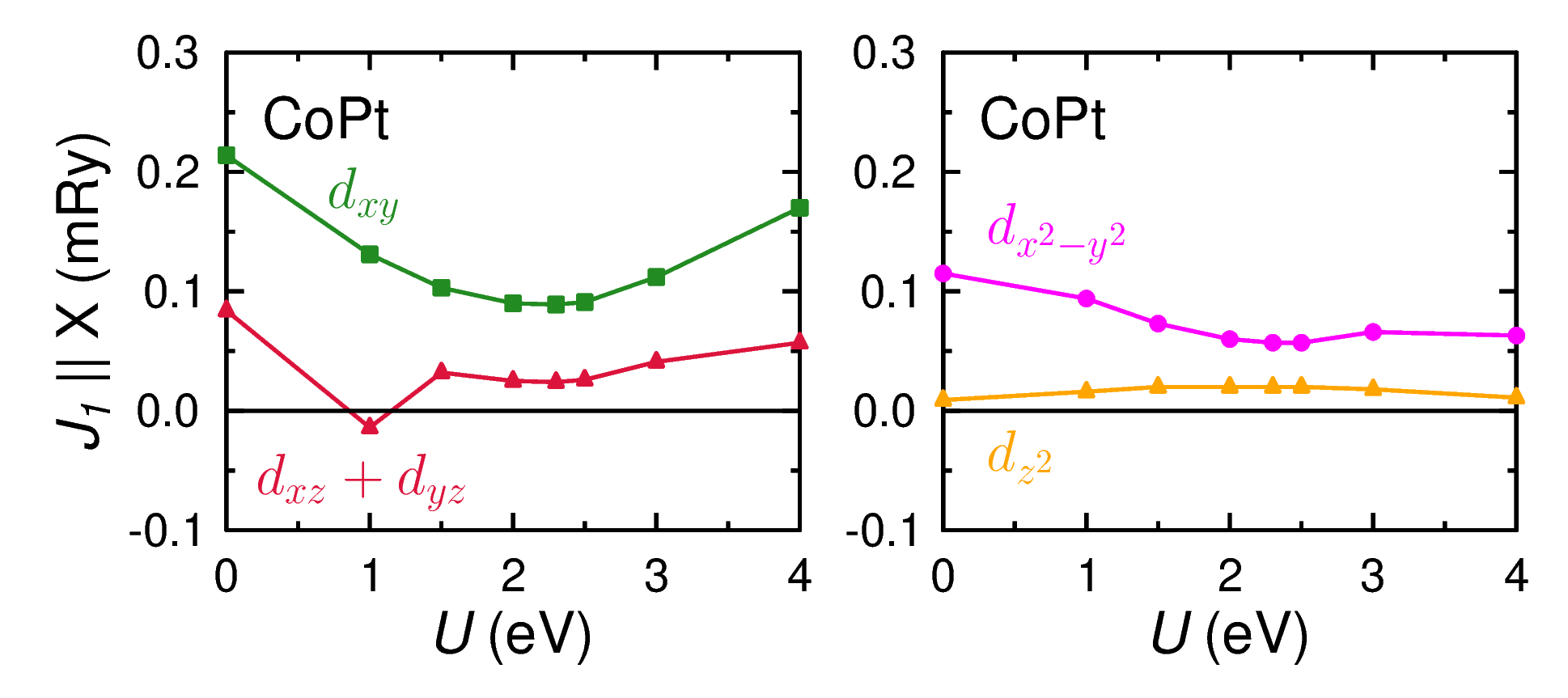}
 \caption{Orbital-resolved Heisenberg exchange interaction for the nearest neighbors in CoPt along the $x$ direction as a function of the correlation strength $U$.
 }
\label{f:CoPt_Jij_orbital_resolved} 
\end{figure}

In this section, we show how a physical intuition for the observed non-trivial changes of the exchange parameters in response to electronic correlations can be obtained by analyzing the individual orbital contributions, on the example of the CoPt bulk alloy. This is a particularly instructive example due to the non-trivial behaviour of the exchange parameters with respect to $U$ (Fig.~\ref{f:CoPt}c). The results for the orbital-decomposed nearest-neighbour exchange are plotted in Fig.~\ref{f:CoPt_Jij_orbital_resolved} where correlation effects for the \textit{d} orbitals are clearly visible. Since these orbitals have different angular distribution and contribute differently to each band of the electronic structure, it is natural that they also contribute in a unique way to the exchange interaction parameters, as is clear from Fig.~\ref{f:CoPt_Jij_orbital_resolved}. It appears that the non-trivial behavior of the Heisenberg exchange in this system (Fig.~\ref{f:CoPt}c) is mostly due to contributions from the $t_{2g}$ orbitals. In general, the interplay between different orbitals depends a lot on the specific system. On the other hand, the orbital decomposition of the DM interaction is more complicated due to the mixing between the spin-up and spin-down orbitals and will be addressed in a future work.

\begin{figure}
\centering
 \includegraphics[width=0.99\columnwidth]{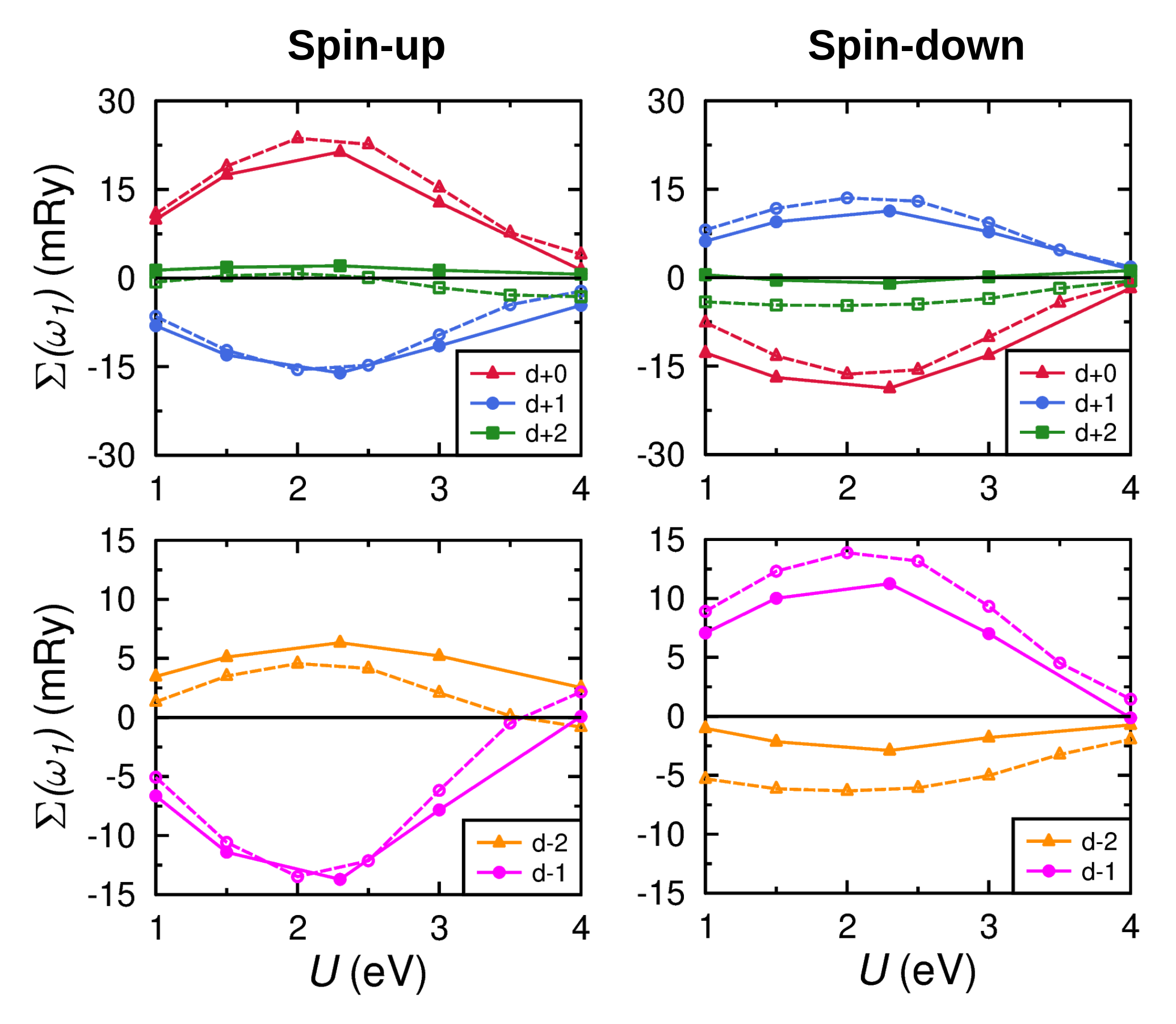}
 \caption{Orbital-resolved real part of the self-energy in CoPt at the first Matsubara frequency as a function of the correlation strength $U$. The diagonal components of $\Sigma(\omega_1)$, in the basis of complex spherical harmonics $d_i$ ($i=-2\ldots+2$ is the magnetic quantum number), are plotted for the ferromagnetic configurations along $x$ (solid lines) and $z$ (dashed lines) directions.
 }
\label{f:CoPt_Sigma_orbital_resolved} 
\end{figure}

It is also helpful to analyze the behavior of the self-energy in relation to the observed changes of the magnetic interactions. In Fig.~\ref{f:CoPt_Sigma_orbital_resolved}, the real part of the self-energy $\Sigma(\omega)$, evaluated at the first Matsubara frequency, is shown for different values of parameter $U$. The double-counting correction is included in these plots, assuring that the self-energy would be zero in the limit $U=0$. One may observe that the diagonal components of the self-energy (decomposed into complex spherical harmonics) shown in Fig.~\ref{f:CoPt_Sigma_orbital_resolved} resemble closely the behavior of the Heisenberg exchange parameters for the \textit{d} orbitals (Fig.~\ref{f:CoPt_Jij_orbital_resolved}). Furthermore, based on the unnumbered equations on page 3 of the manuscript, one can say that the DM interaction depends on the behavior of the self-energy projected on two perpendicular directions, for example, the $x$ and $z$ directions (solid and dashed lines in Fig.~\ref{f:CoPt_Sigma_orbital_resolved}). The qualitative behavior of the self-energy is similar in these two cases, although some of the components ($d_{-2}$ and $d_{+2}$ in Fig.~\ref{f:CoPt_Sigma_orbital_resolved}) show a pronounced anisotropy in the self-energy.

\begin{figure}
\centering
 \includegraphics[width=0.99\columnwidth]{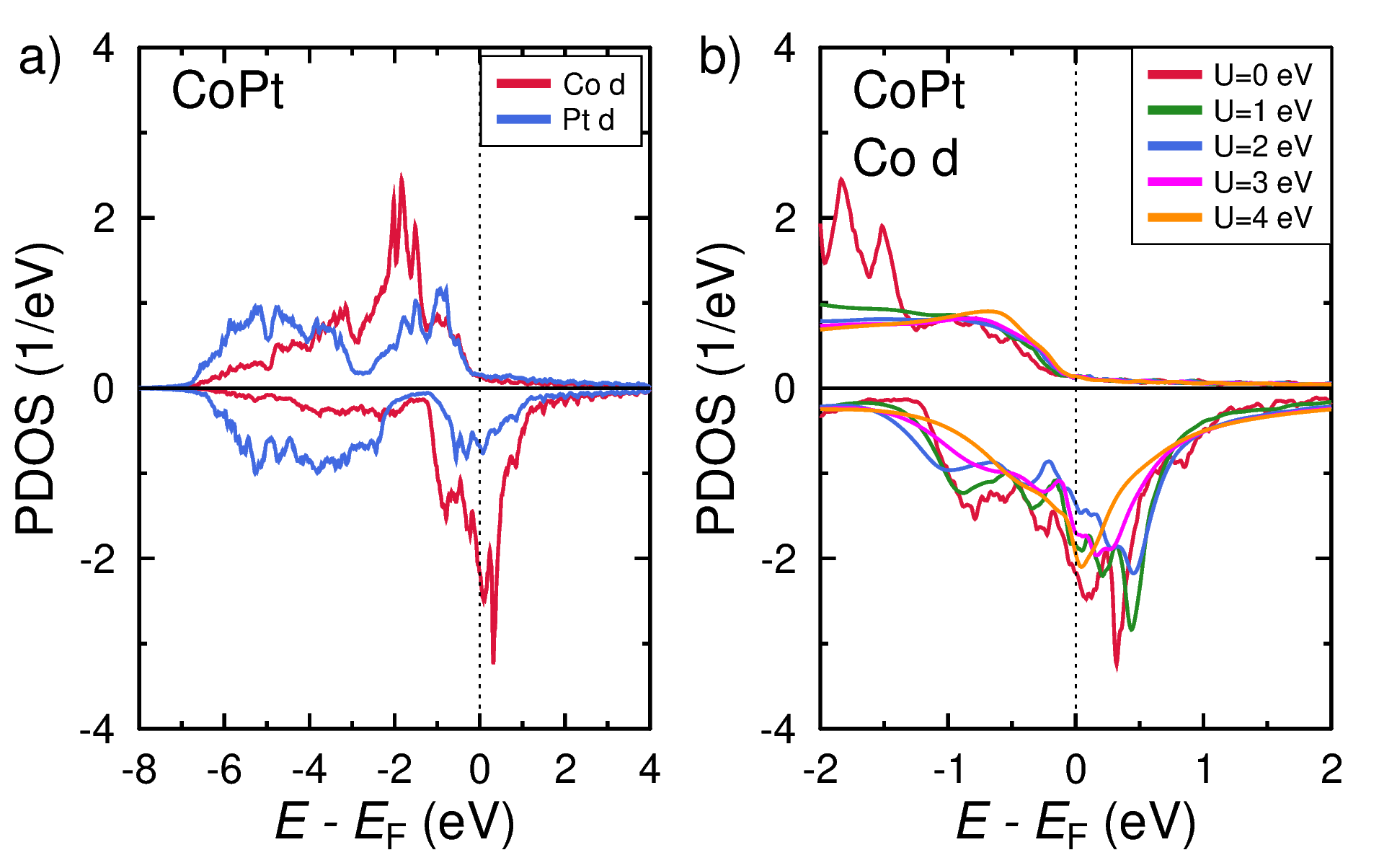}
 \caption{Partial density of states of a) Co \textit{d} and Pt \textit{d} spin-up (positive DOS) and spin-down (negative DOS) orbitals in CoPt for $U=0$ and b) of Co \textit{d} states for different values of the correlation strength $U$.
 }
\label{f:CoPt_PDOS_vs_U}
\end{figure}

Finally, we analyse the partial density of states of the Co 3\textit{d} orbitals, which dominates near the Fermi level (Fig.~\ref{f:CoPt_PDOS_vs_U}a), and how it evolves as a function of correlation strength $U$ (Fig.~\ref{f:CoPt_PDOS_vs_U}b). These data reveal a general broadening of spectral features with increasing value of $U$ as well as a band narrowing for $U > \unit[2]{eV}$. Both of these features are expected when dynamical correlations become more important and, since this band narrowing is accompanied by a reduced exchange splitting, the reduction of interatomic exchange with increasing $U$ value (shown e.g. in Fig.~\ref{f:CoPt}c) is natural. Interestingly, in the range $0 < U < \unit[2]{eV}$, the exchange splitting seems to increase. This is most clearly seen in Fig.~\ref{f:CoPt_PDOS_vs_U}b, where the spin-minority peak at $\unit[0.3]{eV}$ for $U=0$ shifts towards $\unit[0.5]{eV}$ for $U=\unit[1]{eV}$ and $\unit[2]{eV}$. This initial increase in the exchange splitting explains the increasing trend of $J$ and $D$ shown in Fig.~\ref{f:CoPt}c.

Overall, the results in Fig.~\ref{f:CoPt_PDOS_vs_U}b as well as in Fig.~\ref{f:CoPt}c represent a delicate non-linear balance between kinematic (band) effects and the on-site Coulomb repulsion ($U$).

\clearpage

\bibliographystyle{prb-titles}
\bibliography{main}

\end{document}